\documentclass[paper]{JHEP3} 
\usepackage{epsfig,multicol,bbm,amsmath} 
\usepackage{booktabs}
\usepackage{cite}
\newcommand{\gsim}{\buildrel>\over{_\sim}}
\newcommand{\anc}{\rule{0mm}{0mm}}
\newcommand{\befig}{\begin{figure}}
\newcommand{\efig}{\end{figure}}
\newcommand{\betab}{\begin{table}}
\newcommand{\etab}{\end{table}}
\newcommand{\barray}{\begin{array}}
\newcommand{\earray}{\end{array}}
\newcommand{\bit}{\begin{itemize}}
\newcommand{\eit}{\end{itemize}}
\newcommand{\ben}{\begin{enumerate}}
\newcommand{\een}{\end{enumerate}}
\newcommand{\bee}{\begin{equation}}
\newcommand{\eee}{\end{equation}}
\newcommand{\bea}{\begin{eqnarray}}
\newcommand{\eea}{\end{eqnarray}}
\newcommand{\benn}{\begin{displaymath}}
\newcommand{\eenn}{\end{displaymath}}
\newcommand{\beann}{\begin{eqnarray*}}
\newcommand{\eeann}{\end{eqnarray*}}
\newcommand{\nn}{\nonumber}
\newcommand{\Order}{{\cal O}}   
\newcommand{\bk}{\mbox{\boldmath $k$}}
\newcommand{\meV}{\mathrm{meV}}
\newcommand{\eV}{\mathrm{eV}}
\newcommand{\keV}{\mathrm{keV}}
\newcommand{\MeV}{\mathrm{MeV}}
\newcommand{\GeV}{\mathrm{GeV}}
\newcommand{\TeV}{\mathrm{TeV}}
\newcommand{\K}{\mathrm{K}}
\newcommand{\Mpc}{\mathrm{Mpc}}
\newcommand{\km}{\mathrm{km}}
\newcommand{\fb}{\mathrm{fb}}
\newcommand{\seconds}{\mathrm{s}}

\newcommand{\mZ}{\ensuremath{m_{\mathrm{Z}}}}

\newcommand{\refeq}[1]{(\ref{#1})} 
\newcommand{\reffigure}[1]{Figure~\ref{#1}}
\newcommand{\reffigures}[1]{Figures~\ref{#1}}
\newcommand{\reffig}[1]{Fig.~\ref{#1}}
\newcommand{\reffigs}[1]{Figs.~\ref{#1}}
\newcommand{\refsec}[1]{Sec.~\ref{#1}}
\newcommand{\refsecs}[1]{Secs.~\ref{#1}}

\newcommand{\reftable}[1]{Table~\ref{#1}}
\newcommand{\MPl}{\mathrm{M}_{\mathrm{Pl}}}
\newcommand{\fPQ}{f_{\mathrm{a}}}

\newcommand{\Reheating}{\mathrm{R}}
\newcommand{\TR}{T_{\Reheating}}
\newcommand{\TNR}{T_{\mathrm{nr}}}
\newcommand{\Tf}{T_{\mathrm{f}}}

\newcommand{\eq}{\mathrm{eq}}

\newcommand{\had}{\mathrm{had}}
\newcommand{\tot}{\mathrm{tot}}

\newcommand{\gravitino}{{\widetilde{G}}}
\newcommand{\neutralinoone}{{\widetilde{\chi}^0_1}}
\newcommand{\axino}{{\widetilde a}}
\newcommand{\Bino}{{\widetilde B}}
\newcommand{\gluino}{{\widetilde g}}
\newcommand{\slepton}{\ensuremath{\tilde{l}}}
\newcommand{\sleptonone}{\ensuremath{\tilde{l}_1}}

\newcommand{\selectronR}{\ensuremath{\widetilde{\mathrm{e}}_{\mathrm{R}}}}
\newcommand{\selectronL}{\ensuremath{\widetilde{\mathrm{e}}_{\mathrm{L}}}}
\newcommand{\sel}{\ensuremath{\tilde{\mathrm{e}}}}
\newcommand{\electron}{\ensuremath{\mathrm{e}^-}}
\newcommand{\positron}{\ensuremath{\mathrm{e}^+}}

\newcommand{\smuonR}{\ensuremath{\widetilde{\mu}_{\mathrm{R}}}}
\newcommand{\smu}{\ensuremath{\tilde{\mu}}}
\newcommand{\stau}{{\widetilde \tau}}
\newcommand{\stauone}{{\widetilde \tau}_1}
\newcommand{\stautwo}{{\widetilde \tau}_2}
\newcommand{\stauR}{\widetilde{\tau}_{\mathrm{R}}}
\newcommand{\stauL}{\widetilde{\tau}_{\mathrm{L}}}
\newcommand{\st}{\ensuremath{\tilde{\tau}}}
\newcommand{\rhostau}{\ensuremath{\rho_{\tilde{\tau}}}}
\newcommand{\thetastau}{\ensuremath{\theta_{\tilde{\tau}}}}
\newcommand{\fstau}{\ensuremath{f_{\tilde{\tau}}}}

\newcommand{\ax}{\ensuremath{\tilde{a}}}

\newcommand{\Bi}{\ensuremath{\tilde{B}}}
\newcommand{\mgravitino}{m_{\widetilde{G}}}
\newcommand{\maxino}{m_{\widetilde{a}}}
\newcommand{\mstau}{m_{\widetilde{\tau}}}
\newcommand{\mbino}{m_{\widetilde{B}}}
\newcommand{\mtau}{m_{\tau}}
\newcommand{\EM}{\mathrm{em}}
\newcommand{\HAD}{\mathrm{had}}

\newcommand{\NLSP}{\mathrm{NLSP}}
\newcommand{\NTP}{\mathrm{NTP}}
\newcommand{\TP}{\mathrm{TP}}

\newcommand{\X}{\mathrm{WDM}}

\newcommand{\FS}{\mathrm{FS}}

\newcommand{\dm}{\mathrm{dm}}
\newcommand{\CDM}{\mathrm{dm}}
\newcommand{\WDM}{\mathrm{WDM}}
\newcommand{\OmegaM}{\Omega_{\mathrm{m}}}

\newcommand{\OmegaDM}{\Omega_{\mathrm{dm}}}
\newcommand{\DNueff}{\Delta N_{\nu}}
\newcommand{\Nueff}{N_{\nu}}
\newcommand{\rhorad}{\rho_{\mathrm{rad}}}
\newcommand{\Lisix}{^6\mathrm{Li}}
\newcommand{\Liseven}{^7\mathrm{Li}}
\newcommand{\Benine}{^9\mathrm{Be}}
\newcommand{\Hefour}{^4\mathrm{He}}
\newcommand{\Hethree}{^3\mathrm{He}}
\newcommand{\Beeight}{^8\mathrm{Be}}
\newcommand{\Deut}{\mathrm{D}}
\newcommand{\Hyd}{\mathrm{H}}
\newcommand{\qbar}{\bar{q}}
\title{Axinos in Cosmology and at Colliders}
\author{Ayres Freitas\\
        Department of Physics and Astronomy, University of Pittsburgh, PA 15260,
        USA\\
        E-mail: \email{afreitas@pitt.edu}}
\author{Frank Daniel Steffen\\
        Max-Planck-Institut f\"ur Physik,
        F\"ohringer Ring 6, D--80805 Munich, Germany\\
        E-mail: \email{steffen@mpp.mpg.de}}
\author{Nurhana Tajuddin\\
        Institut f\"ur Theoretische Physik, Universit\"at Z\"urich, 
        Winterthurerstrasse 190, CH--8057~Z\"urich, Switzerland\\
        E-mail: \email{nurhana@physik.uzh.ch}}
\author{Daniel Wyler\\
        Institut f\"ur Theoretische Physik, Universit\"at Z\"urich, 
        Winterthurerstrasse 190, CH--8057~Z\"urich, Switzerland\\
        E-mail: \email{wyler@physik.unizh.ch}}
\preprint{MPP-2011-1}
\date{\today}
\abstract{
The axino, the fermionic superpartner of the axion, is a well-motivated 
candidate for cold dark matter if it is the lightest supersymmetric 
particle. Since the axino couples very weakly to the matter multiplets, 
the next-to-lightest supersymmetric particle (NLSP) has a long lifetime, 
which has important consequences for both cosmology and collider 
phenomenology. Assuming that a charged slepton is the NLSP, we calculate 
the complete leading one- and two-loop contributions to its decay. We 
analyze in detail constraints on the parameters space from cosmology
and discuss how this scenario can be probed at colliders.
Scenarios in which both the axino and the gravitino are lighter than
the long-lived charged slepton are also explored with particular
emphasis on cosmological constraints and collider phenomenology. }
\begin{document} 

\section{Introduction}
\label{Sec:Introduction}

There are compelling reasons for physics beyond the standard model
(SM). As the experiments at the Large Hadron Collider (LHC) are taking
data, we could well see the first hints of this new physics soon.
Along with high energy experiments, cosmology provides important
testing grounds for models of new physics and their effects in the
early universe.

Among the extensions of the SM, supersymmetry (SUSY) is in some sense
the most ambitious and appealing one
\cite{Golfand:1971iw,Volkov:1973ix,Wess:1974tw}. It evokes a
completely new symmetry of nature and opens new angles on old puzzles,
such as the hierarchy problem, the unification of the fundamental
forces or even the incorporation of gravity into the SM of particle
physics.

Supersymmetry is particularly attractive in connection with the
Peccei--Quinn (PQ) mechanism~\cite{Peccei:1977hh,Peccei:1977ur}.
Relying on a global U(1)$_{\mathrm{PQ}}$ symmetry that is broken
spontaneously at a high scale $\fPQ$, this mechanism provides a compelling potential solution of the strong CP problem, which
consists of the question of why CP violation in QCD is so small.
The breaking of the PQ symmetry gives rise to the famous axion which
could provide (part of) dark
matter~\cite{Sikivie:2006ni,Kim:2008hd,Graf:2010tv} or influence other
physics because of its lightness and its weak interactions. In fact,
axion searches and a large variety of astrophysical and cosmological
considerations point to the following lower limit on the PQ
scale~\cite{Raffelt:2006cw,Nakamura:2010zzi}:
\begin{equation}
  f_a\gsim 6\times 10^8\,\GeV
  \, .
\label{Eq:f_a_axion}
\end{equation}

Because of the large breaking scale, new physics must be involved in
viable axion models and the coupling of the axion to the SM fields
must be highly suppressed. One class of such models are `hadronic
models' -- also known as KSVZ axion
models~\cite{Kim:1979if,Shifman:1979if} -- where new heavy `quark' fields are introduced that couple to
the axion. Via interactions of gluons (and possibly also of other
gauge fields) with those heavy coloured fields, loop-suppressed axion
couplings to the SM fields emerge. SUSY versions of this type of
models are the ones on which we focus in this work.

In the considered SUSY theories, there must be a fermionic partner to
the axion: the axino~\cite{Tamvakis:1982mw,Frere:1982sg,Kim:1983ia}.%
\footnote{There is also the scalar partner of the (pseudoscalar)
  axion, the saxion, whose late decays could lead to significant
  non-standard entropy
  production~\cite{Kim:1992eu,Lyth:1993zw,Chang:1996ih,Hashimoto:1998ua}.
  For the present investigation, however, we assume a standard thermal
  history and thereby implicitly a saxion mass such that saxion
  effects are negligible.}
Its mass depends on the details of the
model~\cite{Tamvakis:1982mw,Moxhay:1984am,Nieves:1985fq,Rajagopal:1990yx,Goto:1991gq,Chun:1992zk,Chun:1995hc,Asaka:1998ns,Asaka:1998xa}
and can be such that the axino is the lightest supersymmetric particle
(LSP), which is assumed in most of this work together with R-parity
conservation. In such settings, the axino can constitute a
(significant) fraction of dark
matter~\cite{Bonometto:1989vh,Rajagopal:1990yx,Bonometto:1993fx,Covi:1999ty,Covi:2001nw,Covi:2004rb,Brandenburg:2004du,Baer:2008yd,Freitas:2009fb,Strumia:2010aa}.

With the axino being lighter than the SM superpartners, the lightest
sparticle in the minimal supersymmetric standard model (MSSM) is
unstable and can thus be a charged slepton such as the lighter
stau~\cite{Covi:2004rb,Brandenburg:2005he,Steffen:2005cn,Freitas:2009fb,Freitas:2009jb}.
Such scenarios are particularly
attractive since the charged slepton would typically be long-lived.
Its existence could thus be associated with striking signatures of
tracks of heavy quasi-stable charged particles in collider
experiments. Some fraction of the charged sleptons are even expected
to get stopped in the main detectors or the surrounding material, such
that their decays can be
analysed~\cite{Goity:1993ih,Hamaguchi:2004df,Feng:2004yi,Brandenburg:2005he,Martyn:2006as,Hamaguchi:2006vu}.
Long-lived charged particles can also affect big bang nucleosynthesis
(BBN) since the decay products can reprocess primordial
nuclei~\cite{Freitas:2009jb}.  Moreover, being electrically charged,
they themselves can form primordial bound states, which may catalyse
the synthesis of $\Lisix$ and $\Benine$ in the early
universe~\cite{Pospelov:2006sc,Pospelov:2007js,Pospelov:2008ta,Freitas:2009fb}.

Given the theoretical as well as the experimental attractiveness of
axino dark matter scenarios with a long-lived charged slepton, we
present in this paper a comprehensive study of this class of models.
In~\refsec{Sec:HadronicAxionModels} we describe the considered SUSY
hadronic axion models. They typically have two scales, the weak scale
and the breaking scale of the PQ symmetry, which differ by several
orders of magnitude. This leads to possibly large logarithms at the
two-loop level which require detailed analyses.
In~\refsec{Sec:NLSPDecays} we therefore present a two-loop calculation
of the decay rate of the long-lived charged slepton and an analysis of
the various relevant decays. This is necessary to correctly interpret
cosmological constraints and collider studies of axino signatures and the PQ scale $\fPQ$.
While some of our results are presented in previous short
communications~\cite{Freitas:2009fb,Freitas:2009jb}, we provide here a
more complete systematic treatment along with details of the
calculations.
We also note that the applied methods will be useful for other
two-scale models and we have thus assembled the technical details in
the appendix.

In \refsec{Sec:AxinoConstraints} we outline the considered
cosmological settings and analyse the associated cosmological
constraints. These include constraints from the present dark matter
density, from structure formation and from primordial nucleosynthesis.
The phenomenology at the LHC and at a future linear collider such as
the International Linear Collider (ILC) is explored in
\refsec{Sec:Collider}. Here we propose and study four benchmark
scenarios in the constrained MSSM (CMSSM) extended with a hadronic
axion model. Moreover, we address prospects for analysing late charged
slepton decays, for the microscopic measurements of the PQ scale
$\fPQ$ and for distinguishing between the axino LSP and the gravitino
LSP, which is another well-motivated dark matter candidate that allows
for the existence of a long-lived charged slepton;
cf.~\cite{Steffen:2008qp} and references therein.

In \refsec{Sec:AxGrSt} we study settings in which both the
axino and the gravitino are lighter than the long-lived charged
slepton, which is still considered to be the lightest SM superpartner.
Such settings can occur naturally in gauge-mediated SUSY breaking
scenarios~\cite{Asaka:1998ns,Asaka:1998xa}. Here we also provide
results on partial decay widths and the lifetime of the long-lived
charged slepton, on the associated cosmological constraints and on the
possible impact on collider phenomenology.
Conclusions are presented in the final section.

\section{Supersymmetric Hadronic Axion Models}
\label{Sec:HadronicAxionModels}

In line with our previous
studies~\cite{Freitas:2009fb,Freitas:2009jb}, we consider the
KSVZ-type or hadronic axion models~\cite{Kim:1979if,Shifman:1979if} in
a SUSY setting; see
also~\cite{Frere:1982sg,Kim:1983ia,Asaka:1998ns,Asaka:1998xa},
or~\cite{Nilles:1981py,Tamvakis:1982mw,Kim:1983dt} for other
supersymmetrised axion models. Here new heavy quark multiplets
$Q_{1,2}$ are introduced that couple to the axion multiplet $\Phi$.
Only the new (s)quarks $Q_{1,2}$ and $\Phi$ are charged under
U(1)$_{{\rm PQ}}$. The corresponding interactions are described by the
following superpotential
\bea
        W_{{\rm PQ}} = y\Phi Q_1 Q_2
\eea
with the Yukawa coupling $y$. 
Interactions of MSSM fields -- which do not carry PQ charge in
KSVZ-type models -- with the axion multiplet $\Phi$ are mediated via
the $Q_{1,2}$ fields, which are coloured and may also carry other
charges under the SM gauge group.

While our study can easily be generalized, we focus on the case in
which $Q_{1,2}$ are SU(3)$_\mathrm{c}$ triplets and SU(2)$_\mathrm{L}$
singlets, with electrical charge $e_Q$ and hypercharge $e_Y=e_Q$, as
given in \reftable{tab:multiplets}.
%
\TABLE[t]{
\centerline{
\begin{tabular}{lcl}
\hline
chiral multiplet
& \qquad U(1)$_{{\rm PQ}}$ \qquad\qquad
& (SU(3)$_{{\rm c}}$,\,SU(2)$_{{\rm L}}$)$_{{\rm Y}}$
\\ \hline
$\Phi\,\,=\,\,\phi\,\,+\,\sqrt{2}\chi\theta+F_{\Phi}\theta\theta$
& +1    
& ($\bf{1}$,\,$\bf{1}$)$_0$
\\
$Q_1=\widetilde{Q}_1+\sqrt{2}q_1\theta+F_1\theta\theta$                 
& -1/2 
& ($\bf{3}$,\,$\bf{1}$)$_{+e_Q}$
\\
$Q_2=\widetilde{Q}_2+\sqrt{2}q_2\theta+F_2\theta\theta$                 
& -1/2 
& ($\bf{3^*}$,\,$\bf{1}$)$_{-e_Q}$
\\
\hline
\caption{The axion multiplet $\Phi$, the heavy KSVZ quark multiplets $Q_{1,2}$,
and the associated quantum numbers considered in this work. The
complex scalar $\phi = s + ia$ is a combination of the saxion $s$ and
the axion $a$.}
\label{tab:multiplets}
\end{tabular}}}
%
With the 2-component field notation in \reftable{tab:multiplets}, the
4-component fields describing the axino and the heavy KSVZ quark are
given respectively by
\bea
        \axino = \begin{pmatrix}\chi \\ \bar{\chi}\end{pmatrix}
        \quad \mbox{and} \quad
        Q =  \begin{pmatrix} q_1 \\ \bar{q}_2 \end{pmatrix} .
\eea
For the heavy KSVZ (s)quark masses, we use the SUSY limit
\bea
   M_Q = M_{\tilde{Q}_{1,2}} = y \,\langle\phi\rangle= y \, f_a/\sqrt{2} ,
\label{eq:mquark}
\eea
with both $y$ and $f_a$ taken to be real by field redefinitions. The
lower limit on the PQ scale~\eqref{Eq:f_a_axion} then implies, for
$y=\mathcal{O}(1)$, a large hierarchy between the mass of the new KSVZ
fields and the weak and the soft SUSY mass scales:
\bea
        M_{Q/\tilde{Q}_{1,2}} \gg m_Z, m_{\rm SUSY} .
\label{eq:hierarchy}
\eea

For clarification of the definition of
$f_a=\sqrt{2}\langle\phi\rangle$ in the considered models, note that
axion--gluon, axion--photon, axino--gluino--gluon and
axino--bino--photon/$Z$ boson interactions are obtained as described by
the effective Lagrangians
\begin{eqnarray}
   {\cal L}_{a gg(g)} 
   &=& 
   \frac{g_\mathrm{s}^2}{32\pi^2f_a}\,
   a\,G_{\mu\nu}^a\widetilde{G}^{a\mu\nu}
\label{Eq:Laxiongluon}\ ,
\\
   {\cal L}_{a \gamma \gamma} 
   &=& 
   \frac{e^2 C_{a \gamma \gamma}}{32\pi^2f_a}\,
   a\,F_{\mu\nu}\widetilde{F}^{\mu\nu}\ ,
\label{Eq:Laxionphoton}
\\
 {\cal L}_{\axino\gluino g(g)}
  &=& 
  i\,\frac{g_\mathrm{s}^2}{64\pi^2 f_a}\,
  \bar{\axino}\,\gamma_5
  \left[\gamma^{\mu},\gamma^{\nu}\right]\,
  \gluino^a\, G^a_{\mu\nu}\ ,
\label{Eq:L_agG}
\\
 {\cal L}_{\axino\Bino\gamma/Z}
  &=& 
  i\,\frac{(e/\cos\theta_W)^2 C_{a \mathrm{YY}}}{64\pi^2 f_a}\,
  \bar{\axino}\,\gamma_5
  \left[\gamma^{\mu},\gamma^{\nu}\right]\,
  \Bino\,\left(\cos\theta_W F_{\mu\nu}-\sin\theta_W Z_{\mu\nu}\right)\ ,
\label{Eq:L_aBgammaZ}
\end{eqnarray}
once the heavy KSVZ (s)quarks are integrated out. Here $G^a_{\mu\nu}$,
$F_{\mu\nu}$ and $Z_{\mu\nu}$ are the field strength tensors of the
gluon, photon and $Z$ boson, respectively, with the duals
$\widetilde{G}^{a}_{\mu\nu}=\epsilon_{\mu\nu\rho\sigma}G^{a\:\!\rho\sigma}/2$
and
$\widetilde{F}_{\mu\nu}=\epsilon_{\mu\nu\rho\sigma}F^{\rho\sigma}/2$,
the strong coupling $g_\mathrm{s}$, $e=\sqrt{4\pi \alpha}$, and the
Weinberg angle $\theta_W$. The weak hypercharge of the heavy KSVZ
fields $e_Q$ determines $C_{a \mathrm{YY}}=6e_Q^2$ and $C_{a \gamma
  \gamma}$, which becomes
$C_{a\gamma\gamma}=6 e_Q^2 - 2(4+z)/[3(1+z)]$
after chiral symmetry breaking, where $z=m_u/m_d\simeq 0.56$
denotes the ratio of the up and down quark masses.

While the axino mass $\maxino$ depends on the details of the
model~\cite{Tamvakis:1982mw,Moxhay:1984am,Nieves:1985fq,Rajagopal:1990yx,Goto:1991gq,Chun:1992zk,Chun:1995hc,Asaka:1998ns,Asaka:1998xa},
we simply treat it as a free parameter. When focussing on the axino
LSP case, this parameter is bounded from above by the mass of lightest
MSSM superpartner, which is considered to be the lightest charged
slepton in all of this work. For simplicity and concreteness, we focus
mainly on scenarios where this slepton is a dominantly right-handed
lighter stau: $\stauone\approx\stauR$.%
\footnote{When considering prospects for collider phenomenology in
  \refsec{sec:trappedstaus}, CMSSM benchmark scenarios are considered
  in which the lighter stau $\stauone$ has a left-handed component of 
  $\lesssim 10\%$.}
This is often a good approximation at least for small values of the
mixing angle $\tan\beta$ in the Higgs sector (which is the ratio of
the two MSSM Higgs doublet vacuum expectation values).
Moreover, while the weak scale mass hierarchy 
$\mstau\equiv m_{\stauone}<m_{\selectronR}\simeq m_{\smuonR}$ 
is typically predicted in scenarios such as the CMSSM, most of our
results can be directly transferred to the alternative cases in which
the lightest slepton is the right-handed selectron $\selectronR$ or
smuon $\smuonR$.
It will also
be straightforward but require more effort to generalize the
calculations of the decays to cases in which the lightest slepton
$\slepton_1$ has a left-handed component $\slepton_\mathrm{L}$,
including the case of a lighter stau
$\stauone=\cos\thetastau\stauL+\sin\thetastau\stauR$ with
$\cos\thetastau>0$.

For the $\slepton_1=\slepton_\mathrm{R}$ cases, the coupling of the
respective lighter charged slepton to neutralinos is dominated by the
bino coupling. Here we assume for further simplicity that the mixing
in the neutralino sector is such that one of the neutralino states is
an (almost) pure bino. In fact, the cosmological considerations and
collider studies are presented for spectra in which that state is the
lightest neutralino, $\neutralinoone=\Bino$, while the results
presented in \refsecs{Sec:NLSPDecays} and~\ref{Sec:LifetimeBRAxGrSt}
are not restricted to this case.

If SUSY is realised as a gauge symmetry~\cite{Cremmer:1982en}, also
the gravitino $\gravitino$ -- the superpartner of the graviton -- will
exist as the gauge field associated with local SUSY transformations.
Its mass $\mgravitino$ depends on the SUSY breaking mechanism and can
thus often be treated as a free parameter.
For example, $\mgravitino$ can be in the eV range in gauge-mediated
breaking schemes~\cite{Dine:1994vc,Dine:1995ag,Giudice:1998bp} and in
the GeV to TeV range in gravity-mediated breaking
schemes~\cite{Nilles:1983ge,Haber:1984rc,Martin:1997ns}.
Accordingly, the following mass orderings are possible:
(i)~$\maxino<m_{\sleptonone}<\mgravitino$,
(ii)~$\mgravitino<m_{\sleptonone}<\maxino$,
(iii)~$\maxino<\mgravitino<m_{\sleptonone}$ and
(iv)~$\mgravitino<\maxino<m_{\sleptonone}$.
In most of this work we will assume ordering~(i). Here additional
cosmological constraints may occur since gravitinos produced thermally
in the early universe~\cite{Pradler:2006qh} may decay during/after
BBN~\cite{Kohri:2005wn,Jedamzik:2006xz,Cyburt:2009pg}. In
\refsec{Sec:DistAxGrav}, case~(ii) is considered when exploring
collider prospects to distinguish between the axino LSP and the
gravitino LSP. Moreover, a dedicated treatment of the cases~(iii)
and~(iv) is presented in \refsec{Sec:AxGrSt},
where we will also study cosmological constraints associated with the
gravitino.
For those gravitino-related constraints in the cases~(i) and~(ii), we
refer, e.g., to
Refs.~\cite{Kohri:2005wn,Jedamzik:2006xz,Cyburt:2009pg}
and~\cite{Pospelov:2008ta}, respectively.

\section{Decays of Charged NLSP Sleptons into Axino Dark Matter}
\label{Sec:NLSPDecays}

In this section, we outline the calculations and present the results
of the 2-, 3- and 4-body decays of the stau NLSP into the axino. In
the particle physics setting considered in this paper, this is the
2-body decay of the right-handed stau into the tau and the axino,
$\stauR\to\tau\axino$, the 3-body decay into the tau, the axino and a
photon, $\stauR\to\tau\axino\gamma$, and the 4-body decay into the
tau, the axino and a $q\qbar$ pair, $\stauR\to\tau\axino q\qbar$.
Finally, we look at the resulting lifetime of the stau NLSP, which is
governed by the 2-body decay. After the obvious substitutions, 
the results presented in this section are equally valid for the 
alternative $\selectronR$ and $\smuonR$ NLSP cases.

\subsection{2-Body Decays of Charged NLSP Sleptons}
\label{Sec:Axino2Body}

Let us start with the 2-body decay $\stauR\to\tau\axino$.
The relevant Feynman diagrams are shown in Fig.~\ref{fig:2body2loop}.
As $m_{\tau}\ll\mstau$, we work in the limit $m_{\tau}\to 0$.

For the calculation of the two-loop diagrams, we can make use of the
large hierarchy~\eqref{eq:hierarchy} between the mass of the heavy
quark multiplet and the weak scale or the SUSY breaking scale, see
\refeq{eq:mquark}. As a result, one can perform a heavy mass expansion
(HME) in powers of $1/f_a$. Already the leading term $\propto 1/f_a$
of the amplitude provides a precise approximation since the
sub-leading terms are suppressed by many orders of magnitude. After
the HME, only vacuum two-loop integrals (i.e. integrals with zero
external momenta) and one-loop integrals remain in the expansion
coefficients, for which well-known results are available in
literature. More details about the HME of the two-loop diagrams can be
found in Appendix~\ref{Sec:HME}. The methodology was laid out 
previously in~\cite{Schilling:2005dr,Tajuddin:2010dr}.

After a lengthy calculation, we obtain the following result for the
decay rate
\begin{align}
&\Gamma(\stau_{\mathrm R}\to\tau\,\axino) = \frac{\mstau(1-A_{\ax})^2}{16\pi} \,
|A|^2, 
\label{eq:2lamp_Gamma}
\\
&A = 
\begin{aligned}[t]
&
\frac{3\alpha^2 \, e_Q^2}{8 \pi^2 \, c^4_{\rm W}}
\,
\frac{\sqrt{2}\,\mstau}{f_a}  \biggl[
3 \sqrt{A_{\Bi}} \, \log \left(\frac{y^2 f_a^2}{2\,\mstau^2}\right)
- \frac{1}{4}\sqrt{A_{\ax}}
\\
&+ \frac{2(1-A_{\Bi})[\sqrt{A_{\ax}}(1-A_{\Bi}) - \sqrt{A_{\Bi}}(1-A_{\ax})]}
{1-A_{\ax}} \log \frac{A_{\Bi}-1}{A_{\Bi}}
\\
&+ \frac{2c^2_{\rm W} \sqrt{A_{\Bi}}
(A_{\ax}-A_{\Bi})(1-\sqrt{A_{\Bi}/A_{\ax}})}{1-A_{\ax}} 
\log \frac{A_{\Bi}-A_{\ax}}{A_{\Bi}}
\\
&+ \frac{2\sqrt{A_{\ax}}(1-A_{\Bi})^2 - \sqrt{A_{\Bi}}(2-A_{\Bi})(1-A_{\ax}) +
2c^2_{\rm W} (1-A_{\Bi}) 
(A_{\ax}\sqrt{A_{\Bi}} + A_{\Bi}\sqrt{A_{\ax}})}{(1-A_{\ax})(1-A_{\Bi})}
\log A_{\Bi}
\\
&- \frac{s^2_{\rm W}(\sqrt{A_{\ax}}-\sqrt{A_{\Bi}})
[2(\sqrt{A_{\ax}}+\sqrt{A_{\Bi}})^2 - A_Z(1+\sqrt{A_{\ax}A_{\Bi}})]}{1-A_{\ax}}
\, \mstau^2 \, C_0(\mstau^2, \maxino^2, 0, \mstau^2, m_Z^2, \mbino^2)
\\
&- \frac{2c^2_{\rm W}(\sqrt{A_{\ax}}+\sqrt{A_{\Bi}})(A_{\ax}-A_{\Bi})}
{1-A_{\ax}}\, \mstau^2 \, C_0(\mstau^2, \maxino^2, 0, \mstau^2, 0, \mbino^2)
\\
&- \frac{2s^2_{\rm W}[\sqrt{A_{\ax}}(1-A_{\Bi}) - \sqrt{A_{\Bi}}(1-A_{\ax})]}
{1-A_{\ax}}\, m_Z^2 \, C_0(0, \maxino^2, \mstau^2, 0, m_Z^2, \mbino^2)
\\
&- \frac{s^2_{\rm W}\sqrt{A_{\ax}}(2 \sqrt{A_{\ax}A_{\Bi}} + 2 A_{\Bi} - A_Z)}
{1-A_{\ax}}\left(B_0(\maxino^2, m_Z^2, \mbino^2)-B_0(0,0,\mstau^2)\right)
\\
&+ \frac{s^2_{\rm W}(2\sqrt{A_{\Bi}}+2\sqrt{A_{\ax}}-\sqrt{A_{\ax}} A_Z)}
{1-A_{\ax}}\Bigl(B_0(\mstau^2, m_Z^2, \mstau^2)-B_0(0,0,\mstau^2)\Bigr)
\\
&- \frac{2[s^2_{\rm W}\sqrt{A_{\ax}}(1-A_{\Bi})-(4+c^2_{\rm
W})\sqrt{A_{\Bi}}(1-A_{\ax})]}{1-A_{\ax}}
\biggr]
\label{eq:2lamp_A}
\end{aligned}
\end{align}
with
\bea
A_{\ax} \equiv \frac{\maxino^2}{\mstau^2},  \qquad
A_{\Bi} \equiv \frac{\mbino^2}{\mstau^2},  \qquad
A_Z \equiv \frac{m_Z^2}{\mstau^2},
\eea
$c_{\rm W} \equiv \cos \theta_{\rm W}$ and $s_{\rm W} \equiv \sin
\theta_{\rm W}$.
Here $B_0$ and $C_0$ are the usual scalar one-loop self-energy and
vertex functions, respectively~\cite{'tHooft:1978xw}.  We have
arranged the result so as to make the independence of the
regularisation scale (which appears implicitly in the $B_0$ functions)
manifest.
%
\FIGURE[t]{ 
\epsfig{figure=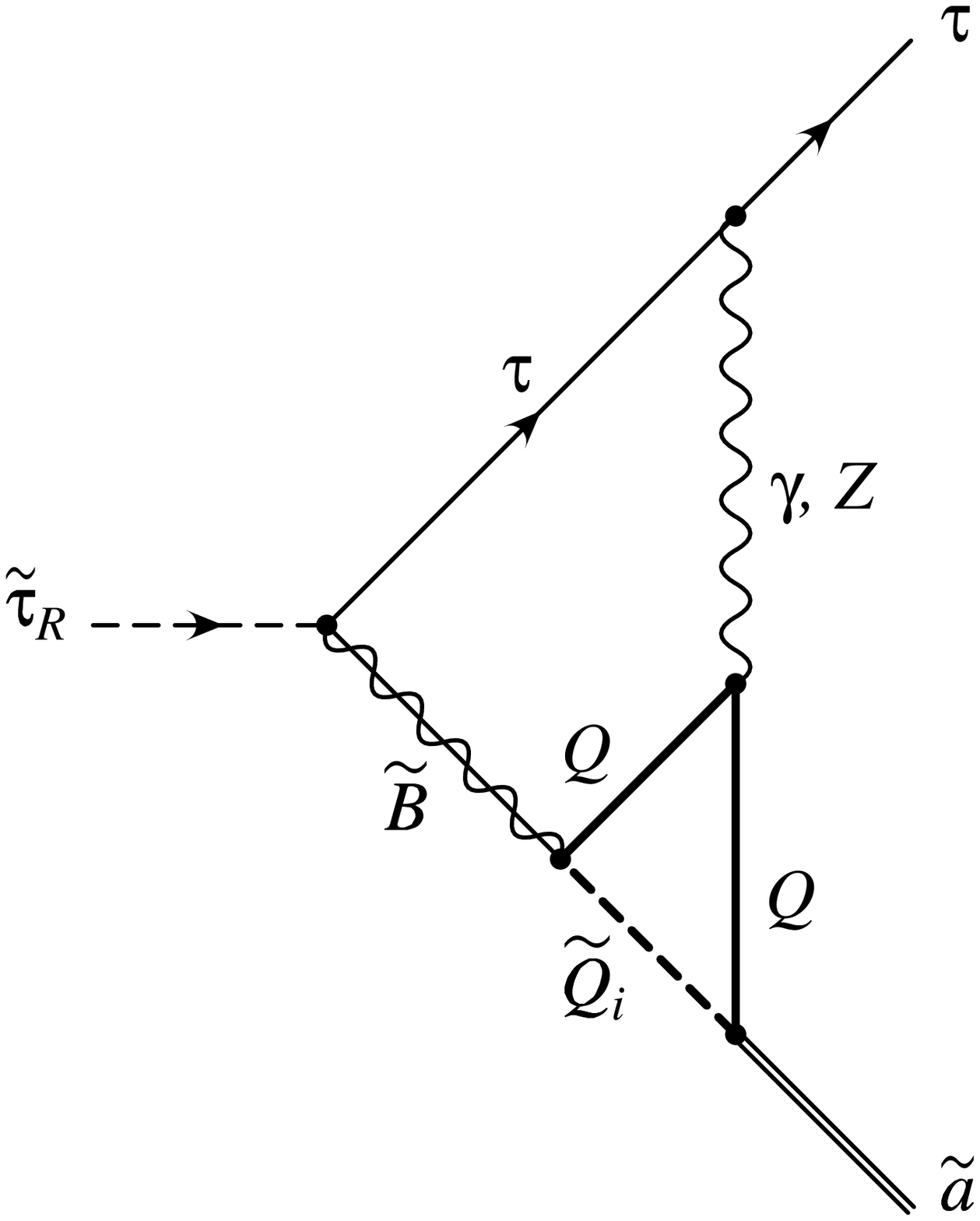, width=3.7cm}%
\epsfig{figure=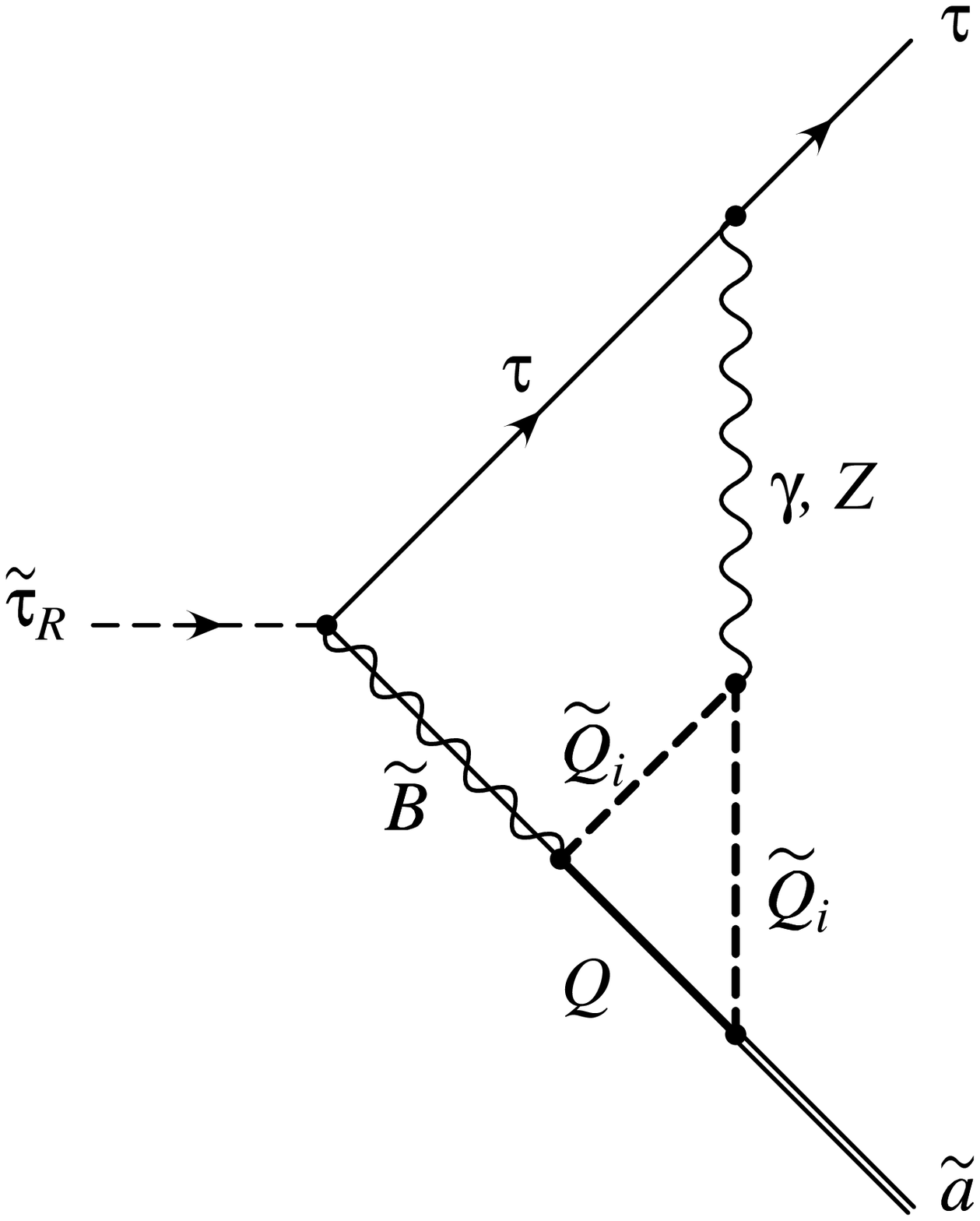, width=3.7cm}%
\epsfig{figure=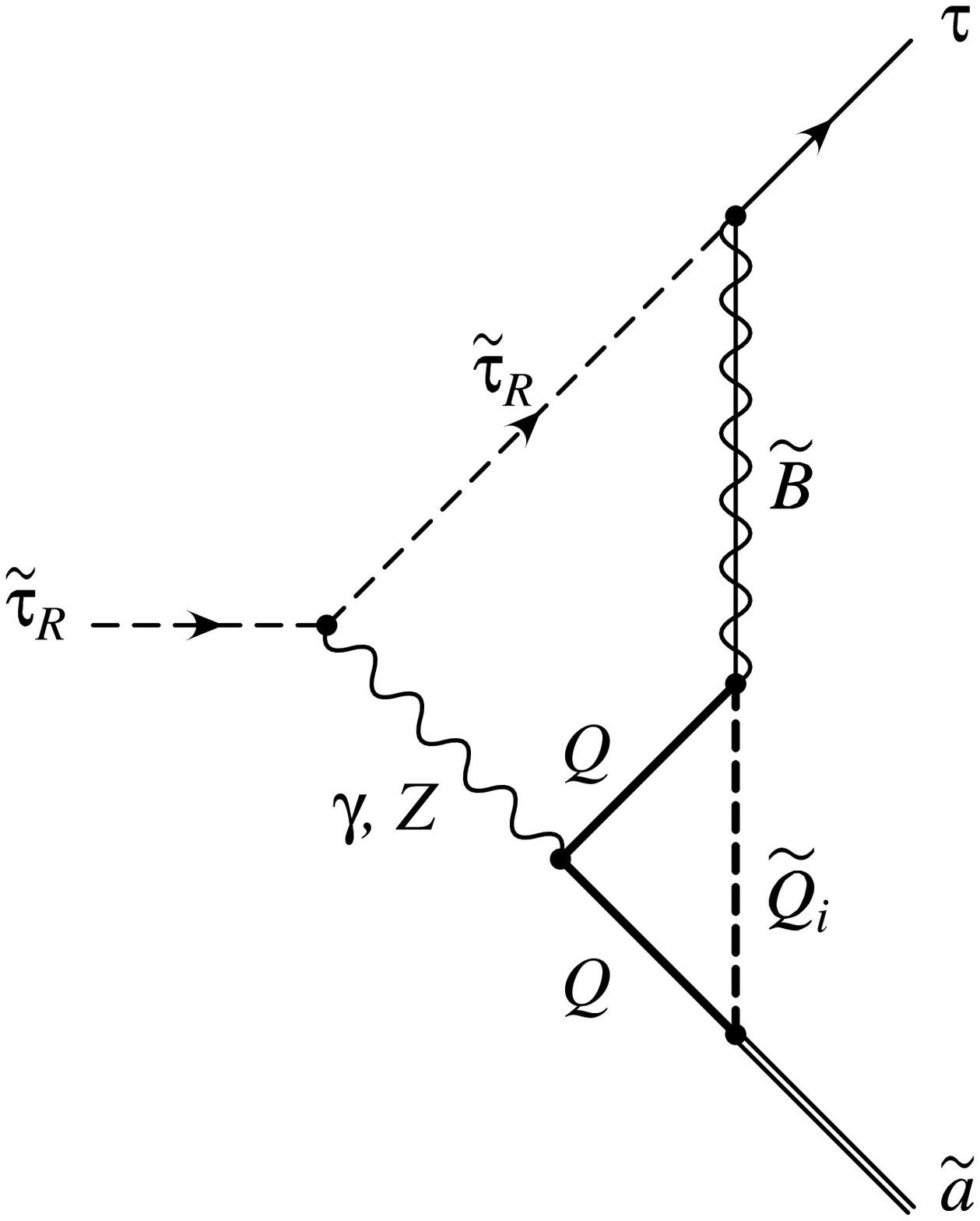, width=3.7cm}%
\epsfig{figure=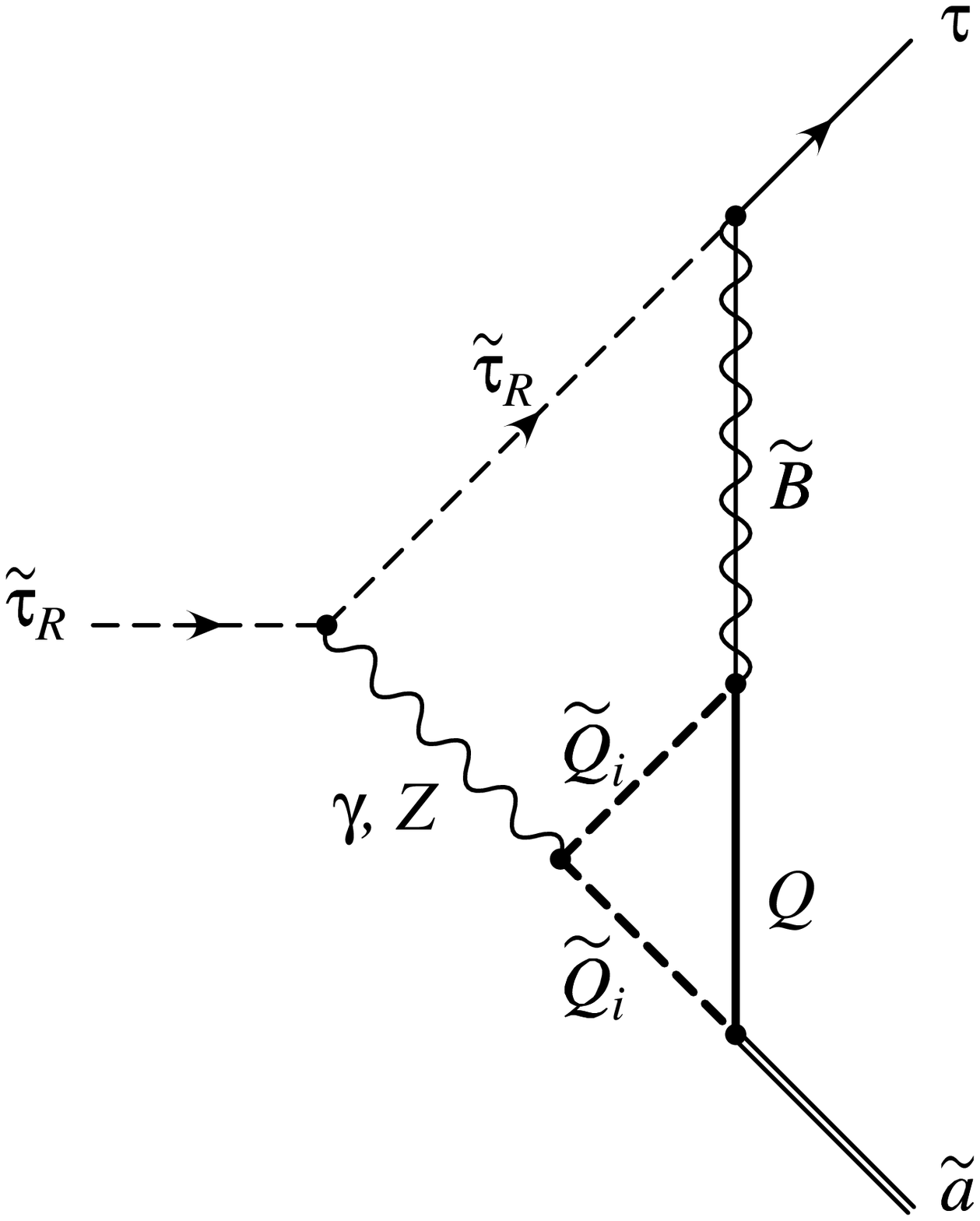, width=3.7cm}%
\caption{Feynman diagrams for the 2-body stau NLSP decay
  $\stau_{\mathrm R}\to\tau\,\axino$ in the considered type of KSVZ
  axion models. The thick solid (dashed) lines indicate heavy (s)quark propagators.}
\label{fig:2body2loop}}

A striking feature of \eqref{eq:2lamp_Gamma} is the appearance of the
logarithm term $\propto \log [y^2 f_a^2/(2\mstau^2)]$ in the first
line of \refeq{eq:2lamp_A}. For typical values, $y \approx 1$, $f_a
\sim 10^9 \dots 10^{16}$ GeV, and $\mstau \sim 100 \dots 1000$~GeV,
the logarithm becomes large, $\log (y^2 f_a^2/2\mstau^2) \sim 26 \dots
63$, and dominates the numerical result. We will further elaborate on
this aspect below, and in particular in
\refsec{Sec:StauNLSPLifetime}.

Before proceeding, let us mention the sensitivity of
$\Gamma(\stau_{\mathrm R}\to\tau\,\axino)$ to the two large scales
$f_a$ and $M_{Q/\tilde{Q}_{1,2}}$. Depending on $y$, those scales may
differ; cf.~\eqref{eq:mquark}. This is an interesting property of the
$\stauR\tau\axino$ vertex governed by the two-loop diagrams shown in
\reffig{fig:2body2loop}.
From the one-loop diagrams leading to the effective vertices described
by \eqref{Eq:Laxiongluon}--\eqref{Eq:L_aBgammaZ}, only a sensitivity
to the scale $f_a$ appears.
Phenomenological studies of both types of vertices may thereby probe
the coupling $y$ that relates the scales $f_a$ and
$M_{Q/\tilde{Q}_{1,2}}$ as can be seen in \eqref{eq:mquark}.

\subsection{3-Body Decays of Charged NLSP Sleptons}
\label{Sec:Axino3Body}

Now we turn to the 3-body decay
$\stau_{\mathrm R}\to\tau\axino\gamma$.
As before, the tau mass will be neglected throughout. The 3-body decay
occurs already at one-loop order, see \reffig{fig:3body}~(a), where
the heavy quarks and squarks run in the loop indicated by the blob.
However, the two-loop contributions shown in \reffig{fig:3body}~(b-j)
can be numerically important since they could generate a large
logarithmic factor $\log[y^2 f_a^2/(2\mstau^2)]$, similar to the one
in \eqref{eq:2lamp_Gamma}. This factor can partially compensate the
loop suppression factor $\alpha$.

\FIGURE[t]{ 
\raisebox{6mm}{\epsfig{figure=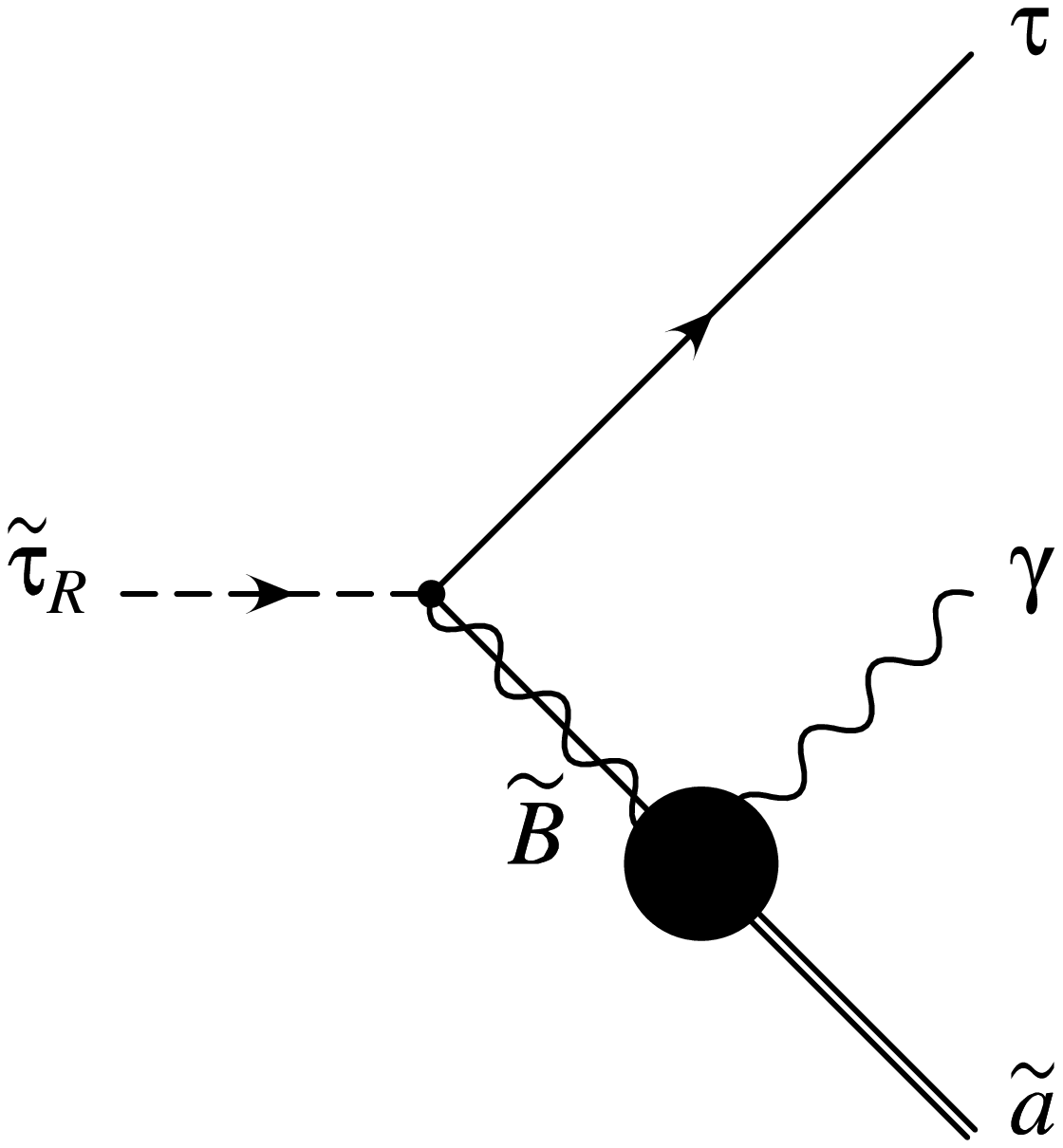, width=3cm}}\hfill
\epsfig{figure=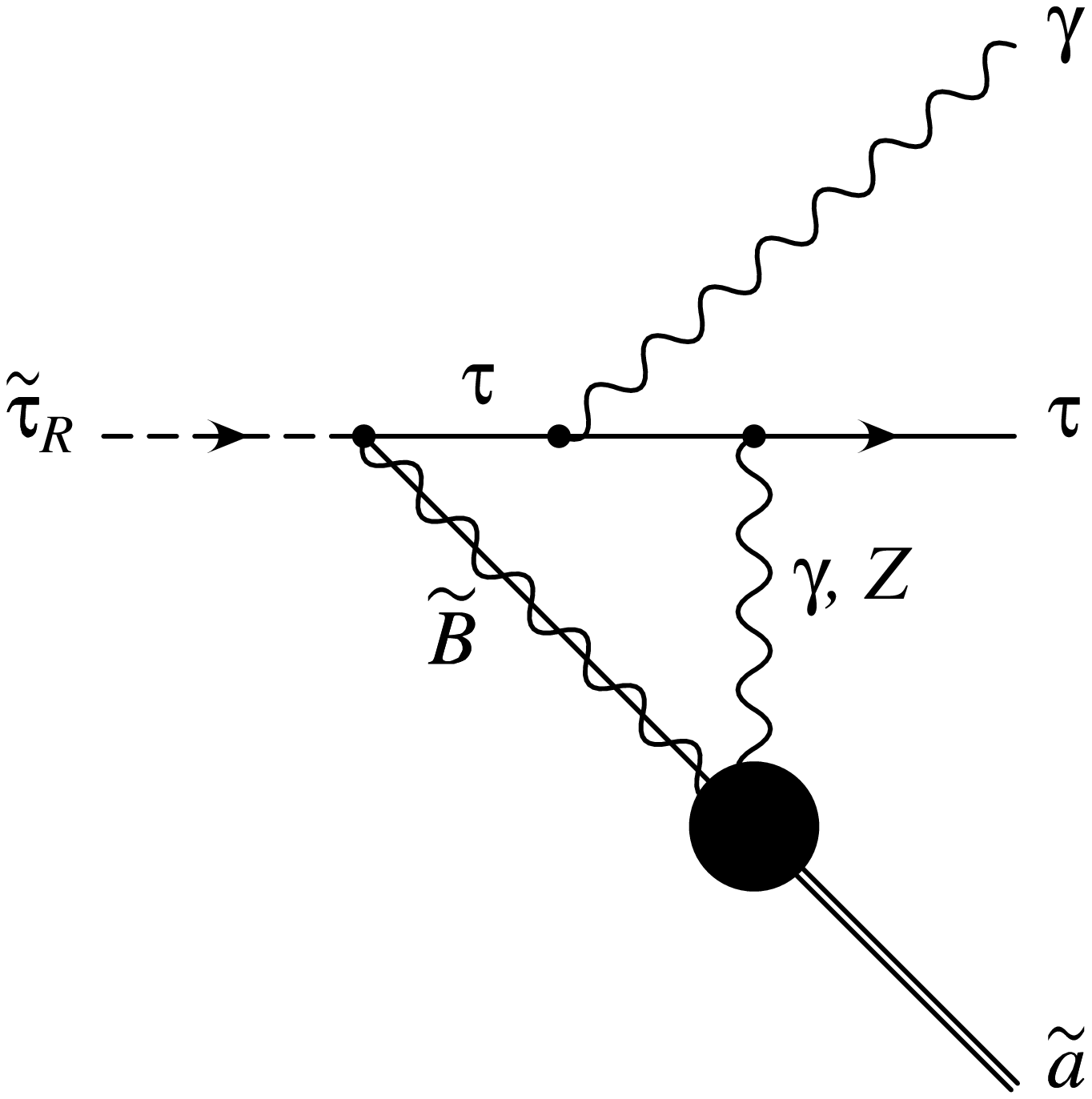, width=3.7cm}\hfill
\epsfig{figure=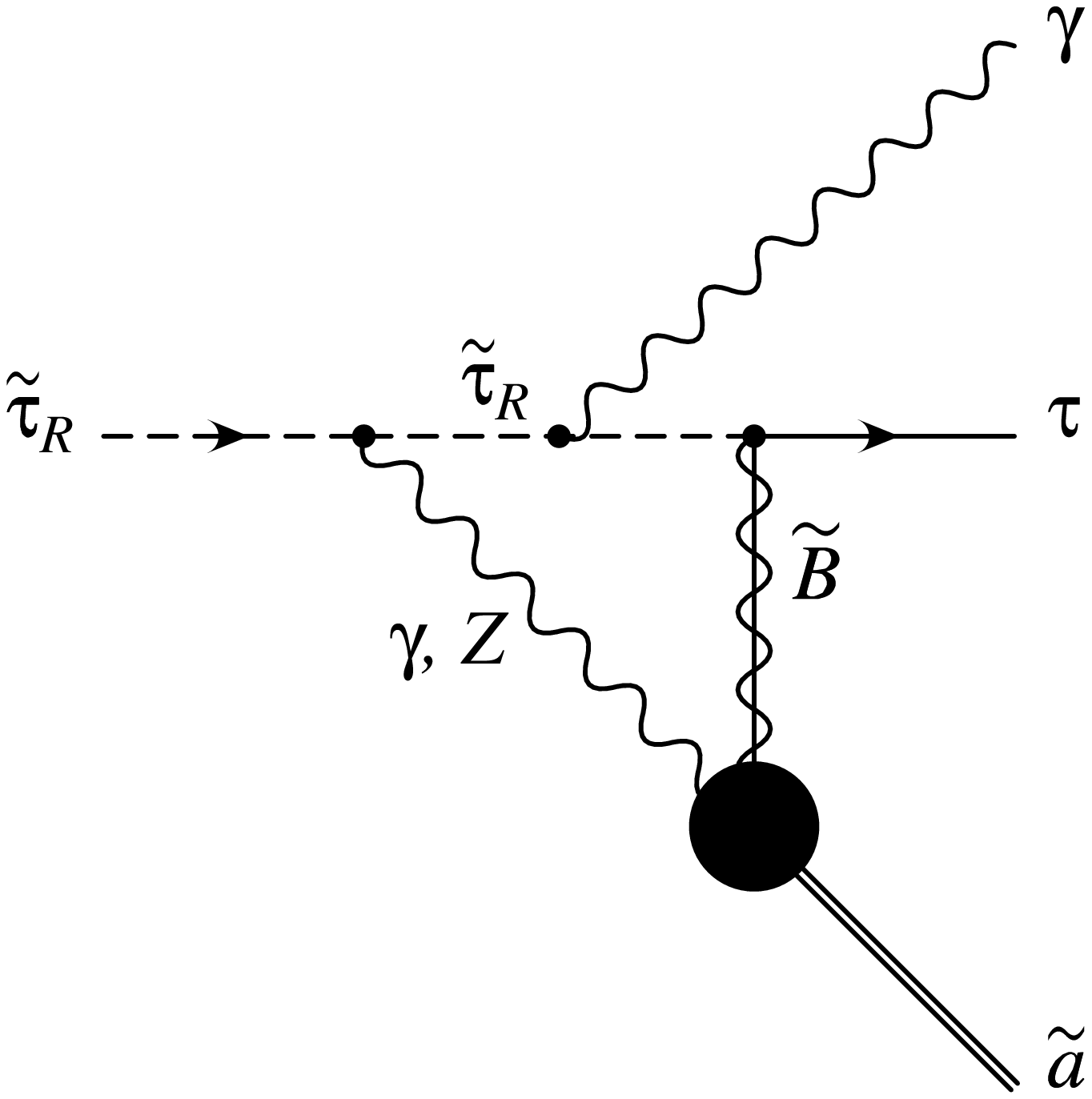, width=3.7cm}\hfill
\epsfig{figure=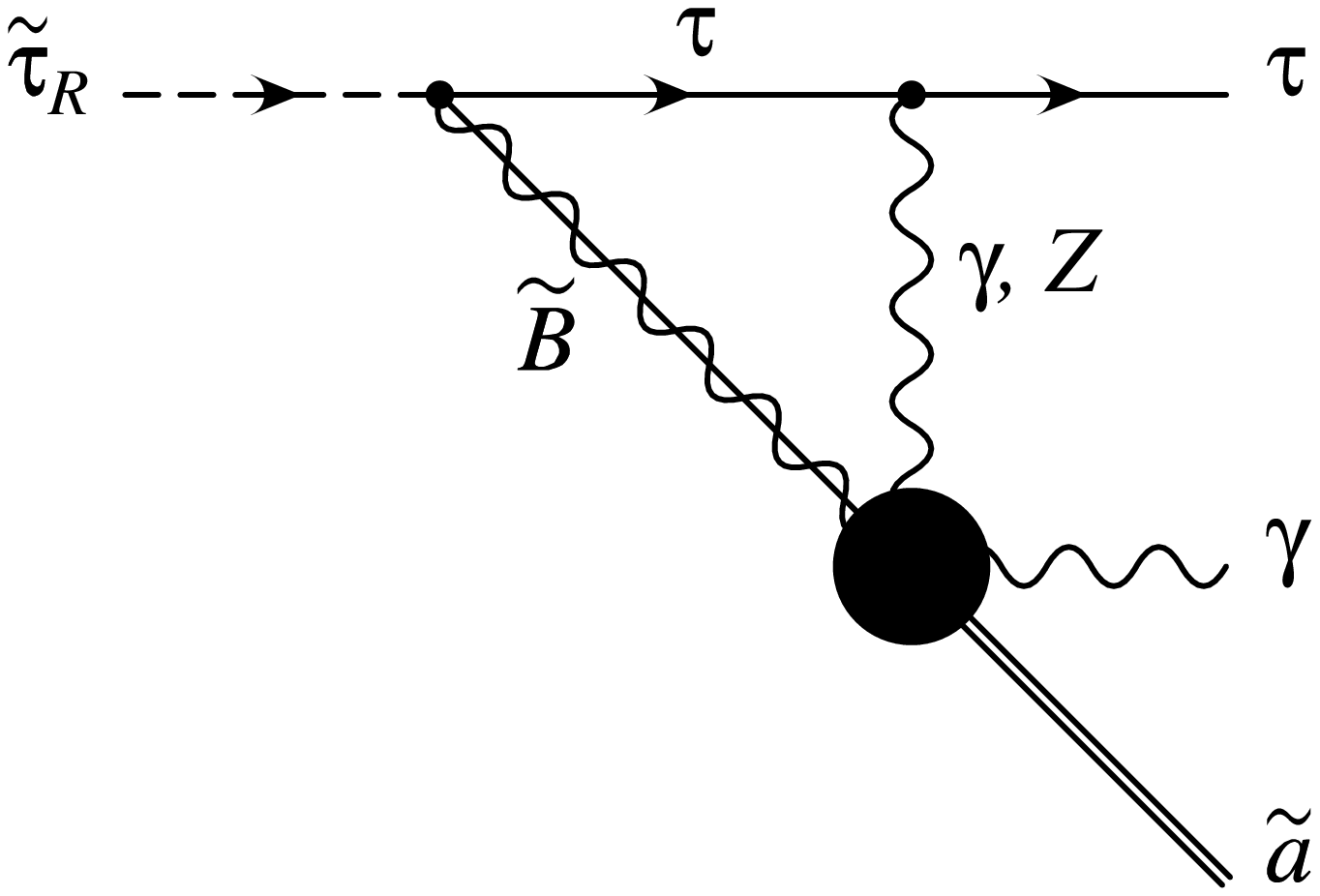, width=3.7cm}\\[-2ex]
\makebox[3cm][c]{(a)}\hfill
\makebox[3.7cm][c]{(b)}\hfill
\makebox[3.7cm][c]{(c)}\hfill
\makebox[3.7cm][c]{(d)}\\[1ex]
\anc\hspace{3cm}\hfill
\epsfig{figure=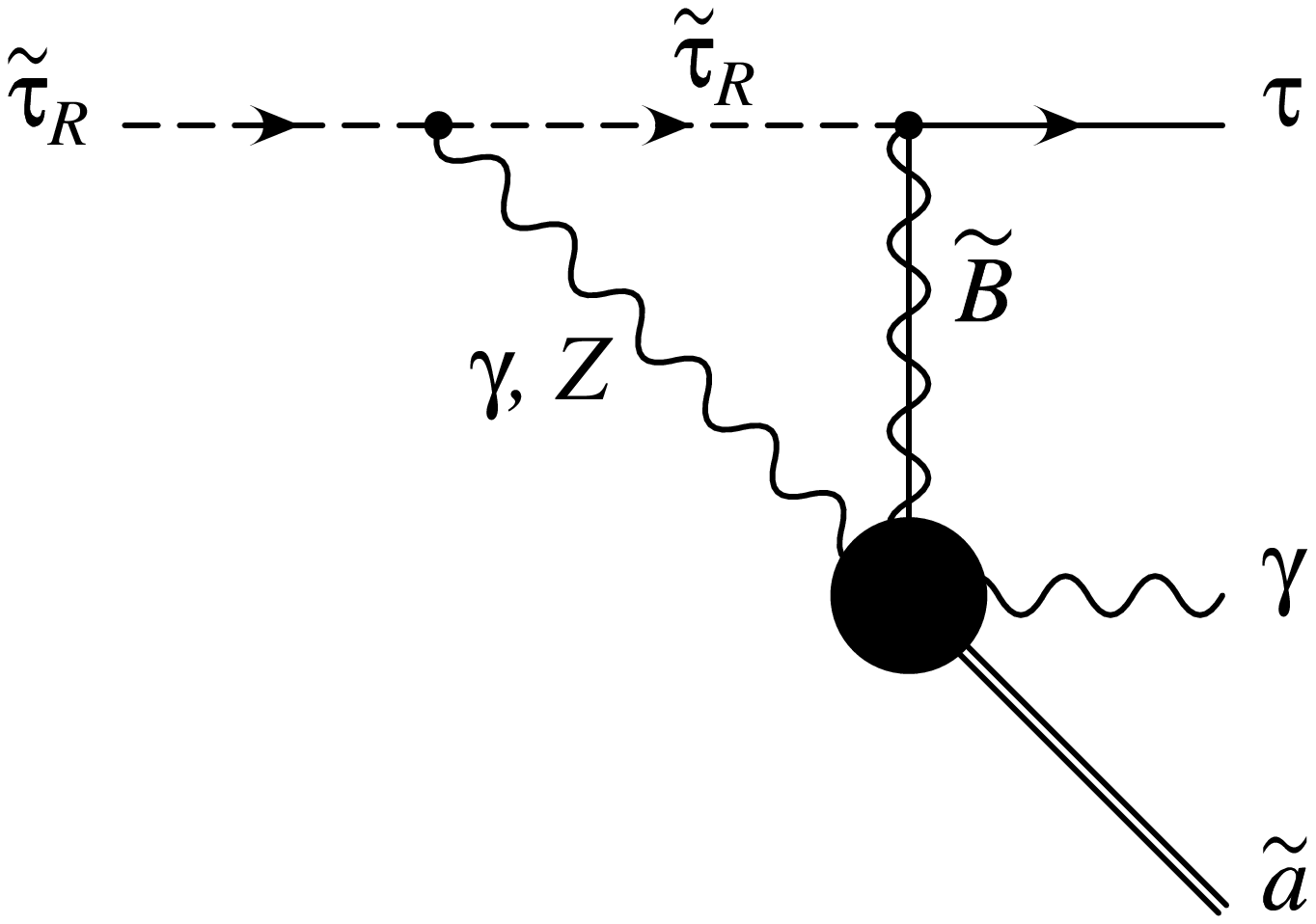, width=3.7cm}\hfill
\epsfig{figure=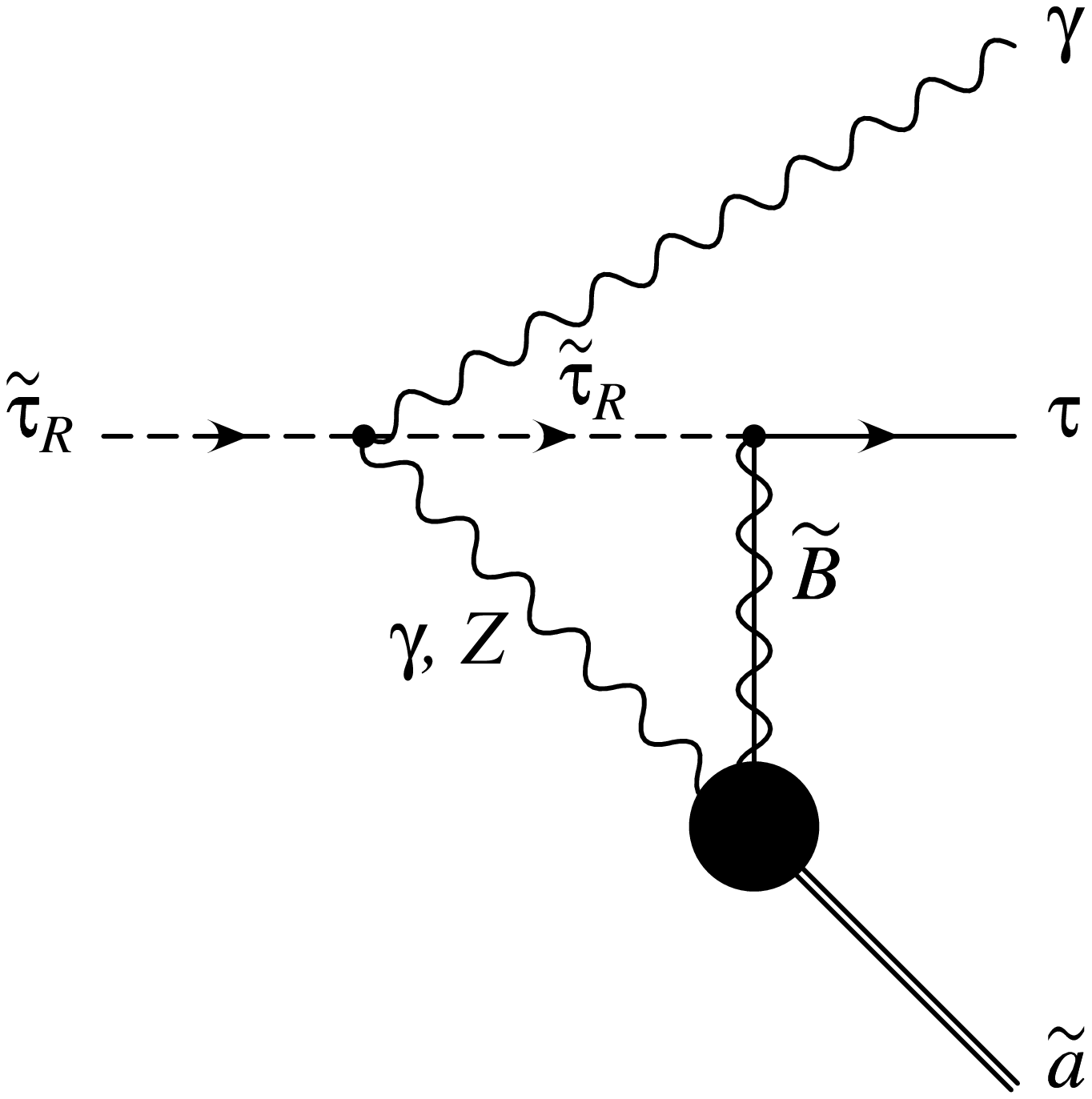, width=3.7cm}\hfill
\epsfig{figure=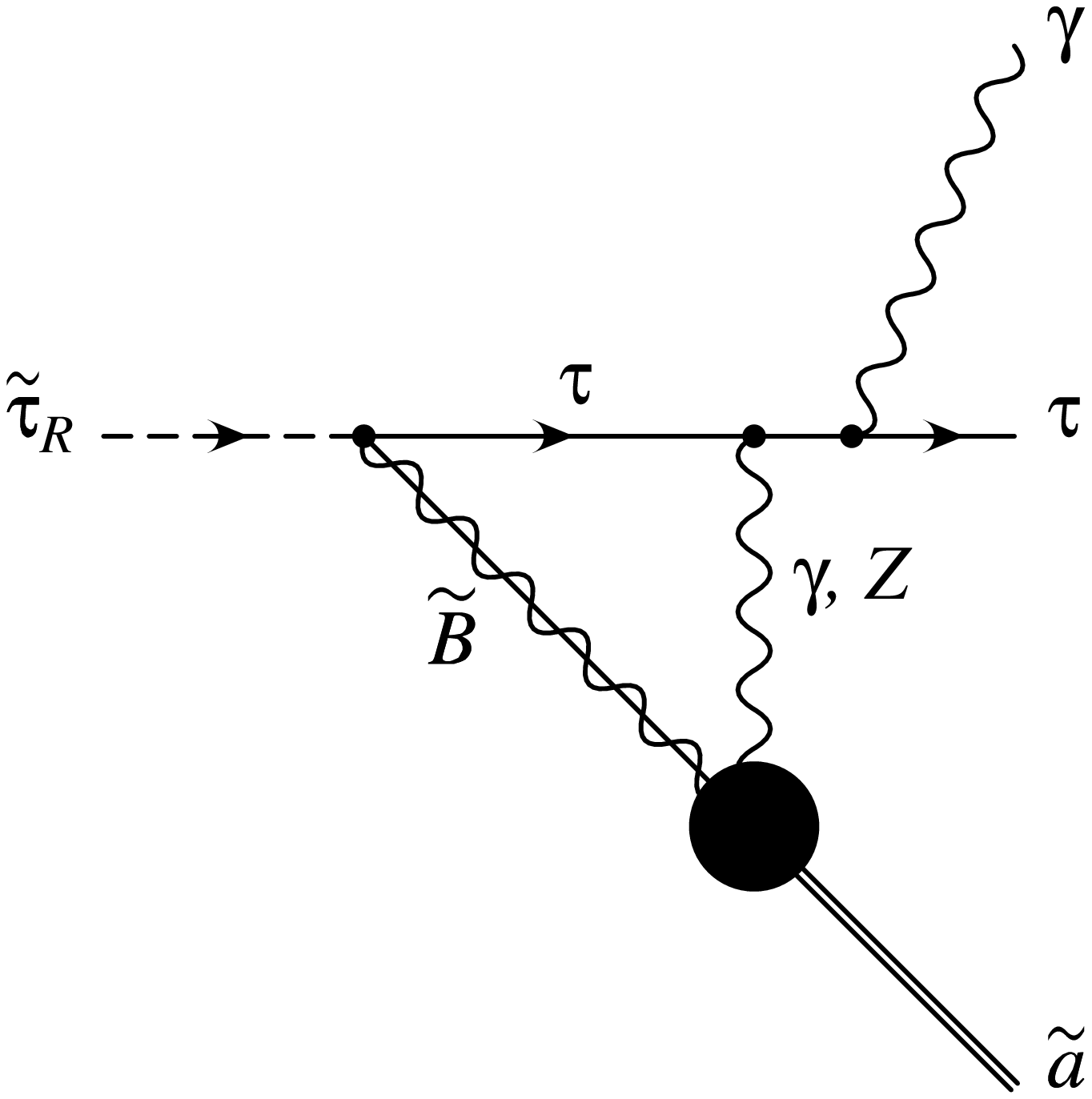, width=3.7cm}\\[-2ex]
\anc\hspace{3cm}\hfill
\makebox[3.7cm][c]{(e)}\hfill
\makebox[3.7cm][c]{(f)}\hfill
\makebox[3.7cm][c]{(g)}\\[1ex]
\anc\hspace{3cm}\hfill
\epsfig{figure=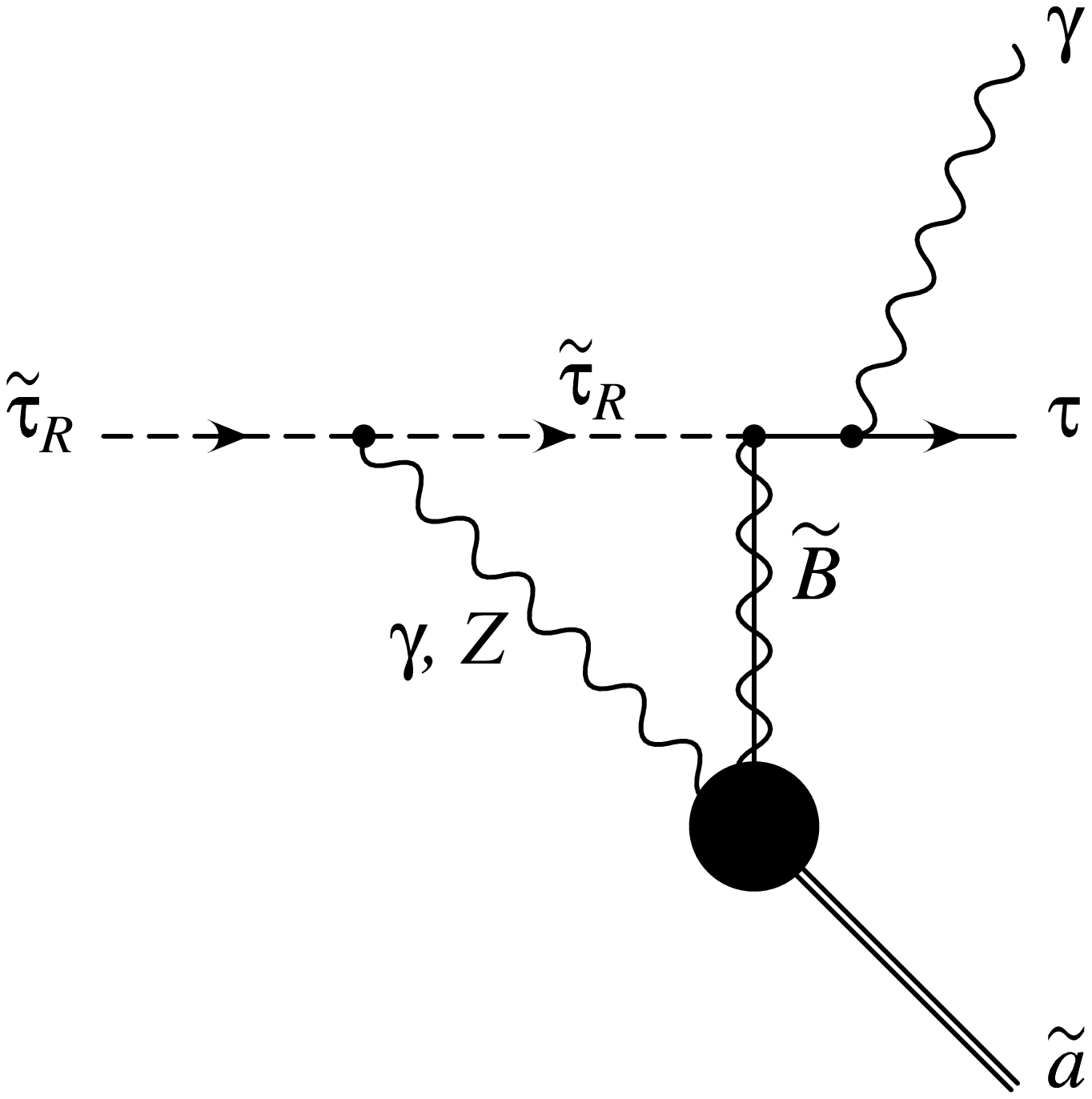, width=3.7cm}\hfill
\epsfig{figure=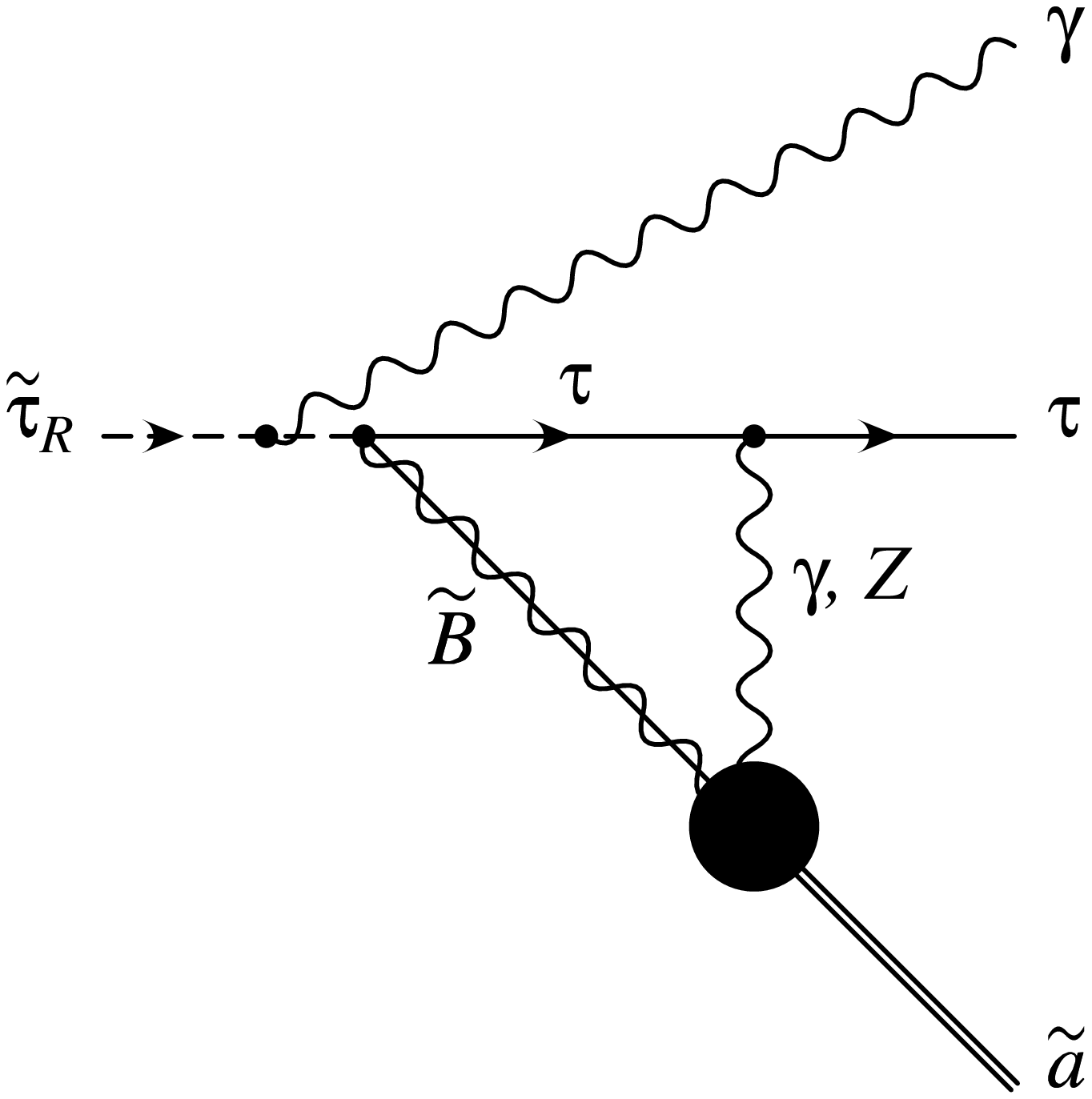, width=3.7cm}\hfill
\epsfig{figure=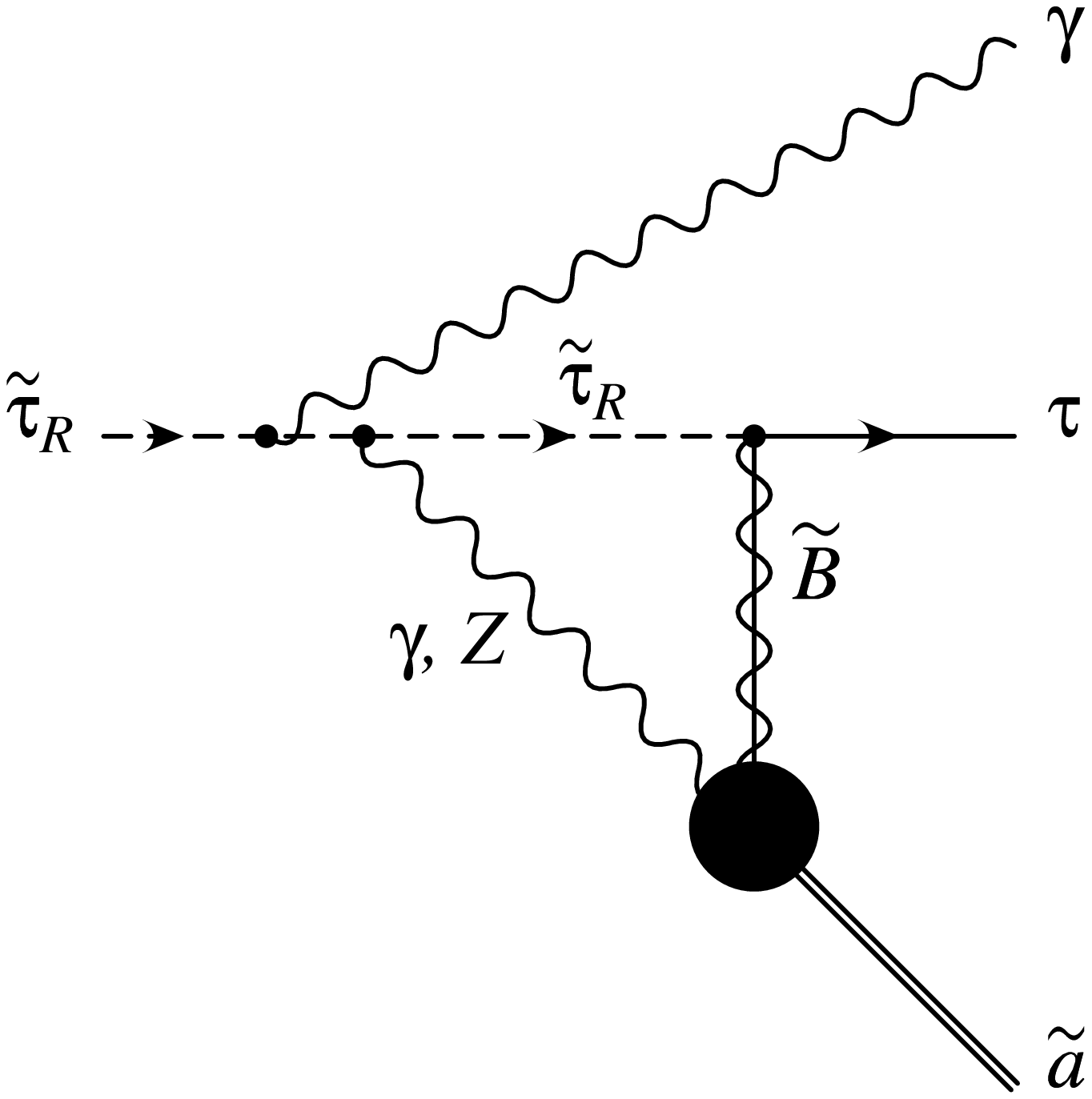, width=3.7cm}\\[-2ex]
\anc\hspace{3cm}\hfill
\makebox[3.7cm][c]{(h)}\hfill
\makebox[3.7cm][c]{(i)}\hfill
\makebox[3.7cm][c]{(j)}
\caption{Feynman diagrams for the 3-body stau NLSP decay
  $\stau_{\mathrm R}\to\tau\axino\gamma$ in KSVZ axion models, at
  one-loop~(a) and two-loop~(b-j) level.  The blob indicates the heavy
  (s)quark loops.}
\label{fig:3body}}

Again we calculate the associated matrix element of the
diagrams using an asymptotic expansion for large (s)quark masses,
\eqref{eq:mquark}, and retaining only the leading term $\propto
1/f_a$. The complete result for the two-loop contribution is rather
lengthy. However, since the 3-body decay rate is suppressed with
respect to that of the 2-body decay, it is sufficiently accurate to
only keep the dominant logarithm term
$\propto \log [y^2 f_a^2/(2\mstau^2)]$ 
of the two-loop part.  The resulting differential decay rate including
the one- and two-loop amplitudes reads
\begin{align}
\label{eq:3amp}
&\frac{{\rm d}^2\Gamma(\stau_{\mathrm R}\to\tau\axino\gamma)}{{\rm
d}x_\gamma\,{\rm d}\cos\theta} = 
\frac{9}{64\pi^4} \, 
\frac{\alpha^3 \, e_Q^4}{c^4_{\rm W}} \, \frac{\mstau^3}{f_a^2} \,
\frac{x_\gamma (1-A_{\ax}-x_\gamma)}{[2-x_\gamma(1-\cos\theta)]^2}
\\
&\times \!\!\!\! \begin{aligned}[t] \biggl [ 
& \frac{2 x_\gamma^2 (1-A_{\ax}-x_\gamma) 
[1+\cos\theta + A_{\ax}(1-\cos\theta)][1+\cos\theta + A_{\Bi}(1-\cos\theta)]}
{[2A_{\ax} - A_{\Bi}(2-x_\gamma+x_\gamma\cos\theta) + x_\gamma(1+\cos\theta)]^2}
\\
 +&\frac{3 \alpha}{\pi c^2_{\rm W}} \, \log\!\left(\frac{y^2 f_a^2}{2 \mstau^2}\right)
 \frac{[A_{\Bi}(1-A_{\ax})+\sqrt{A_{\ax}A_{\Bi}}(1-A_{\ax}-x_\gamma)]
(1+\cos\theta)+A_{\ax}A_{\Bi}x_\gamma(1-\cos\theta)}
{2A_{\ax} - A_{\Bi}(2-x_\gamma+x_\gamma\cos\theta) + x_\gamma(1+\cos\theta)}
\\
 +&\frac{9 \alpha^2}{8\pi^2 c^4_{\rm W}} \, \log^2\!\left( \frac{y^2 f_a^2}{2 \mstau^2}\right)
A_{\Bi} \frac{2(1-A_{\ax})(1-A_{\ax}-x_\gamma)(1+\cos\theta) +
x_\gamma^2[2-(1-A_{\ax})(1-\cos\theta)]}{x_\gamma^2(1-\cos\theta)(1-A_{\ax}-x_\gamma)}
\biggr ],
\end{aligned}
\nonumber
\end{align}
where 
\bea
   x_\gamma = 2 E_\gamma/\mstau \ ,
\eea
is the photon energy $E_\gamma$ divided by its maximum value and
$\theta$ is the angle between the photon and tau in the rest frame of
the stau.

It is interesting to note that only the contributions from the
diagrams in Fig.~\ref{fig:3body}~(f--j) are logarithmically enhanced.
In these diagrams the photon is radiated either from the external tau
or stau line or from a four-vertex, so that the loop integral is
essentially the same as for the 2-body decay. If the photon is
radiated off the heavy (s)quark loop, as in \reffig{fig:3body}~(d,e),
an additional heavy (s)quark propagator is introduced in this loop. In
fact, the leading term in the HME of these diagrams is not ${\cal
  O}(1/f_a)$ but ${\cal O}(1/f_a^2)$ and thus strongly suppressed.
Finally, the absence of a large logarithm $\log (y^2
f_a^2/2\,\mstau^2)$ in the diagrams \reffig{fig:3body}~(b,c) can be
understood as follows: a logarithmic dependence on the heavy (s)quark
mass corresponds to an ultraviolet (UV) divergence if the (s)quark
mass becomes infinite, $M_{Q/\tilde{Q}_{1,2}} = y \, f_a /\sqrt{2}\to
\infty$. When the (s)quark loop is integrated out,
\reffig{fig:3body}~(b,c) turn into one-loop diagrams with four
propagators in the loop. One can see from simple power counting that
these one-loop terms are UV finite so that the full two-loop
contribution from \reffig{fig:3body}~(b,c) cannot generate any large
logarithms.

In~\cite{Brandenburg:2005he}, the 2- and 3-body decay rates were
calculated using an effective theory where heavy (s)quark loops were
integrated out; cf.~\eqref{Eq:L_aBgammaZ}. 
As mentioned above, this method leads to a logarithmic
divergence in the result. As in~\cite{Covi:2004rb}, that logarithmic
divergence was regulated with the cutoff $f_a$, and only the dominant
contributions were kept. Moreover, the authors
of~\cite{Brandenburg:2005he} introduced a factor $\xi$ and a mass
scale $m$ to parametrise the uncertainty associated with this cutoff
procedure. Our direct two-loop result now allows us to make a direct
connection with the parameters of the underlying model, yielding
agreement between \refeq{eq:3amp} and the 3-body result
of~\cite{Brandenburg:2005he} for $\xi=1$, $m=\sqrt{2}\mstau/y$ and,
as mentioned below~\eqref{Eq:L_aBgammaZ},
$C_{\rm aYY}=6 e_Q^2$. The comparison of the 2-body results can be
found in~\cite{Freitas:2009fb} and will also be addressed in
\refsec{Sec:StauNLSPLifetime}.

The term in the last line of \eqref{eq:3amp} with the squared
logarithm has an infrared (IR) divergence for $x_\gamma \to 0$ and a
collinear divergence for $\cos\theta \to 1$. These divergencies would
be cancelled by the virtual three-loop correction to the 2-body decay
channel. However, the calculation of this three-loop contribution is a
formidable task and beyond the scope of this paper. Furthermore, as
will be shown below, the branching ratio of the 3-body mode is small
and has a minor effect on the total lifetime. Therefore, the
three-loop contribution to the 2-body decay mode will have a similarly
small effect and can be safely neglected. Nevertheless, our
calculation of the 3-body decay mode provides a meaningful description
of the energy and angular distributions of the photon (as long as
$x_\gamma$ and $1-\cos\theta$ do not become too small), which can be
used for analysing this decay at colliders, see \refsecs{Sec:Collider}
and~\ref{Sec:ColliderPhenomenologyAxGrSt} below.

Because of the soft and collinear divergencies, we consider the
integrated 3-body rate with cuts on the scaled photon energy,
$x_\gamma > x_\gamma^{\rm cut}$, and on the photon angle, $\cos\theta
< 1- x_\theta^{\rm cut}$,
\begin{equation}
\Gamma(\stau_{\mathrm R}\to\tau\axino\gamma;
x_\gamma^{\rm cut};x_\theta^{\rm cut}) 
\equiv
\int_{x_\gamma^{\rm cut}}^{1-A_{\ax}} {\rm d}x_\gamma 
\int_{-1}^{1-x_\theta^{\rm cut}} {\rm d}\cos\theta \, 
\frac{{\rm d}^2\Gamma(\stau_{\mathrm R}\to\tau\axino\gamma)}
{{\rm d}x_\gamma\,{\rm d}\cos\theta}.
\label{Eq:3BodyWidthwCuts}
\end{equation}
\reffigure{fig:3br} shows the branching ratio of the integrated 3-body
decay rate,
\begin{align}
&{\rm BR}(\stau_{\mathrm R}\to\tau\axino\gamma;x_\gamma^{\rm cut};x_\theta^{\rm cut})
\equiv \frac{\Gamma(\stau_{\mathrm R}\to\tau\axino\gamma;x_\gamma^{\rm
cut};x_\theta^{\rm cut})}{\Gamma_{\rm tot}^{\stau_{\mathrm R}}},
\label{eq:3bodyBR}
\intertext{with}
&\Gamma_{\rm tot}^{\stau_{\mathrm R}} \equiv \Gamma(\stau_{\mathrm R}\to\tau\axino) +
\Gamma(\stau_{\mathrm R}\to\tau\axino\gamma;x_\gamma^{\rm
cut};x_\theta^{\rm cut}).
\end{align}
%
\FIGURE[t]{ 
\epsfig{figure=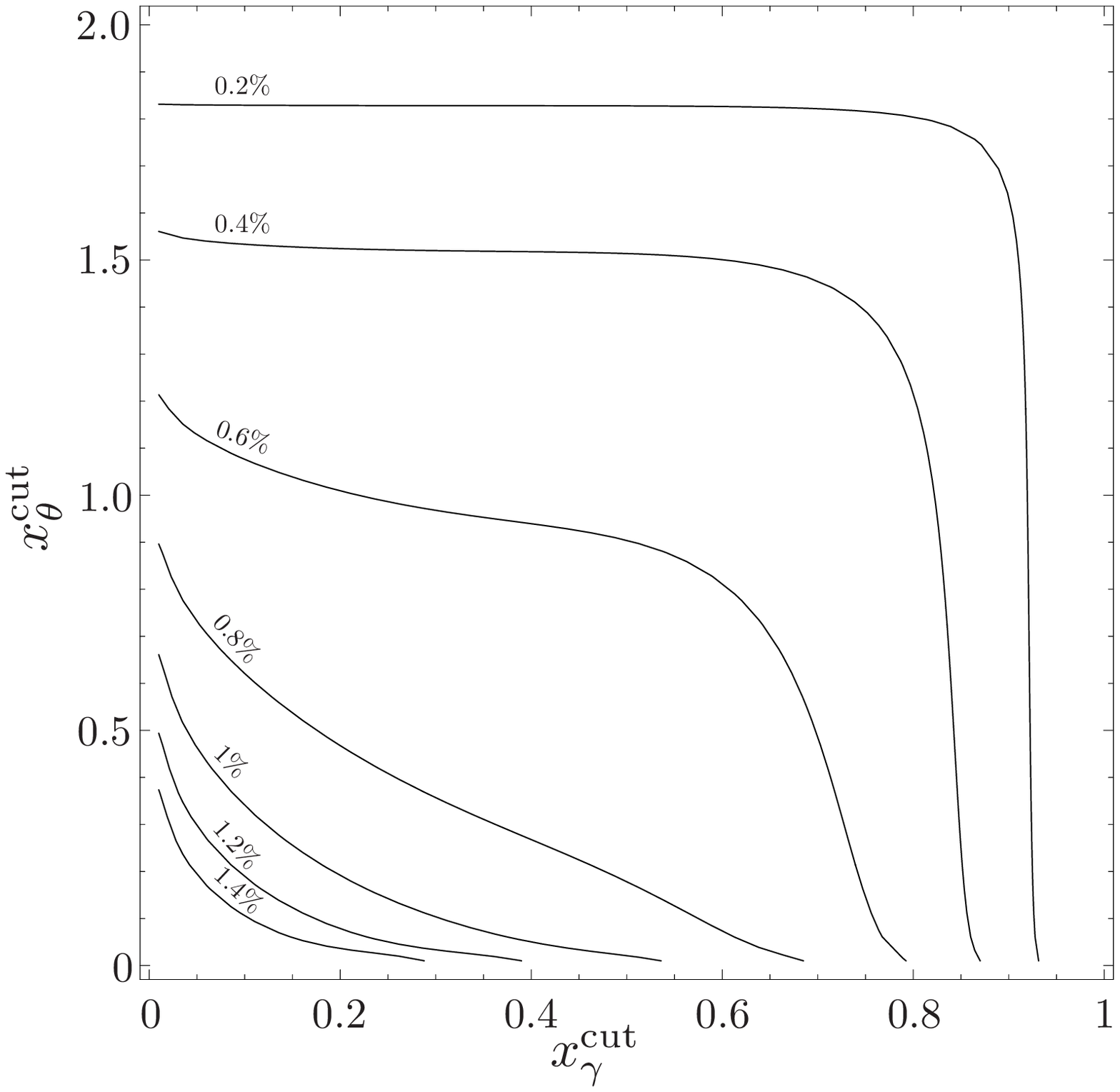, height=6.8cm}
\hfill
\epsfig{figure=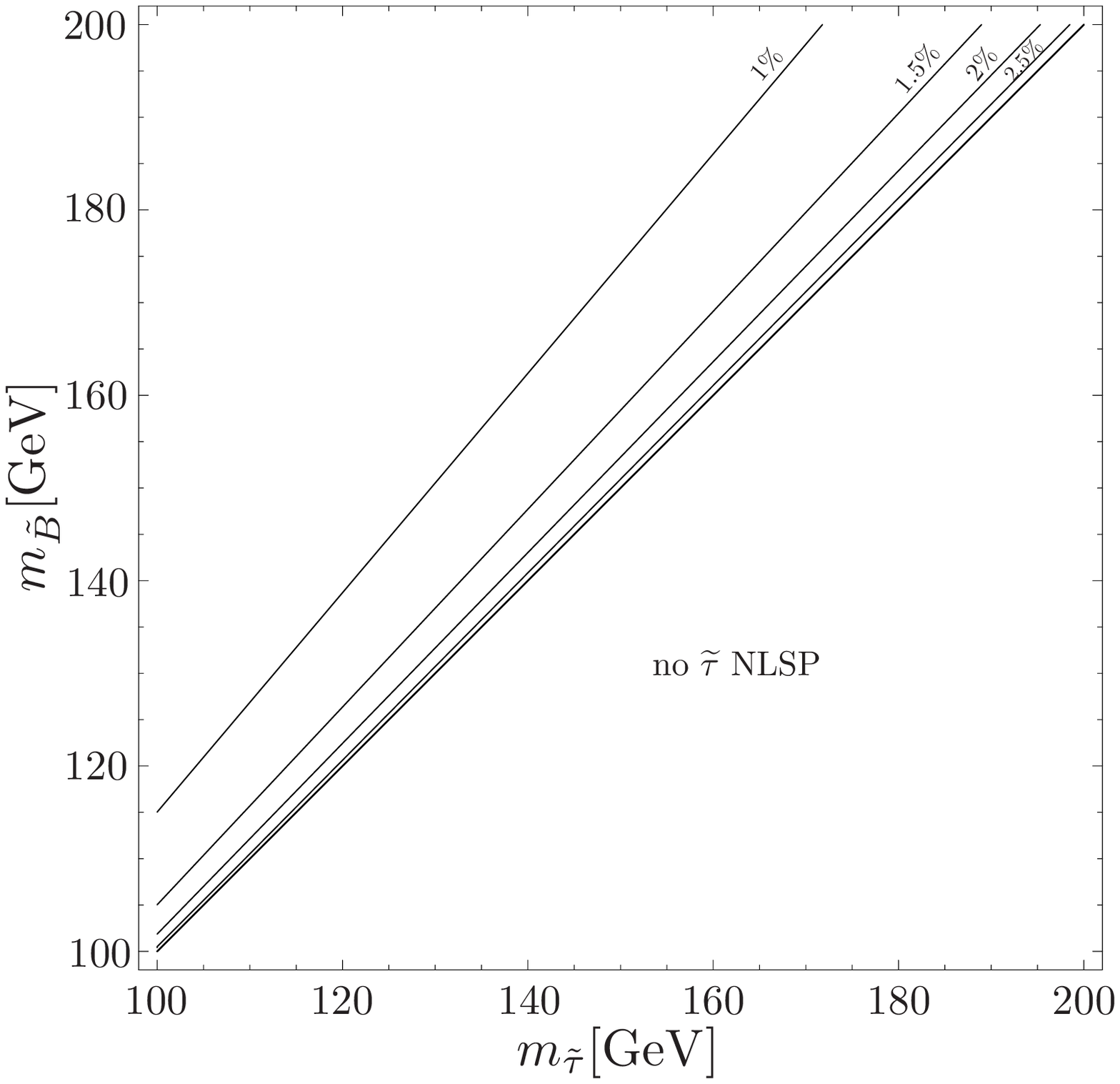, height=6.8cm}
\\
\hspace*{3cm}
\makebox[2cm][c]{(a)}\hfill
\hspace*{4cm}
\makebox[2cm][c]{(b)}\hfill
\caption{Branching ratio of the 3-body decay $\stau_{\mathrm
    R}\to\tau\axino\gamma$ as a function of (a)~the soft and collinear
  cut parameters $x_\gamma^{\rm cut}$ and $x_\theta^{\rm cut}$ and (b)~the 
  masses $\mstau$ and $\mbino$. In both plots the other parameters
  are set to $\maxino=10$~GeV, $f_a = 10^{11}$~GeV, $|e_Q|=1/3$,
  $y=1$, as well as $\mstau = 100$~GeV, $\mbino=110$~GeV for the left
  figure, and $x_\gamma^{\rm cut}=x_\theta^{\rm cut}=0.1$ for the
  right figure.}
\label{fig:3br}}
%
As can be seen from \reffig{fig:3br}~(a), within reasonable ranges of
$x_\gamma^{\rm cut}$ and $x_\theta^{\rm cut}$, the 3-body decay
contributes only ${\cal O}(1\%)$ to the total decay rate. The latter
is therefore well approximated by the 2-body contribution.  For
$A_{\tilde{B}} \equiv \mbino^2/\mstau^2 \to1$, the branching ratio
\refeq{eq:3bodyBR} grows rapidly as shown in \reffig{fig:3br}~(b) for
$x_\gamma^{\rm cut}=x_\theta^{\rm cut}=0.1$. This is because the
dominant, logarithmically-enhanced part of
$\Gamma(\stauR\to\tau\axino)\propto\mstau\,\mbino^2$ while
$\Gamma(\stauR\to\tau\axino\gamma)\propto\mstau^3$.  However, the
branching ratio always stays below about 3.5\% for
$\mstau\lesssim5\,\TeV$ and $x_\gamma^{\rm cut}=x_\theta^{\rm
  cut}=0.1$. As long as $\maxino^2/\mstau^2 \ll 1$, the influence of
$\maxino$ on the branching ratio is small.  The product $y f_a$
appears only in the logarithm but determines the relative contribution
of the one- and two-loop amplitudes to the 3-body decay rate. For the
parameters in \reffig{fig:3br}, both the one- and two-loop amplitudes
are of roughly equal importance. Varying $y f_a$ by one order of
magnitude up or down changes the 3-body branching ratio by 2--10\%.
Nevertheless, this factor is important when considering the prospects
for distinguishing between the axino LSP and the gravitino LSP, 
as will be discussed in the last two paragraphs 
of \refsec{Sec:DistAxGrav} below.

\FIGURE[t]{ 
\epsfig{figure=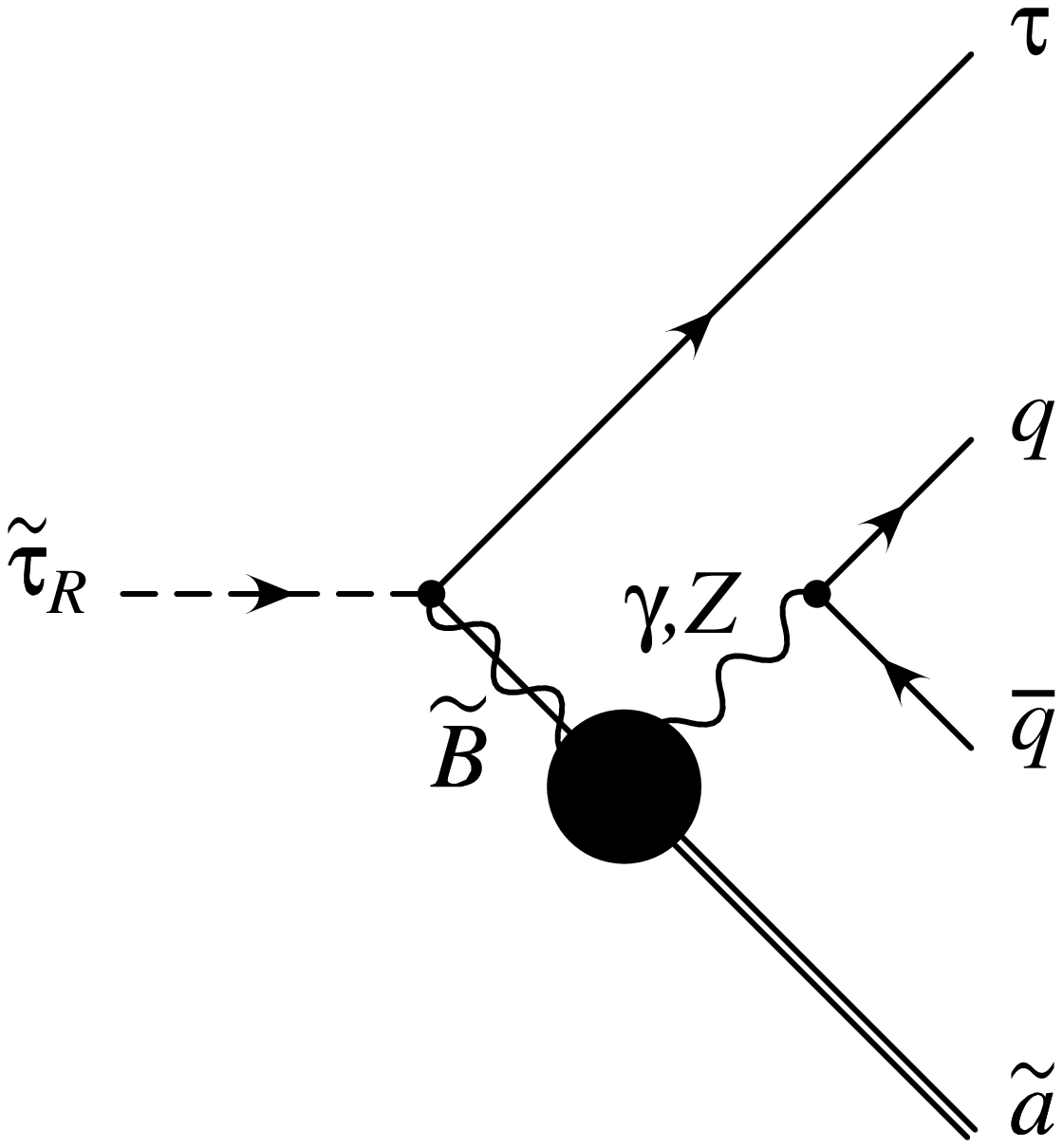, width=4cm}
\caption{One of the Feynman diagrams of the 4-body stau NLSP decay
  $\stau_{\mathrm R}\to\tau\axino q\qbar$ in KSVZ axion models. The
  blob indicates the heavy (s)quark loop.}
\label{fig:4body}}

\subsection{4-Body Decays of Charged NLSP Sleptons}
\label{Sec:Axino4Body}

The 4-body decay modes $\stau_{\mathrm R}\to\tau\axino\gamma\gamma$,
$\stau_{\mathrm R}\to\tau\axino l^+ l^-$ and $\stau_{\mathrm
  R}\to\tau\axino q\bar{q}$ have a negligible impact on the total stau
decay rate since they are suppressed by an additional factor $\alpha$
compared to the 3-body decay. However, for $\tau_\stau>100$~s,
hadronic energy
injection~\cite{Reno:1987qw,Dimopoulos:1987fz,Dimopoulos:1988ue,Kohri:2001jx,Kawasaki:2004qu,Jedamzik:2004er,Jedamzik:2006xz,Cyburt:2009pg}
from $\stauR\to\tau\axino q\bar{q}$ decays can affect the abundances
of primordial light nuclei significantly, see
\refsecs{Sec:NucleosynthesisConstraints}
and~\ref{Sec:CosmologicalConstraintsAxGrSt}. The evaluation of BBN
constraints from this effect requires the calculation of the hadronic
4-body decays.

The Feynman diagrams for $\stau_{\mathrm R}\to\tau\axino q\bar{q}$ can
be deduced from the diagrams in \reffig{fig:3body} by splitting the
final state photon into a $q\bar{q}$ pair. In addition there are
diagrams where this photon is replaced by a $Z$ boson.

The one-loop contribution, see \reffig{fig:4body}, has been calculated
with the same HME techniques explained in the previous sections.  For
the two-loop amplitude, we again only retain the terms $\propto
\log(y^2 f_a^2/2\mstau^2)$. Following the arguments of the previous
section, no such logarithmic term is generated from diagrams where the
$(\gamma \to q\bar{q})$ or $(Z \to q\bar{q})$ is emitted from inside
the loops. One has to consider only the two-loop diagrams where the
$\gamma,Z$ is emitted from the external stau or tau lines or is
attached via a quartic coupling to the external and internal stau and
the internal $\gamma,Z$ (see also Fig.~1 of~\cite{Freitas:2009jb}). As
a result, one can directly use the result for the 3-body decay without
having to recompute the complete amplitude. As the only necessary
modifications, the stau-$\gamma$/$Z$ or tau-$\gamma$/$Z$ splitting
functions have to be added and the stau or tau momentum in the loop
integrals moved off-shell. In our calculation we include the width of
the $Z$ boson by introducing a complex mass in the propagator, $m_Z^2
\to m_Z^2 - i m_Z \Gamma_Z$.

The differential decay rate as a function of the $q\bar{q}$ invariant
mass $m_{q\qbar}$ is given by
\begin{equation}
\begin{gathered}
\frac{{\rm d}\Gamma(\stau_{\mathrm R}\to\tau\axino q\bar{q})}{{\rm
d}m_{q\qbar}} = \frac{1}{32\mstau^2}\,\frac{1}{(2\pi)^8} \,
\int {\rm d} m_{\tau V} \int {\rm d}\Omega_{\ax} \int {\rm d}\Omega'_V \int
{\rm d}\Omega''_{q} \\ \hspace{15em}
\times |\bk_{\ax}| \, |\bk'_V| \, |\bk''_q| \; 
|{\cal M}(\stau_{\mathrm R}\to\tau\axino q\bar{q})|^2,
\end{gathered}
\end{equation}
where $m_{\tau V}$ is the invariant mass of the final state $\tau$ and
the intermediate $V=\gamma,Z$ boson, $\Omega_{\ax}$ and $\bk_{\ax}$
are the solid angle and 3-momentum of the axino in the $\stau_{\mathrm
  R}$ rest frame, $\Omega'_V$ and $\bk'_V$ are the solid angle and
3-momentum of the $V=\gamma,Z$ boson in the $\tau V$ rest frame, and
$\Omega''_{q}$ and $\bk''_q$ are the solid angle and 3-momentum of the
final state quark in the $V=\gamma,Z$ rest frame.

One of the solid angle integrations, say $\int {\rm d} \Omega_{\ax}$,
is trivial and simply corresponds to a redefinition of the coordinate
system. One additional azimuthal angle integration is also trivial
since the incoming stau carries no polarisation. The integration over
the angle of the final state $q\bar{q}$ pair yields, for massless
quarks,
\begin{equation}
\int {\rm d} \Omega''_{q} \; \left[ \bar{u}(k_q) \gamma^\mu (v + a \gamma_5)
u(k_{\bar{q}}) \right] \left[ \bar{u}(k_{\bar{q}}) \gamma^\nu (v + a \gamma_5)
u(k_q) \right] = \frac{16\pi}{3} \left( -g_{\mu\nu} k_V^2 + k_V^\mu k_V^\nu \right)
(v^2+a^2),
\label{eq:qqsum}
\end{equation}
where $v$ and $a$ are the respective vector and axial couplings and 
$k_V = k_q + k_{\bar{q}}$. The remaining integrations 
over $\cos\theta'_V$ and $m_{\tau V}$,
\begin{equation}
\begin{gathered}
\frac{{\rm d}\Gamma(\stau_{\mathrm R}\to\tau\axino q\bar{q})}{{\rm
d}m_{q\qbar}} = \frac{1}{128(2\pi)^6\mstau^2} \,
\frac{\lambda^{1/2}(\mstau^2,\maxino^2,m_{\tau V}^2)}{2\mstau}\,
\frac{m_{\tau V}^2 - m_{q\qbar}^2}{2m_{\tau V}} \,
\frac{m_{q\qbar}}{2} \\
\hspace{15em} \times \int_{-1}^1 {\rm d} \cos\theta'_V \,
\int_{m_{q\qbar}}^{\mstau-\maxino} {\rm d}m_{\tau V} \; |\hat{{\cal M}}|^2,
\end{gathered}
\label{Eq:dGammadmqqbar}
\end{equation}
need to be performed numerically. Here $\lambda(a,b,c)\equiv
a^2+b^2+c^2-2(ab+ac+bc)$, $\theta'_V$ is the polar angle of
$\Omega'_V$, and $|\hat{{\cal M}}|^2$ is the squared matrix element
with the external quark current replaced by expression
\eqref{eq:qqsum}.

In \reffig{fig:4br} 
%
\FIGURE[t]{ 
\centerline{\epsfig{figure=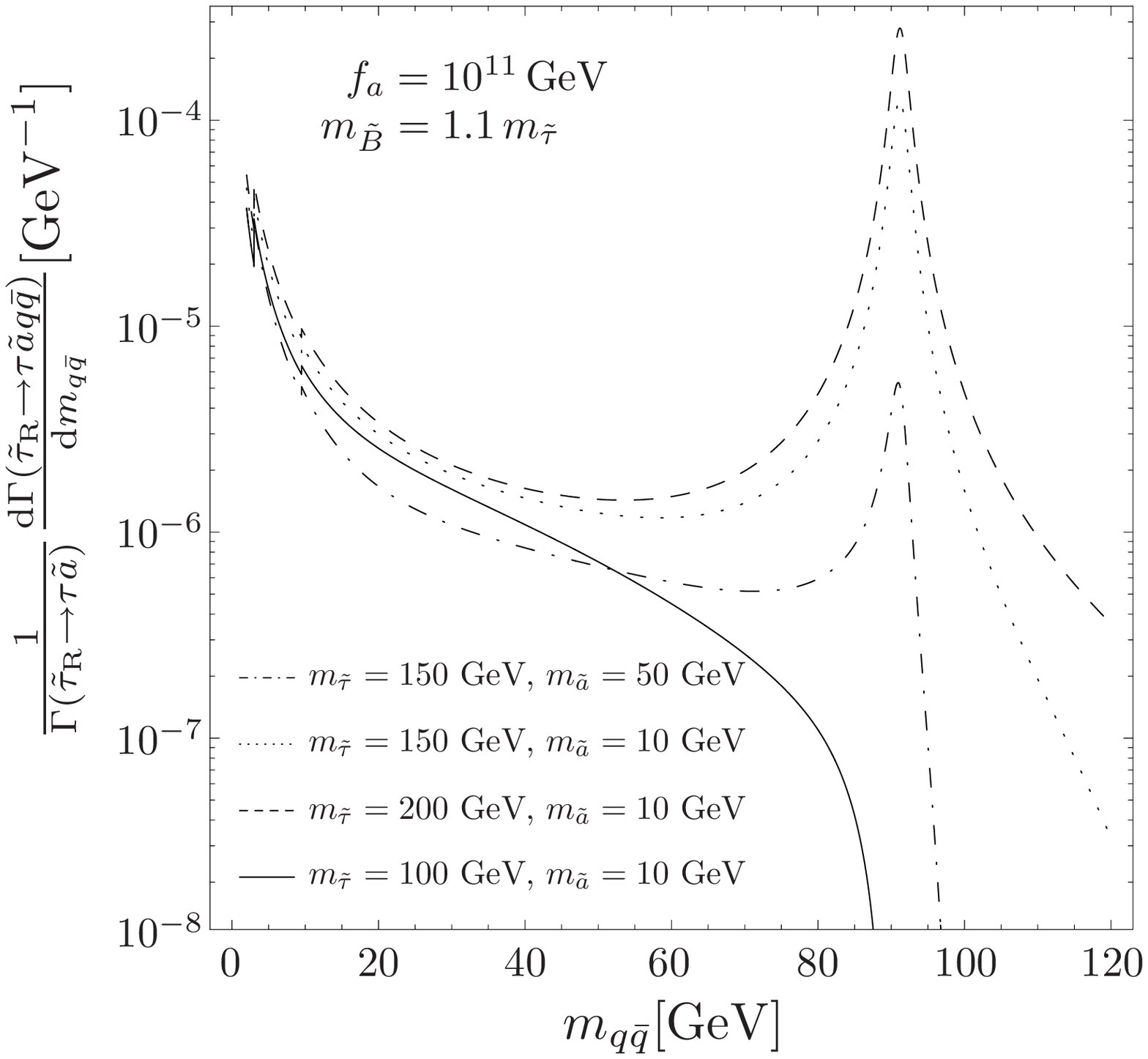, width=8cm}}
\caption{Energy spectrum of quark-antiquark pairs emitted from the
  4-body decay $\stau_{\mathrm R}\to\tau\axino q\bar{q}$ with an
  invariant mass $m_{q\qbar}$, normalised to the 2-body decay rate,
  $\Gamma(\stauR\to\tau\axino)$. The shown quantity is independent of
  $e_Q$. The other parameters are set to $\mbino=1.1 \,\mstau$, $f_a =
  10^{11}$~GeV and $y=1$.}
\label{fig:4br}}
%
we show the obtained energy spectrum of the quark-antiquark pair
stemming from the 4-body decay for several choices of the stau and
axino masses, $\mbino = 1.1\,\mstau$ and $f_a=10^{11}\,\GeV$. It is
normalised to the 2-body decay width given by \refeq{eq:2lamp_Gamma}.
As evident from the figure the effect of the $Z$ boson resonance
becomes important only for $\mstau-\maxino\gsim 100$~GeV.

Since pions decay fast, the major impact of hadronic energy release on
primordial nucleosynthesis stems from nucleons that hadronise from the
$q\bar{q}$ pair (see \refsec{Sec:NucleosynthesisConstraints}).
Therefore we consider only $q\bar{q}$ pairs with $m_{q\qbar} > 2$~GeV.
Moreover, $c\bar{c}$ and $b\bar{b}$ pairs only contribute for
invariant masses above the $J/\Psi$(1S) resonance with $m_{J/\Psi,\rm
  1S} = 3.0$~GeV and the $\Upsilon$(1S) resonance with
$m_{\Upsilon,\rm 1S} = 9.5$~GeV, respectively. The contribution from
$t\bar{t}$ pairs is strongly suppressed due to the large threshold
$m_{t\bar{t}} \gsim 350$~GeV. In the matrix element all quarks are
taken to be massless. This approximation is justified by the behaviour
of the cross section for $e^+e^-\to q\bar{q}$ as a function of the
centre of mass energy, which is well described by step function
thresholds for the heavy quarks.

\subsection{Lifetime of Charged NLSP Sleptons in Axino Dark Matter
  Scenarios}
\label{Sec:StauNLSPLifetime}

The results on the branching ratios of the 3- and 4-body decays
obtained in \refsecs{Sec:Axino3Body} and~\ref{Sec:Axino4Body} show
that the stau NLSP lifetime
$\tau_\stau=1/\Gamma_{\mathrm{tot}}^{\stau_{\mathrm{R}}}$ 
can be very well estimated by the 2-body partial decay width alone,
given by \refeq{eq:2lamp_Gamma}: 
$\Gamma_{\mathrm{tot}}^{\stauR} \approx \Gamma(\stau_{\mathrm R}\to\tau\axino)$.
Indeed, this decay width is dominated by the leading logarithmic (LL)
term of the 2-body decay,
\bee
   \Gamma(\stau_{\mathrm R}\to\tau\axino)_{\rm LL} = 
   \frac{81\alpha^4 \, e_Q^4}{512 \pi^5 \, c^8_{\rm W}}
   \, \frac{\mstau}{f_a^2} (1-A_{\ax})^2 \,\mbino^2 \,
   \log^2 \frac{y^2 f_a^2}{2\mstau^2}\,.
\label{eq:2bLL}
\eee
The mass of the axino enters only through a kinematic factor in this
leading term, and the lifetime quickly becomes independent of
$\maxino$ if the latter is sufficiently smaller than $\mstau$, i.e.
when $A_{\ax}$ is small enough. Moreover, it is useful to look at the
LL expression \refeq{eq:2bLL} to see analytically the dependency of
the lifetime on the parameters of the underlying model. It is highly
sensitive to the charge of the heavy (s)quarks, 
$\tau_{\stau}\propto 1/e_Q^4$,
and less so to the coupling $y$.

\reffigure{fig:lldiffmstfa}
%
\FIGURE[t]{ 
\centerline{\epsfig{figure=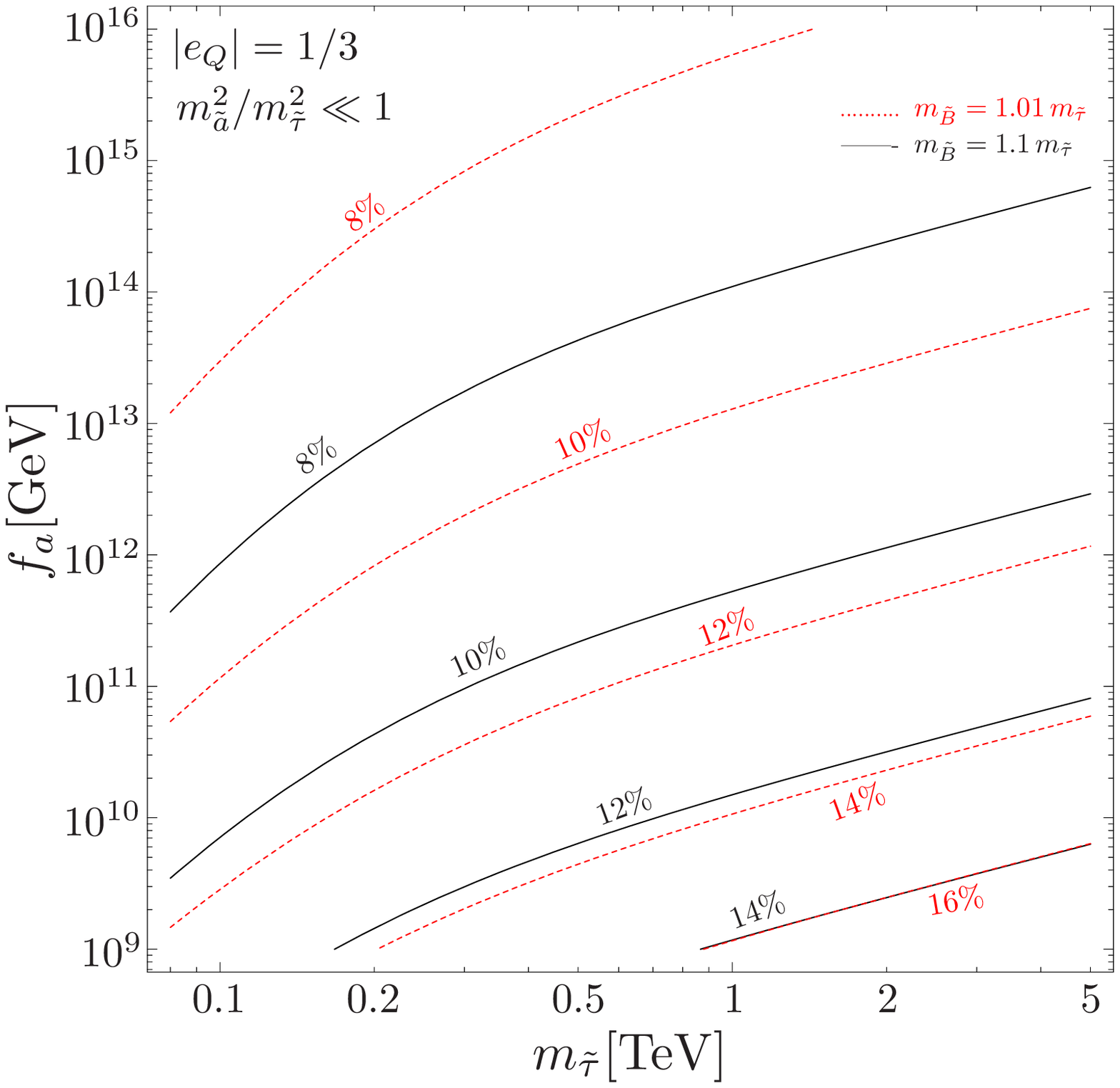, width=8cm}}
\caption{The discrepancy between the LL approximation \eqref{eq:2bLL}
  and full expression \eqref{eq:2lamp_Gamma} for the 2-body decay of
  the stau NLSP, $1- \Gamma(\stau_{\mathrm R}\to\tau\axino)_{\rm LL} /
  \Gamma(\stau_{\mathrm R}\to\tau\axino)$, for $\mbino=1.1\,\mstau$
  (solid lines) and $\mbino=1.01\,\mstau$ (dotted lines, red). We have
  set $|e_Q|=1/3$, $y=1$ and taken the limit $\maxino^2/\mstau^2\ll
  1$.}
\label{fig:lldiffmstfa}}
%
shows the discrepancy between the LL approximation \refeq{eq:2bLL} and
the full expression \refeq{eq:2lamp_Gamma}, 
$1-\Gamma(\stau_{\mathrm R}\to\tau\axino)_{\rm LL} / \Gamma(\stau_{\mathrm R}\to\tau\axino)$.
This is maximised for small $f_a$ and $\maxino$, and large $\mstau$
and $\mbino$ but small $\mbino/\mstau$.
We can also compare our LL result \eqref{eq:2bLL} to Eq.~(2) in
Ref.~\cite{Brandenburg:2005he}, where a cut-off
method~\cite{Covi:2004rb} had been used to calculate the 2-body decay.
Fixing the mass scale to $m=100~\GeV$ as suggested
in~\cite{Brandenburg:2005he}, the discrepancy is insensitive to $f_a$
as, in the leading logarithm, it enters in the same way in both cases.
However, due to this fixed mass scale used
in~\cite{Brandenburg:2005he}, its discrepancy to the exact full result
\eqref{eq:2lamp_Gamma} can be much larger than that of the LL
result~\eqref{eq:2bLL} in our calculation. For this case, the error
increases with increasing $\mstau$ and $\mbino$, and decreasing $f_a$.
In general, it represents an overestimation of the exact result and
the discrepancy reaches more than 35\% for $\mstau=1~\TeV$,
$\mbino=2\,\mstau$ and $f_a=10^9\,\GeV$.

The stau lifetime (including the non-LL terms) is plotted in
\reffig{fig:LifetimesqrtAb}
%
\FIGURE[t]{ 
\centerline{\epsfig{figure=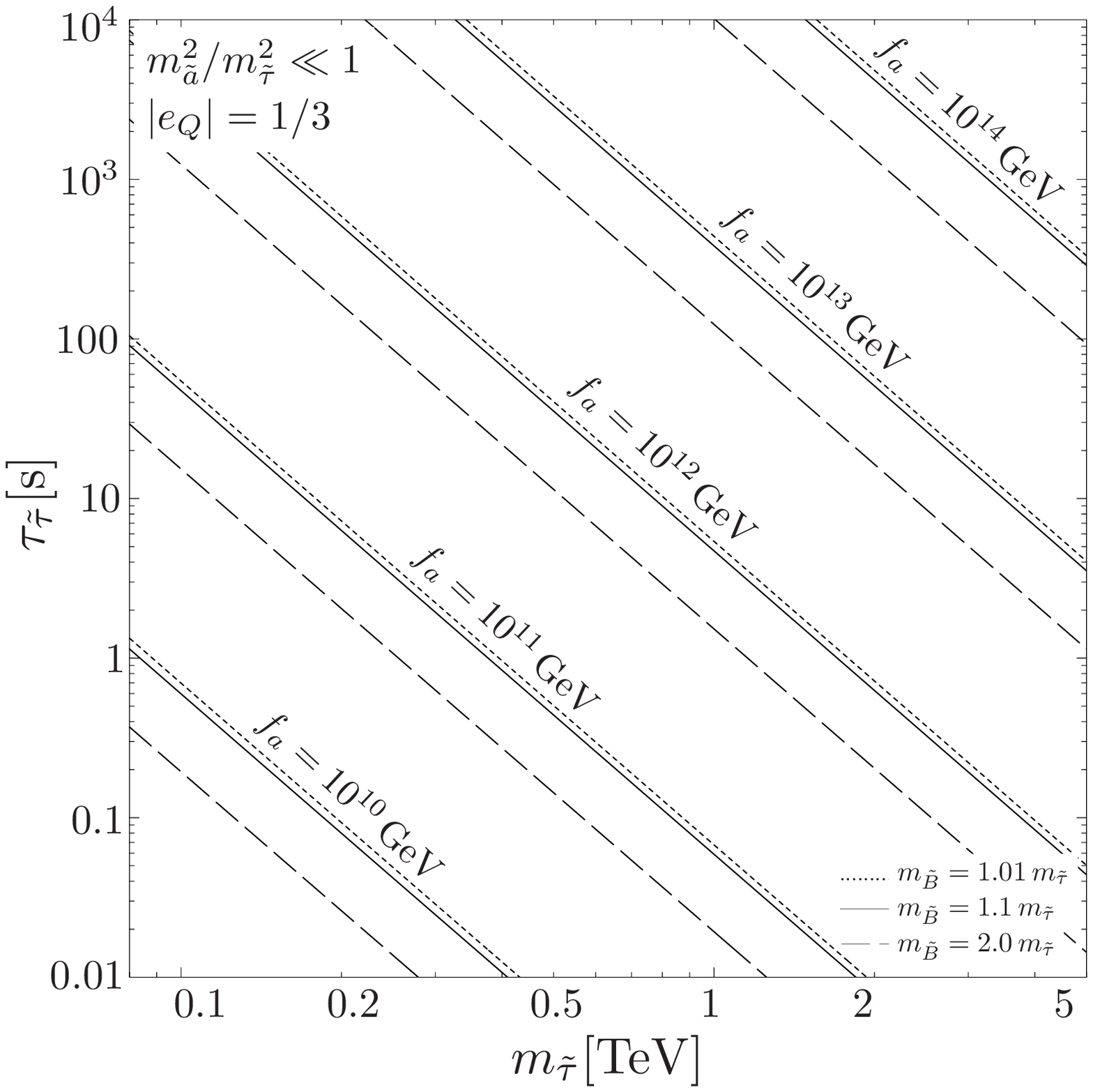, width=8cm}}
\caption{The lifetime of the stau NLSP as a function of $\mstau$ for
  $\maxino^2/\mstau^2\ll 1$, $|e_Q|=1/3$, $y=1$ and $f_a$ values
  ranging from $10^{10}$ to $10^{14}~\GeV$. The solid, dashed and
  dotted lines indicate $\tau_{\stau}$ for $\mbino=1.1\,\mstau$,
  $\mbino=2\,\mstau$ and $\mbino=1.01\,\mstau$, respectively.}
\label{fig:LifetimesqrtAb}}
%
as a function of its mass $\mstau$. It illustrates the sensitivity of
the lifetime not only to $\mstau$ but also to $f_a$ and $\mbino$. We
have taken the limit of $\maxino^2/\mstau^2\ll 1$ where the effect of
a changing $\maxino$ becomes negligible. For stau masses in the range
that would be accessible at the LHC, and the allowed $f_a$
range~\eqref{Eq:f_a_axion}, the lifetime can range from about
$10^{-4}\,\seconds$ to $10^4\,\seconds$. A lifetime larger than
$10^4\,\seconds$ can be excluded by catalysed big bang nucleosynthesis
(CBBN) as will be discussed in
\refsec{Sec:NucleosynthesisConstraints}.
In the analyses that follow, we use the full expression of the 2-body
decay to calculate the stau NLSP lifetime. For demonstrative purposes
we often set $|e_Q|=1/3$ and $y=1$. We also only show our results for
$\mstau\geq 80~\GeV$~\cite{Amsler:2008zzb}, the limit set from
searches for long-lived staus at the Large Electron Positron (LEP)
Collider.

\section{Cosmological Constraints on Axino Dark Matter Scenarios}
\label{Sec:AxinoConstraints}

In this section we use the results for the stau NLSP decays obtained
above to systematically study cosmological constraints on axino dark
matter scenarios. First we focus on the relic axino density, which is
constrained by the present dark matter density. Then we will explore
constraints from structure formation, where upper limits on the
present free-streaming velocities of axino dark matter emerge in order
to respect the power spectrum observed at small scales. Lastly, we
consider BBN which can considerably constrain the parameter space of
the considered SUSY hadronic axion models.
We will close this section with a comparative discussion of the
cosmological constraints.

For concreteness, we focus also in this section on the stau NLSP case.
Nevertheless, all addressed constraints -- except those associated
with photodissociation, which require a minor modification but will
turn out to be subleading anyhow -- apply also directly to the
alternative $\selectronR$ and $\smuonR$ NLSP cases after obvious
substitutions such as $\mstau\to m_{\selectronR/\smuonR}$.

For the following considerations, we assume a standard thermal history
with a post-inflationary reheating temperature in the range of
$\Tf^{\stau} < \TR < f_a$, where $\Tf^{\stau}\sim\mstau/25$ is the
freeze-out temperature of the staus, that is, the temperature at which
the staus decouple from the thermal bath in the early universe. By
considering $\TR < f_a$, we focus on scenarios in which the PQ
symmetry is not restored during or after reheating. In fact, we assume
that the PQ symmetry was broken before inflation and not restored
afterwards. For large $f_a$, the relic axion density $\Omega_a$ is
then governed by the initial misalignment angle $\Theta_i$ of the
axion field with respect to the CP-conserving position;
see~\cite{Sikivie:2006ni,Kim:2008hd} and references therein. This
allows us to keep the presented dark matter density constraints
conservative by assuming $\Omega_a\ll\OmegaDM$ which is possible even
for $f_a$ above $10^{14}\,\GeV$ since $\Theta_i$ can be sufficiently
small. 
We also assume implicitly properties of the saxion -- the bosonic
partner of the axino that appears in addition to the axion -- such that
its potential cosmological
effects~\cite{Kim:1992eu,Lyth:1993zw,Chang:1996ih,Hashimoto:1998ua,Kawasaki:2007mk}
are negligible in this study.
In more general settings, the saxion can be a late decaying particle
and as such be associated with significant entropy
production~\cite{Kim:1992eu,Lyth:1993zw,Chang:1996ih,Hashimoto:1998ua}
and thereby with a non-standard thermal history.
This can affect the cosmological constraints~\cite{Kawasaki:2007mk}
including those derived below, which rely on the hypothesis of a
standard thermal history.

\subsection{Constraints from the Present Dark Matter Density}
\label{Sec:OmegaDMConstraints}

In this section we discuss constraints imposed by the dark matter
density~\cite{Spergel:2006hy,Amsler:2008zzb}
\begin{align}
\Omega_{\CDM}^{3\sigma}h^2=0.105^{+0.021}_{-0.030} \ ,
\label{Eq:OmegaDM}
\end{align}
as obtained from the measurements of the cosmic microwave background
(CMB) anisotropies by the Wilkinson Microwave Anisotropy Probe (WMAP)
satellite, with $h=0.73^{+0.04}_{-0.03}$ denoting the Hubble constant
in units of $100\,\km\,\seconds^{-1}\,\Mpc^{-1}$. Note that this
nominal $3\sigma$ range is derived assuming a six-parameter flat
$\Lambda$CDM model. If the fit is performed in the context of a more
general model, a larger range is possible even with additional data
from other cosmological probes~\cite{Hamann:2006pf}.%
\footnote{The updated values extracted from WMAP 5- and 7-year
  data~\cite{Dunkley:2008ie,Larson:2010gs} have also been published
  and are consistent with the above range inferred from the WMAP
  3-year data set~\cite{Spergel:2006hy}.}
The dark matter density~\eqref{Eq:OmegaDM} constrains the relic axino
density,
\begin{align}
   \Omega_{\axino}
   =
   \Omega_{\axino}^{\mathrm{therm}/\TP} 
   + \Omega_{\axino}^{\NTP}
   \leq
   \Omega_{\CDM}
   - \Omega_a,
   \label{Eq:OmegaConstraint}
\end{align}
where we include contributions of thermal relic (therm) or thermally
produced (TP) axinos and of non-thermally produced (NTP) axinos from
the stau NLSP decays discussed above.

The freeze-out temperature of the axino $\Tf^{\axino}$ can be very
high due to its extremely weak interactions. For example,
$\Tf^{\axino}\sim 10^{9}\,\GeV$ ($10^{11}\,\GeV$) has been found for
$f_a=10^{11}\,\GeV$
($10^{12}\,\GeV$)~\cite{Rajagopal:1990yx,Asaka:2000ew,Brandenburg:2004du},
which is thereby similar to the one calcuated for
axions~\cite{Masso:2002np,Sikivie:2006ni,Graf:2010tv}.

For a reheating temperature $\TR>\Tf^{\axino}$, axinos were initially
in thermal equilibrium with the primordial plasma, and decoupled while
still relativistic, at $\Tf^{\axino} \gg \maxino$.  Accordingly, the
resulting thermal relic axino density is given
by~\cite{Rajagopal:1990yx,Asaka:2000ew,Covi:2001nw,Brandenburg:2004du}
\begin{align}
   \Omega_{\axino}^{\mathrm{therm}}h^2 
   &= 0.05 \left(\frac{230}{g_{*S}(\Tf^{\axino})}\right) 
   \left(\frac{\maxino}{100~\eV}\right) \ , 
\label{eq:AxinoDensityTherm} \\
   &\approx \maxino/(2~\keV)\ , 
\label{eq:AxinoDensityThermapprox}
\end{align}
where $g_{*S}(\Tf^{\axino})$ is the number of effectively massless
degrees of freedom at decoupling. As $\Tf^{\axino}$ is typically very
high, we can consider all particles in the MSSM as relativistic at
this time, and with the addition of the axion and the axino, we can
use $g_{*S}(\Tf)\approx 230$, which leads to
\eqref{eq:AxinoDensityThermapprox} above. The
constraint~\eqref{Eq:OmegaConstraint} thus translates into
$\maxino\lesssim 0.2~\keV$~\cite{Rajagopal:1990yx,Covi:2001nw}
even in the most conservative case with 
$\Omega_{\axino}^{\NTP}+\Omega_a\ll\Omega_{\CDM}$.
As we will see in \refsec{Sec:SmallScaleStructure}, this places
thermal relic axinos in the `hot' dark matter category, where even
more restrictive constraints apply.
It should however be noted that an axion cold dark matter (CDM)
scenario, $\Omega_a\approx\Omega_{\CDM}$, in which thermal relic
axinos with $\maxino\ll 0.2~\keV$ provide a tolerable contribution to
hot dark matter is a viable option~\cite{Brandenburg:2004du}; see also
\refsec{Sec:SmallScaleStructure} below.
Interestingly, most of the BBN constraints discussed in
\refsec{Sec:NucleosynthesisConstraints} will also apply to such
scenarios with a very light axino LSP and a thermal relic stau NLSP.

For $\TR<\Tf^{\axino}$, axinos were never in thermal equilibrium with
the thermal bath but can still be produced via scattering of coloured
particles in the hot MSSM
plasma~\cite{Asaka:2000ew,Covi:2001nw,Brandenburg:2004du,Gomez:2008js,Strumia:2010aa}.
Working within the framework of SUSY QCD, hard thermal loop (HTL)
resummation and the Braaten-Yuan prescription \cite{Braaten:1991dd}
have been used to calculate the rate of this thermal axino production
in a gauge-invariant way and consistent to leading order in the strong
coupling
$\alpha_{\mathrm{s}}=g^2_{\mathrm{s}}/4\pi$~\cite{Brandenburg:2004du}.
Considering $\TR$ as the initial temperature of the
radiation-dominated epoch, this rate leads to the following expression
for the relic density~\cite{Brandenburg:2004du,Freitas:2009fb}
\bea 
   \Omega_{\axino}^{\TP}h^2 
   = 
   5.5\,g^6_{\mathrm{s}}(\TR) 
   \ln\left( \frac{1.211}{g_{\mathrm{s}}(\TR)}\right) 
   \bigg(\frac{\maxino}{0.1\,\GeV}\bigg)
   \left(\frac{10^{11}\,\GeV}{\fPQ}\right)^2 
   \left(\frac{\TR}{10^4\,\GeV}\right) 
   \ ,
\label{Eq:AxinoDensityTP}
\eea 
where we evaluate $g_s(\TR)=\sqrt{4\pi\alpha_s(\TR)}$ according to its
1-loop renormalisation group running within the MSSM from
$\alpha_s(\mZ)=0.1176$ at $\mZ=91.1876~\GeV$.
We work with~\eqref{Eq:AxinoDensityTP} since it emerges from a
gauge-invariant treatment that is consistent to leading order in
$g_s$.  Note however that this treatment relies on the weak coupling
limit, $g_{\mathrm{s}}\ll 1$, and thus on high temperatures, $T\gg
10^4\,\GeV$~\cite{Brandenburg:2004du}.%
\footnote{For alternative approaches that consider also thermal axion
  production at lower temperatures,
  see~\cite{Gomez:2008js,Strumia:2010aa} where~\cite{Strumia:2010aa}
  also includes an additional squark-squark-gluino-axino vertex and a
  different $\TR$ definition.}

By confronting $\Omega_{\axino}^{\TP}$ with the
constraint~\eqref{Eq:OmegaConstraint}, one obtains upper limits on
$\TR$ for given $\maxino$ and $f_a$ (or equivalently lower limits on
$f_a$ for given $\maxino$ and
$\TR$)~\cite{Covi:2001nw,Brandenburg:2004du,Choi:2007rh,Kawasaki:2007mk,Baer:2008yd,Freitas:2009fb,Strumia:2010aa}.
For~\eqref{Eq:AxinoDensityTP} and
$\Omega_{\axino}^{\NTP}+\Omega_a\ll\Omega_{\CDM}$, we have shown these
limits in Fig.~1 of Ref.~\cite{Freitas:2009fb}.
For example, insisting on a high reheating temperature of
$\TR\gsim 10^9\,\GeV$,
which is desirable when considering the viability of standard thermal
leptogenesis, and on axinos being cold dark matter, 
$\maxino\gtrsim 100\,\keV$ (explored more in \refsec{Sec:SmallScaleStructure}),
we are forced to consider fairly high $f_a$ values of 
$f_a>3\times 10^{12}\,\GeV$ \cite{Freitas:2009fb}.  
Note however, that in this paper we do not constrain ourselves to any
specific baryogenesis scenario or range of $f_a$ values other
than~\eqref{Eq:f_a_axion}.

Along with the thermal relic or thermally produced axinos, there are
also non-thermally produced axinos from out-of-equilibrium NLSP
decays.
Due to the highly suppressed coupling of the axino to MSSM particles,
the stau NLSP typically decays after its freeze-out, i.e., when
$T\ll\Tf^{\stau}$. We thus have to consider only the yield of the stau
NLSP after freeze-out,
$Y_{\st}\equiv n_{\st}/s$, 
where $s$ is the total entropy density of the Universe and $n_{\st}$
the number density that the stau NLSP would have today, if it had not
decayed.
As each stau decays into one axino, the relic density of the axino
from this non-thermal production is then given by the following
expression~\cite{Bonometto:1989vh,Bonometto:1993fx,Covi:1999ty,Covi:2001nw,Covi:2004rb}
\bea 
   \Omega_{\axino}^{\NTP}h^2 
   = 
   \maxino\, Y_{\st}\, s(T_0) h^2 / \rho_{\mathrm{c}} \ ,
\label{Eq:AxinoDensityNTP}
\eea 
where 
$\rho_{\mathrm{c}}/[s(T_0) h^2] = 3.6~\eV$
as obtained from the critical density 
$\rho_{\mathrm{c}}/h^2=8.1\times10^{-47}\,\GeV^4$, 
the present temperature 
$T_0=2.73\,\K\equiv 0.235~\meV$ 
and the number of effectively massless degrees of freedom governing
the entropy density today
$g_{*S}(T_0)=3.91$. 

The stau NLSP yield $Y_{\st}$ can be calculated numerically and is
model-dependent~\cite{Asaka:2000zh,Fujii:2003nr,Pradler:2006hh,Berger:2008ti,Ratz:2008qh,Pradler:2008qc}.
For the purposes of our work, with the focus on the $\stauR$ NLSP
setting, we consider the following three characteristic
approximations:
\bea 
   Y_{\stau} 
   \simeq 
   \kappa \times 10^{-12} \left(\frac{m_{\stau}}{1~\TeV}\right), 
   \quad 
   \kappa = 0.7,\,1.4,\,2.0
\label{Eq:Ystau}
\eea 
where $\kappa$ quantifies representative differences in the yield due
to possible mass degeneracies of the stau NLSP to other sparticles and
hence the extent of coannihilation.
The value $\kappa=0.7$ corresponds to the case with
$\mbino=1.1\,\mstau$ and $\mstau\ll m_{\sel,\smu}$, in which
primordial stau annihilation involves only (anti-)staus in the initial
state~\cite{Asaka:2000zh}.%
\footnote{The bino mass $\mbino=1.1\,\mstau$ considered in
  Ref.~\cite{Asaka:2000zh} represents a typical mass splitting in
  regions with $\mbino>\mstau$ encountered in scenarios such as the
  constrained MSSM (CMSSM).}
The yield associated with $\kappa=1.4$ is encountered if there is
either additional stau--slepton coannihilation corresponding to
$\mstau\lesssim m_{\sel,\smu}<1.1\,\mstau$~\cite{Asaka:2000zh} or
additional stau--bino coannihilation corresponding to
$\mstau\lesssim\mbino<1.1\,\mstau$ (see $Y_{\stau}$ contours close to
the dashed line in the right panel of Fig.~3
in~\cite{Pradler:2006hh}).
For an approximate degeneracy of $\mstau$ with both $m_{\sel,\smu}$
and $\mbino$, simultaneous stau--slepton--bino coannihilation can lead
to an even larger yield with $\kappa=2.0$ in \eqref{Eq:Ystau} (see
$Y_{\stau}$ contours close to the dashed line in the left panel of
Fig.~3 of~\cite{Pradler:2006hh}).

Putting the above together, we can rewrite~\eqref{Eq:OmegaConstraint}
to obtain the dark matter constraint on the stau abundance prior to
decay, $Y_{\stau}\leq Y_{\stau\,\CDM}^{\max}$ with
\bea 
   Y_{\stau\,\CDM}^{\max} 
   = 
   4.5\times 10^{-11} 
   \left(\frac{\Omega_{\dm}^{\max} 
       - \Omega_{\axino}^{\TP} 
       - \Omega_a}{0.126/h^2} 
   \right) 
   \left(\frac{10~\GeV}{\maxino}\right)\ .
\label{Eq:DMconstraint}
\eea
\reffigure{Fig:DMConstraint} 
%
\FIGURE[t]{
\centerline{\epsfig{figure=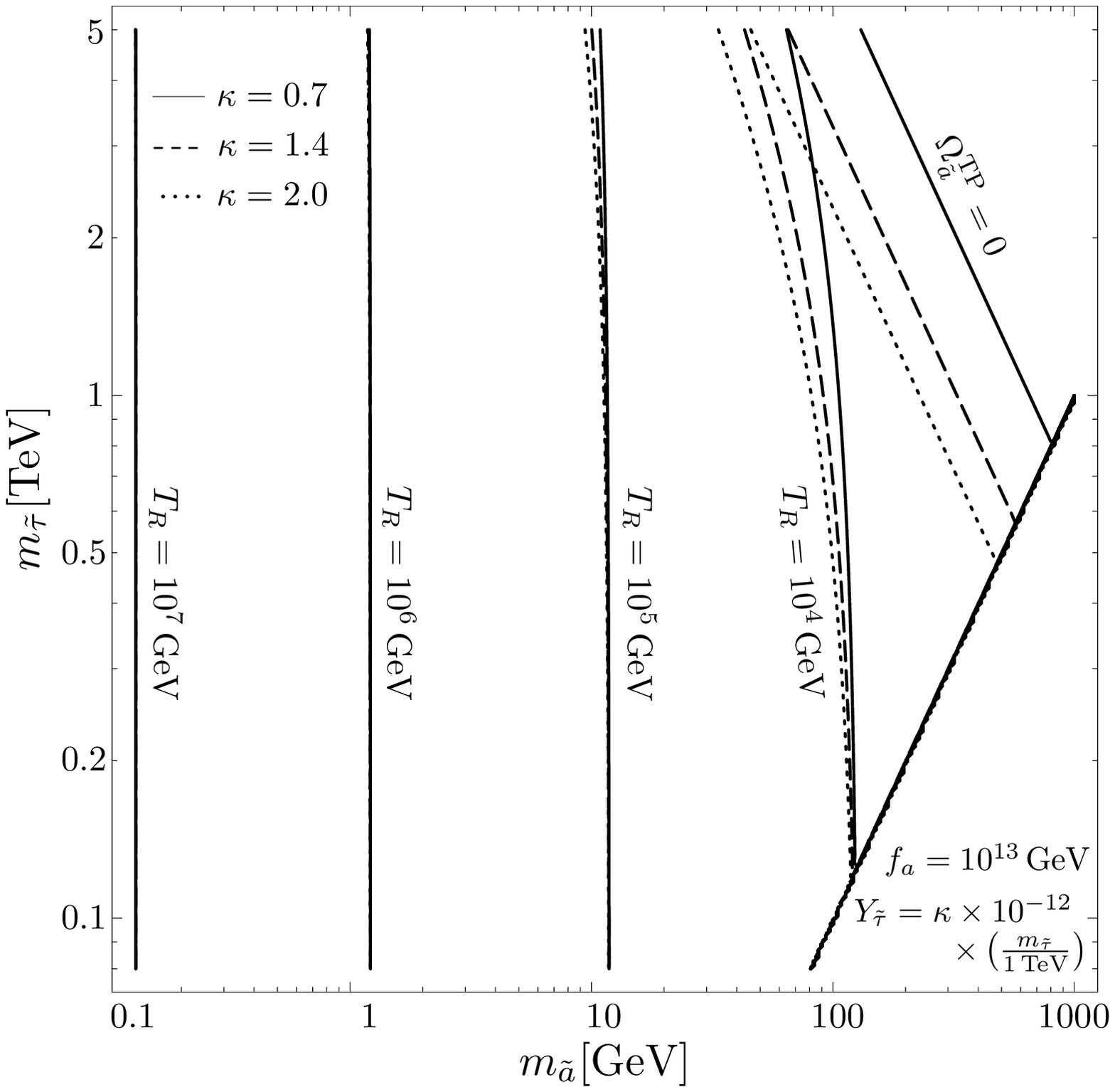, width=8.cm}}
\caption{Constraint imposed by
  $\Omega_{\axino}^{\TP}h^2+\Omega_{\axino}^{\NTP}h^2 \leq 0.126$ for
  $f_a=10^{13}\,\GeV$, $\TR=10^7$, $10^6$, $10^5$, $10^4\,\GeV$ and
  $\TR$ such that $\Omega_{\axino}^{\TP}$ is negligible. The solid,
  dashed and dotted lines are obtained for the thermal relic stau
  yield \eqref{Eq:Ystau} with $\kappa=0.7$, $1.4$ and $2.0$,
  respectively. For given $\TR$ and $\kappa$, the region to the right
  of the corresponding line is disfavoured by the dark matter
  constraint.}
\label{Fig:DMConstraint}}
%
shows the corresponding constraints on $\maxino$ and $\mstau$ that
emerge when~\eqref{Eq:Ystau} is confronted
with~\eqref{Eq:DMconstraint} for $\Omega_{\dm}^{\max}h^2=0.126$,
$\Omega_a\ll\Omega_{\dm}$, $f_a=10^{13}\,\GeV$ and $\TR=10^7$, $10^6$,
$10^5$, $10^4\,\GeV$ and a sufficiently small $\TR$ value, where
$\Omega_{\axino}^{\TP}$ can be neglected.
The solid, dashed and dotted lines are obtained respectively for
$\kappa=0.7$, $1.4$ and $2.0$. The disfavoured region is the one to
the right of the respective curve.
Differences with respect to $\kappa$ become visible/significant only
for $\maxino>10~\GeV$, which illustrates that $\Omega_{\axino}^{\NTP}$
is negligible for $\maxino\lesssim 10~\GeV$ in the considered $\mstau$
range.
By increasing (decreasing) $f_a$ by one order of magnitude, the $\TR$
value at a given curve will increase (decrease) by two orders of
magnitude.
Note that the curves can also be read as the upper limits on $\TR$
that are imposed by the dark matter
constraint~\eqref{Eq:OmegaConstraint} for given $\maxino$, $\mstau$,
and $\kappa$ when $\Omega_a\ll\Omega_{\dm}$.

In the coming sections, we will take the most conservative point of
view with
$\Omega_{\axino}^{\TP}+\Omega_a\ll\Omega_{\dm}$
(i.e., with sufficiently small $\TR$ and $\Theta_i$)
when showing the dark matter constraint.
By confronting~\eqref{Eq:DMconstraint} with $Y_{\stau}$, as described
by \refeq{Eq:Ystau}, the associated constraint will then disfavour
$\maxino$ and $\mstau$ combinations that are located to the
right/above of the grey bands in \reffigs{fig:FSvel} and~\ref{fig:MB}
below.
Note however that the behaviour of the dark matter constraint for
sizeable $\TR$ leading to sizeable $\Omega_{\axino}^{\TP}$ can easily
be inferred from \reffig{Fig:DMConstraint}.

One can also consider a more phenomenologically-driven approach where
one assumes that all of the observed dark matter density is composed
of axinos from NLSP decays,
$\Omega_{\dm}\simeq\Omega_{\tilde{a}}^{\NTP}$.  
Irrespective of the freeze-out behaviour of the stau NLSP, this
requires the following yield
\bea 
   Y_{\stau} 
   = 
   3.8\times 10^{-11} 
   \left(\frac{\Omega_{\dm}}{0.105/h^2}\right) 
   \left(\frac{10~\GeV}{\maxino}\right)
\label{Eq:YSWIMP} ,
\eea
(and again $\TR$ and $\Theta_i$ sufficiently small such that
$\Omega_{\axino}^{\TP}+\Omega_a\ll\Omega_{\dm}$).
For small values of $\maxino$, typical values of the thermal relic
stau yield, as described by~\eqref{Eq:Ystau}, cannot
reach~\eqref{Eq:YSWIMP}. The higher yield can however be accounted for
by additional non-standard sources of staus such as late-decaying
scalar fields.
In the coming sections, we explore cosmological constraints also for
scenarios with $Y_{\stau}$ described by~\eqref{Eq:YSWIMP}. 

Before proceeding, let us stress that there exists a lower limit on
$\maxino$ below which~\eqref{Eq:YSWIMP} is not consistent with the
assumption of a standard thermal history. Indeed, that assumption
implies that the energy density of the non-relativistic stau NLSPs
prior to decay should not exceed the one of radiation:
$\rhostau\leq\fstau\,\rhorad$ with a fraction $\fstau\sim 0.1$ or
smaller.
Imposing this requirement at the time of the stau NLSP decay,
$t=\tau_{\stau}$, we find
\begin{align}
  \maxino
  \geq 
  & 
  \,\frac{33.2~\MeV}{\fstau} \, 
  \left(\frac{50}{\log[y^2 f_a^2/(2\mstau^2)]}\right)
  \left(\frac{1/3}{e_Q}\right)^2 \!
  \left(\frac{f_a}{10^{13}\,\GeV}\right)\! 
\nonumber\\ 
   & \times
   \left(\frac{\mstau}{100~\GeV}\right)^{1/2} \!
   \left(\frac{100~\GeV}{\mbino}\right) \! 
   \left(\frac{g_{*S}(\tau_{\stau})}{3.91}\right) 
   \left(\frac{3.36}{g_{*}(\tau_{\stau})}\right)^{3/4} ,
\label{eq:maxinoStdCosmology}
\end{align}
as obtained for $\maxino\ll\mstau$ when using the
yield~\eqref{Eq:YSWIMP} and the LL approximation of $\tau_{\stau}$ in
the sudden decay approximation in which all $\stauR$'s are considered
to decay at $\tau_{\stau}$.
Here $g_*(\tau_{\stau})$ and $g_{*S}(\tau_{\stau})$ are the effective
number of relativistic degrees of freedom governing the energy density
and entropy density, respectively, at the time the stau NLSP decays
into the axino LSP.
For example, if the staus decay after electrons and positrons become
non-relativistic, $g_{*S}(\tau_{\stau})=g_{*S}(T_0)=3.91$ and
$g_{*}(\tau_{\stau})=g_{*}(T_0)=3.36$.
Note that~\eqref{eq:maxinoStdCosmology} with $\fstau\sim 0.1$ is
respected for the $\maxino$ range considered in \reffigs{fig:FSvel}
and~\ref{fig:SWIMPMB} below.

\subsection{Constraints from Structure Formation}
\label{Sec:SmallScaleStructure}

Let us now address the constraints on axino dark matter scenarios
imposed by the matter power spectrum and related studies.
Depending on details of their primordial origin, axinos can fall into
the categories of cold, warm, and hot dark matter.
As warm dark matter (WDM) or hot dark matter (HDM), they can be
associated with significant suppression of structures on scales below
a potentially sizeable comoving free-streaming scale $\lambda_{\FS}$.
Studies of cosmic structures can thus provide upper limits on
$\lambda_{\FS}$ or, equivalently, on the present free-streaming
velocity $v_{\mathrm{FS}}^0$ of axino dark matter.
We compile various limits on $v_{\mathrm{FS}}^0$ obtained in WDM
studies and confront those limits with the free-streaming velocities
obtained for thermal relic axinos, thermally produced axinos, and
axinos from stau NLSP decays.

In WDM investigations of cosmic structure formation, it is typically
assumed that all dark matter consists of one species with mass
$m_{\X}$ that was once in thermal equilibrium with the primordial
plasma and freezes out while relativistic at $\Tf^{\X}\gg m_{\X}$.
Correspondingly, for a Majorana fermion of spin 1/2, $\Omega_{\X} h^2$
is given by~\eqref{eq:AxinoDensityTherm} after the obvious
substitutions, where $g_{*S}(\Tf^{\X})$ is now fixed by the
requirement $\Omega_{\X}=\Omega_{\CDM}$ for a given $m_{\X}$.
With an initial root mean squared momentum at freeze-out of
$\langle p(\Tf^{\X}) \rangle = 3.151\,\Tf^{\X}$,
the present root mean squared free-streaming velocity is governed by
$m_{\X}$:
\bea
(v_{\mathrm{FS}}^{\mathrm{rms},0})^{\X}
= 0.75 \,\frac{\mathrm{km}}{\mathrm{s}}
\left(\frac{\Omega_{\dm}h^2}{0.105}\right)^{1/3}
\left(\frac{100~\eV}{m_{\X}}\right)^{4/3} 
\ .
\label{eq:vFS0WDM}
\eea
Using this expression, we obtain the upper limits on
$v_{\mathrm{FS}}^0$ listed in Table~\ref{tab:WDMlimits} from the
respective WDM constraints provided in the literature in terms of a
lower limit on $m_{\X}$.
In Table~\ref{tab:WDMlimits} we give each limit also in terms of an
upper limit on the free-streaming scale 
\begin{align}
  \lambda_{\FS}
  \simeq \!\!
  \int_{0}^{t_{\eq}} \!\! dt \frac{v(t)}{a(t)}
  = 2 v^0_{\FS}\, t_{\eq} (1+z_{\eq})^2
  \log\!\left(\!
    \sqrt{1+\frac{1}{(v^{0}_{\FS})^2(1+z_{\eq})^2}}+\frac{1}{v^0_{\FS}(1+z_{\eq})}
  \!\right)
\label{Eq:LambdaFS_WDM} 
\end{align}
with the cosmic scale factor $a(t)$, the dark matter velocity
$v(t)=|\vec{p}(t)|/\sqrt{|\vec{p}(t)|^2 + m_{\X}^2}$ and its present
free-streaming velocity $v_{\mathrm{FS}}^0$, which is assumed to be
non-relativistic.
The time and redshift at matter--radiation equality are given
respectively by $z_\eq\simeq 3200$ and 
$t_{\eq}\simeq 1.8 \times 10^{12}\,\seconds$ 
for a matter density parameter of
$\OmegaM h^2=0.133$~\cite{Nakamura:2010zzi}.
Note that the relation~\eqref{Eq:LambdaFS_WDM} between $\lambda_{\FS}$
and $v_{\mathrm{FS}}^0$ is obtained in the limit that the considered
dark matter population was produced (or decoupled from the primordial
plasma) at $t_i$ much before radiation--matter equality, i.e., for
$t_i\ll t_{\eq}$~\cite{Lin:2000qq,Hisano:2000dz,Steffen:2006hw}.
%
\TABLE[t]{
  \caption{A selection of constraints on WDM particles from
    observations and numerical simulations. The lower limits on
    $m_{\X}$ are taken from the corresponding references. The values
    for $(v_{\FS}^{\mathrm{rms},0})^{\mathrm{max}}$ and
    $(\lambda_{\FS})^{\mathrm{max}}$ are derived from
    $m_{\WDM}^{\mathrm{min}}$ using \eqref{eq:vFS0WDM} and
    \eqref{Eq:LambdaFS_WDM}, respectively, with $z_\eq\simeq 3200$ and
    $t_{\eq}\simeq 1.8 \times 10^{12}\,\seconds$. The limits
    from~\cite{Polisensky:2010rw,Viel:2005qj,Seljak:2006qw,Viel:2006kd,Viel:2007mv}
    were quoted as $2\sigma$ limits, and a $3\sigma$ difference with
    respect to observations was reported in~\cite{Narayanan:2000tp},
    while confidence levels were not quantified
    in~\cite{Maccio':2009rx,Dalcanton:2000hn,Boyanovsky:2007ay}.}
\begin{tabular}{l l l l c}
\toprule
& $m_{\WDM}^{\mathrm{min}}$ 
& $(v_{\FS}^{\mathrm{rms},0})^{\mathrm{max}}$ 
& $(\lambda_{\FS})^{\mathrm{max}}$ 
&  \\
Probe 
& [keV]
& [km/s]                                                                                                
& [Mpc]
&  Reference \\
\hline 
\\
Number of MW satellites 
& 2.1 
& 0.01 
& 0.2 
& \cite{Polisensky:2010rw} 
\\
Luminosity function of MW satellites 
& 1 
& 0.03 
& 0.4 
& \cite{Maccio':2009rx} 
\\
Phase space density in dSphs 
& $0.7$ 
& $0.06$ 
& 0.5 
& \cite{Dalcanton:2000hn} 
\\
Phase space density in dSphs 
& 0.6--1.5 
& 0.02--0.07 
& 0.2--0.7 
& \cite{Boyanovsky:2007ay} 
\\
Lyman-$\alpha$ forest ($z\simeq 3$)
& $0.75$ 
& $0.05$ 
& 0.5 
& \cite{Narayanan:2000tp} 
\\
Lyman-$\alpha$ forest ($z\sim 2.5$)
& $0.55$ 
& $0.08$ 
& 0.7 
& \cite{Viel:2005qj} 
\\
Lyman-$\alpha$ forest ($2\lesssim z \lesssim 4$) 
& 2--2.4 
& 0.01 
& 0.1 
& \cite{Seljak:2006qw,Viel:2006kd} 
\\
Lyman-$\alpha$ forest ($2.0<z<6.4$) 
& 4 
& 0.005 
& 0.07 
& \cite{Viel:2007mv} 
\\
\bottomrule
\label{tab:WDMlimits}
\end{tabular}}

Table~\ref{tab:WDMlimits} lists representative limits inferred from
various astrophysical observations and studies of structure formation
in WDM cosmologies.%
\footnote{For a recent review with references to further WDM studies,
  we refer to~\cite{Primack:2009jr}.}
 Included are constraints from the number of Milky Way (MW)
satellites~\cite{Polisensky:2010rw} and their luminosity
function~\cite{Maccio':2009rx}. Here insights into the number and
properties of satellite galaxies around our MW are taken into account
which became accessible due to the results of the Sloan Digital Sky
Survey (SDSS) only recently.
Another source for the $m_{\WDM}^{\mathrm{min}}$ constraint is the
maximum observed phase space density in dwarf spheroidal galaxies
(dSphs)~\cite{Dalcanton:2000hn,Boyanovsky:2007ay}.
Moreover, one of the most sensitive probes is the Lyman-$\alpha$
forest. Associated studies give some of the most constraining limits
on the thermal WDM
mass~\cite{Narayanan:2000tp,Viel:2005qj,Seljak:2006qw,Viel:2006kd,Viel:2007mv}.
The data are available over a large range of redshifts $z$ and allow
us to study the power spectrum down to very small scales.  In
particular, Ref.~\cite{Viel:2007mv} uses High Resolution Echelle
Spectrometer (HIRES) data in combination with data from SDSS to place
a lower limit on the WDM mass of $m_{\WDM}\gtrsim 4~\keV$. As can be
seen in~\reftable{tab:WDMlimits}, this $4~\keV$ limit is about a
factor of two (or more) larger than the limits coming from other
observations.
The spread in the $m_{\WDM}$ limits reflects the fact that WDM constraints 
are subject to ongoing research and to potential systematic uncertainties.
In the following we will put some focus on 
\begin{align}
m_{\WDM}\gtrsim 1~\keV 
& \quad\equiv\quad 
v_{\FS}^{\mathrm{rms},0}\lesssim 0.03~\km/\seconds
\quad\,\,\,\mbox{(conservative)},
\label{eq:vFSconservative}
\\
m_{\WDM}\gtrsim 4~\keV
& \quad\equiv\quad 
v_{\FS}^{\mathrm{rms},0}\lesssim  0.005~\km/\seconds
\quad\mbox{(most restrictive)}.
\label{eq:vFSrestrictive}
\end{align}
These two limits will also indicate the range into which most of the
limits fall that are given in~\reftable{tab:WDMlimits}. Moreover,
future studies such as investigations of WDM constraints from cosmic
shear~\cite{Markovic:2010te} are expected to be sensitive to
$m_{\WDM}$ limits in that region.

For thermal relic axinos which freeze-out with a mean squared momentum
of
$\langle p(\Tf^{\axino}) \rangle = 3.151\,\Tf^{\axino}$,
the root mean squared value of their present free-streaming velocity
is given by
\bea
(v_{\mathrm{FS}}^{\mathrm{rms},0})_{\axino}^{\mathrm{therm}}
= 0.57 \, \frac{\mathrm{km}}{\mathrm{s}}
\left(\frac{230}{g_{*S}(\Tf^{\axino})}\right)^{1/3}
\left(\frac{100~\eV}{m_{\axino}}\right) 
\ .
\label{eq:vFS0ThermAxino}
\eea
Thus, the axino mass limit derived from the dark matter density
constraint in \refsec{Sec:OmegaDMConstraints},
$\maxino\lesssim 0.2~\keV$~\cite{Rajagopal:1990yx,Covi:2001nw},
implies 
$(v_{\mathrm{FS}}^{\mathrm{rms},0})_{\axino}^{\mathrm{therm}}\gtrsim
0.3~\km/\seconds$
which is an order of magnitude above the conservative
limit~\eqref{eq:vFSconservative}, i.e., thermal relic axinos would
suppress small-scale structure too much when providing all the dark
matter.  They are thus considered too `hot' and are disfavoured as the
dominant component of today's dark matter density.
Nevertheless, they can provide hot dark matter in addition to
neutrinos and can coexist with some other species providing CDM such
as axions~\cite{Brandenburg:2004du}.
For such a setting, we find
\bea
\maxino\lesssim 37~\eV
\label{eq:AxinoHDMConstraint}
\eea
by translating the gravitino mass limit $\mgravitino \lesssim 16~\eV$
obtained at the $2\sigma$ level for mixed models with CDM and thermal
relic gravitinos with $g_{*S}(\Tf^{\gravitino})\simeq 100$
in~\cite{Viel:2005qj} to the mixed case with CDM and thermal relic
axinos with $g_{*S}(\Tf^{\axino})\simeq 230$.
In fact, a comparison of the Lyman-$\alpha$ limits listed
Table~\ref{tab:WDMlimits} indicates that a more restrictive $\maxino$
limit could emerge once such a mixed scenario is confronted with the
data sets considered in~\cite{Seljak:2006qw,Viel:2006kd,Viel:2007mv}.

For thermally produced axions, we estimate their mean squared momentum
at the reheating temperature $\TR$, where their production is most
efficient, as
$\langle p(\TR) \rangle = 3.151\,\TR$.
Although never in thermal equilibrium, this choice is motivated by the
fact that these axinos are produced in scattering of particles that
are in thermal equilibrium with the primordial plasma.
We estimate
$(v_{\mathrm{FS}}^{\mathrm{rms},0})_{\axino}^{\mathrm{TP}}$
accordingly by \eqref{eq:vFS0ThermAxino} after the substitution
$\Tf^{\axino}\to\TR$.
As mentioned in the previous section, we can use $g_{*S}\approx 230$
for large $\TR$. 
Clearly, for $\maxino\gtrsim 100\,\keV$, the present free-streaming
velocity of these thermally produced axinos is well below the
constraints shown in \reftable{tab:WDMlimits}, which places them into
the category of CDM.
For $12~\keV\lesssim\maxino<100\,\keV$, thermally produced axions can
be considered as WDM but will still be compatible with the most
restrictive limit in \reftable{tab:WDMlimits}:
$(v_{\FS}^{\mathrm{rms},0})_{\axino}^{\TP}\lesssim 0.005~\km/\seconds$.
The more conservative limit~\eqref{eq:vFSconservative} would even
allow for $\maxino$ as light as about $2~\keV$.
The lower limit on $\maxino$ can become relevant for models of
inflation and baryogenesis in axino dark matter scenarios with
$\Omega_{\axino}^{\TP}\simeq\Omega_{\CDM}$: the lighter $\maxino$
allowed by structure formation the higher will be the $\TR$ value
allowed by the dark matter constraint~\eqref{Eq:OmegaConstraint} for a
given $f_a$; see also \reffig{Fig:DMConstraint} above and Fig.~1 in
Ref.~\cite{Freitas:2009fb}.

For axinos produced non-thermally in stau NLSP decays, the present
free-streaming velocity $(v_{\FS}^{0})_{\axino}^{\NTP}$ can be
inferred from the 2-body stau decay calculated
in~\refsec{Sec:NLSPDecays}. Since the 2-body decay is the dominant
decay channel, most non-thermally produced axinos have an initial
momentum given by
\bea
|\vec{p}_{\axino}(t_i)| 
= \frac{\mstau^2-\maxino^2-m_{\tau}^2}{2\,\mstau} 
\ .
\label{Eq:paxino}
\eea
By considering the redshift of~\eqref{Eq:paxino} due to the expansion
of the universe, one then obtains their present free-streaming
velocity~\cite{Jedamzik:2005sx},
\bea
(v_{\FS}^{0})_{\axino}^{\NTP} 
= 
\frac{T_0}{\maxino}\,
\frac{\mstau^2-\maxino^2-m_{\tau}^2}{2\mstau}
\left(\frac{g_{*S}(T_0)}{g_{*S}(\tau_{\stau})}\right)^{1/3}
\left( 
  \frac{4\pi^2 g_*(\tau_{\stau}) 
    \tau_{\stau}^2}{90\, \MPl^2} 
\right)^{1/4} 
\ ,
\label{Eq:vFS0axinoNTP}
\eea
which is valid in the usual case that they are produced during the
radiation-dominated epoch, $t_i\ll t_{\eq}$, and that they are
non-relativistic today, 
$|\vec{p}_{\axino}(t_0)|=\maxino (v_{\FS}^{0})_{\axino}^{\NTP}$.
Here $\MPl=2.4\times10^{18}\,\GeV$ is the reduced Planck mass.
Note that we use the sudden decay approximation in which all
staus are considered to decay at time $t_i=\tau_{\stau}$.%
\footnote{For treatments considering the phase space distribution
  beyond the sudden decay approximation, see
  Refs.~\cite{Bonometto:1989vh,Bonometto:1993fx,Kawasaki:1992kg,Kang:1993xz,Pierpaoli:1994am,Covi:2001nw,Kaplinghat:2005sy}.}
In the following, we will use our full result~\eqref{eq:2lamp_Gamma}
to calculate $(v_{\FS}^{0})_{\axino}^{\NTP}$ in the limit 
$m_{\tau}\to 0$.
Nevertheless, the main features are already described by the following
estimate obtained with the LL approximation~\refeq{eq:2bLL},
\begin{align}
  (v_{\FS}^{0})_{\axino}^{\NTP}  
  \approx 
  & 
  \,\, 1.9 \,\frac{\km}{\seconds} 
  \left(\frac{50}{\log[y^2 f_a^2/(2\mstau^2)]}\right)
   \left(\frac{1/3}{e_Q}\right)^2 \!
   \left(\frac{f_a}{10^{13}\,\GeV}\right)\! 
   \left(\frac{1\,\GeV}{\maxino}\right) \! 
\nonumber\\ 
   & \times
   \left(\frac{\mstau}{100~\GeV}\right)^{1/2} \!
   \left(\frac{100~\GeV}{\mbino}\right) \! 
   \left(\frac{3.91}{g_{*S}(\tau_{\stau})}\right)^{1/3} 
   \left(\frac{g_{*}(\tau_{\stau})}{3.36}\right)^{1/4} .
\label{eq:vFS0LL}
\end{align}
While a related estimate was given in~\cite{Jedamzik:2005sx},
\eqref{eq:vFS0LL} illustrates explicitly the dependence on the
logarithmic factor and on $e_Q$.

Although axinos produced in primordial stau NLSP decays have a
non-thermal distribution, we will now confront
$(v_{\FS}^{0})_{\axino}^{\NTP}$ with the
limits~\eqref{eq:vFSconservative} and~\eqref{eq:vFSrestrictive}
inferred from WDM studies (cf.~\reftable{tab:WDMlimits}) assuming a
thermal distribution.
This approach is motivated by the results of~\cite{Lin:2000qq} which
show that velocity limits derived with \eqref{eq:vFS0WDM} from
$m_{\WDM}$ limits can, to a reasonable extent, also be applied to the
case with a monochromatic spectrum.

\reffigure{fig:FSvel}
%
\FIGURE[t]{
\centerline{\epsfig{figure=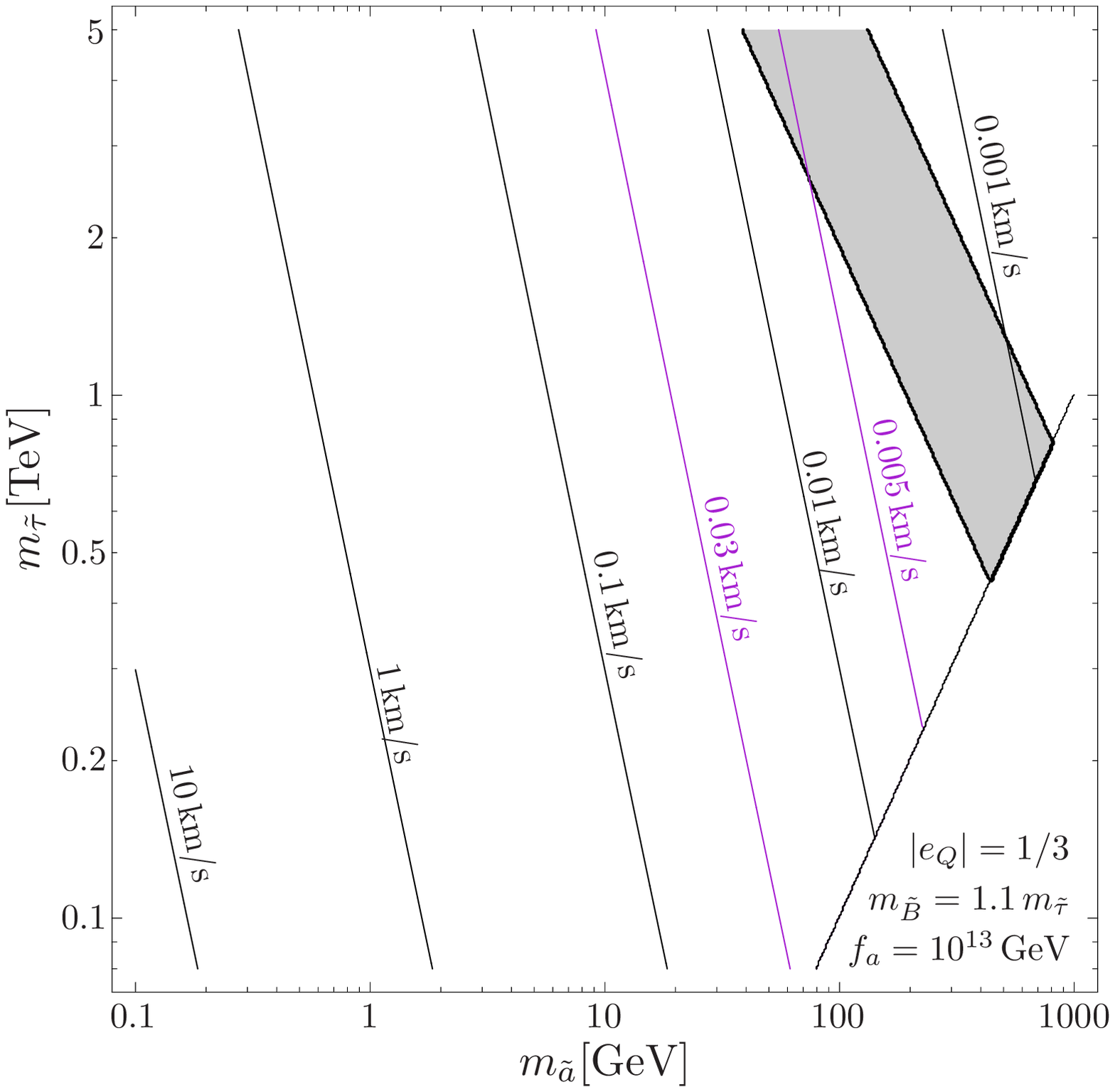, width=8cm}}
\caption{Contours of the present free-streaming velocity of axinos
  from stau NLSP decays, $(v_{\FS}^{0})_{\axino}^{\NTP}$, for
  $\mbino=1.1\,\mstau$, $f_a=10^{13}\,\GeV$, $|e_Q|=1/3$ and $y=1$.
  For the case in which $\Omega_{\CDM}$ is provided by
  $\Omega_{\axino}^{\NTP}$, the conservative limit
  ($0.03~\km/\seconds$) and the most restrictive limit
  ($0.005~\km/\seconds$) listed in \reftable{tab:WDMlimits} disfavour
  the regions to the left of the respective (purple) lines. The stau
  NLSP yield~\eqref{Eq:YSWIMP} satisfies
  $\Omega_{\axino}^{\NTP}\!\!=\Omega_{\CDM}$ by construction
  everywhere in the shown parameter space. The grey band indicates the
  region with $\Omega_{\axino}^{\NTP}\!\!=\Omega_{\CDM}^{3\sigma}$ for
  the yield \eqref{Eq:Ystau} with $\kappa=0.7$ or $1.4$.}
\label{fig:FSvel}}
%
 shows contours of the present free-streaming velocity of axinos from
stau NLSP decays $(v_{\FS}^{0})_{\axino}^{\NTP}$.
For $\Omega_{\axino}^{\NTP}\!\!=\Omega_{\CDM}$, as obtained with the
stau NLSP yield~\eqref{Eq:YSWIMP}, the regions to the left of the
(purple) lines labeled with $0.03~\km/\seconds$ and
$0.005~\km/\seconds$ are disfavoured by the respective
limits~\eqref{eq:vFSconservative} and~\eqref{eq:vFSrestrictive}.
The grey band indicates the region in which the axino density
$\Omega^{\NTP}_{\axino}$ agrees with the dark matter density range
$\Omega_{\dm}^{3\sigma}$ given in~\eqref{Eq:OmegaDM}, when using the
thermal relic stau yield \eqref{Eq:Ystau} with $\kappa=0.7$ or $1.4$,
where the latter accounts for potential stau--slepton coannihilation.
We see that scenarios within this band respect
$(v_{\FS}^{0})_{\axino}^{\NTP}< 0.01~\km/\seconds$
(corresponding to $m_{\X}>2~\keV$ listed in~\reftable{tab:WDMlimits})
and that the most restrictive Lyman-$\alpha$ forest constraint,
$(v_{\FS}^{\mathrm{rms},0})_{\axino}^{\NTP}\lesssim 0.005~\km/\seconds$,
starts to disfavour some of that region only for $\mstau\gtrsim
3\,\TeV$.
As is evident from~\eqref{eq:vFS0LL}, the velocity contours are
sensitive to $e_Q$ and $f_a$: An increase in the charge of the heavy
(s)quarks to $|e_Q|=1$ reduces the velocities by one order of
magnitude while an increase in $f_a$ by an order of magnitude shifts
the contours in the opposite way by the same amount.
Thus, the constraints become less restrictive for higher $e_Q$ and lower
$f_a$ values.

\subsection{Constraints from Primordial Nucleosynthesis}
\label{Sec:NucleosynthesisConstraints}

For the baryon-to-photon ratio 
$\eta$ ($\equiv n_{\mathrm{B}}/n_\gamma$)
inferred from studies of the CMB, standard BBN (SBBN) predicts the
primordial abundances of deuterium, helium, and lithium in good
overall agreement with observations~\cite{Nakamura:2010zzi}. When
dealing with particle physics scenarios beyond the SM, one therefore
has to check for non-standard processes that may spoil this success.
In our case, the outcome of BBN may be affected in the three following
ways:

\begin{itemize}
  
\item[(1)] Axinos may contribute to the radiation density at the onset
  of BBN prior to $e^{\pm}$ annihilation, $t\sim 1~\seconds$, in a way
  analogous to the case of extra light neutrino species. The
  resulting speed-up of the Hubble expansion rate would then lead to a
  more efficient output of $\Hefour$.  The inferred primordial
  $\Hefour$ abundance thereby implies upper limits on the equivalent
  number of additional neutrinos $\DNueff$;
  cf.~\cite{Simha:2008zj,Izotov:2010ca,Aver:2010wq} for recent
  studies. (Prior to $e^{\pm}$ annihilation, $\DNueff\equiv\Nueff-3$,
  where the effective number of light neutrino species $\Nueff$
  parametrises the number of light relativistic species and their
  contribution to the total relativistic energy density, i.e.,
  $\rhorad=\pi^2[2+7\,\Nueff/4]\,T^4/30$ for $T\simeq 1~\MeV$.)  While
  the resulting bounds on $\DNueff$ are still somewhat uncertain, we
  will calculate $\DNueff$ potentially contributed in the form of
  axinos at $T\simeq 1~\MeV$.  Moreover, we will show how an upper
  limit on $\DNueff$ translates into a constraint on the considered
  axino LSP scenarios with a stau NLSP described by~\eqref{Eq:YSWIMP}.

\item[(2)] In late stau NLSP decays, energetic SM particles can be
  generated that could scatter off and/or dissociate the primordial
  nuclei. If injected at $t\gtrsim 100~\seconds$, emitted hadrons can
  reprocess the primordial light elements significantly via
  hadrodissociation~\cite{Reno:1987qw,Dimopoulos:1987fz,Dimopoulos:1988ue,Kohri:2001jx,Jedamzik:2004er,Jedamzik:2006xz,Cyburt:2009pg}.
  At later times, $t\gtrsim 10^4\,\seconds$, also photodissociation
  caused by electromagnetic energy injection can become
  efficient~\cite{Holtmann:1998gd,Cyburt:2002uv,Kawasaki:2004yh,Kawasaki:2004qu,Jedamzik:2006xz,Cyburt:2009pg}.
  For the considered scenarios, we have already published first
  results on the associated BBN constraints from both hadronic and
  electromagnetic energy release in a Letter~\cite{Freitas:2009jb}.
  This section expands on our previous results by exploring those
  constraints in more detail.

  For $t\lesssim 100~\seconds$, the decay products lose their energy
  very efficiently and are quickly stopped via electromagnetic
  interactions. While dissociation of nuclei is not important at these
  times, additional slow-moving hadrons can take part in
  proton-neutron interconversion
  processes~\cite{Kawasaki:2004yh,Kawasaki:2004qu} which could
  increase the abundances of $\Hefour$ and $\Deut$.
  Without a dedicated numerical treatment of such non-standard BBN
  processes at hand, we do not consider the associated limits in this
  work. As in gravitino LSP scenarios with a stau
  NLSP~\cite{Kawasaki:2008qe}, we anyhow expect those limits to be
  milder than the ones associated with hadro/photodissociation.

\item[(3)] If the stau NLSP is sufficiently long-lived,
  $\tau_\stau>10^3\,\seconds$, negatively charged staus can form bound
  states with the primordial nuclei and thereby catalyse the formation
  of $\Lisix$ and $\Benine$ in the early
  universe~\cite{Pospelov:2006sc,Pospelov:2007js,Pospelov:2008ta}.
  Together with restrictive observationally inferred limits on the
  primordial abundances of $\Lisix$ and $\Benine$, this catalysed BBN
  (CBBN) has allowed us to constrain the parameter space of the
  considered models considerably.
  While CBBN constraints on axino LSP scenarios with the stau NLSP
  were first presented in our Letter~\cite{Freitas:2009fb}, they will
  also be addressed in this section.

\end{itemize}

Let us first turn to the BBN constraint on $\DNueff$ $(=\Nueff-3)$
governed by the observationally inferred primordial $\Hefour$
abundance (for given $\eta$).
For example, for a $\Hefour$ mass fraction $Y_{\mathrm{P}}=0.240\pm
0.006$, favoured contours (95\%~CL) with $\eta/10^{-10}=5.7\pm 0.8$
and $N_{\nu}=2.4\pm 0.8$ were obtained in Ref.~\cite{Simha:2008zj}.
However, larger
$Y_{\mathrm{P}}=0.2565\pm 0.001\,(\mathrm{stat})\,\pm 0.005 \,(\mathrm{syst})$~\cite{Izotov:2010ca}
and
$Y_{\mathrm{P}}=0.2561\pm 0.011$~\cite{Aver:2010wq}
have been obtained recently, which point to larger $N_{\nu}$, such as
$N_{\nu}=3.68^{+0.80}_{-0.70}$ (95\%~CL) for $\eta/10^{-10}=6.47$ and
a neutron lifetime of $885\pm0.9~\seconds$~\cite{Izotov:2010ca}.
This exemplifies the uncertainty of the BBN bound on $\DNueff$ 
with, e.g., $\DNueff\sim 1$ being still compatible with observations. 
With this in mind, we now show that the current restrictions on $\DNueff$ 
do not impose (significant) constraints for most of the parameter space.

The equivalent number of additional neutrinos $\DNueff$ provided in
the form of axinos depends on their primordial origin, the axino mass
$\maxino$ and the cosmic time at which $\DNueff$ is considered.
At $T\simeq 1~\MeV$, prior to the epoch of $e^{\pm}$ annihilation,
thermal relic axinos have the temperature
$T_{\axino}=[10.75/g_{*S}(\Tf^{\axino})]^{1/3} T\simeq 0.36~\MeV$
since they decouple when $g_{*S}(\Tf^{\axino})\simeq 230$.
A thermal axino population with $\maxino \ll 0.36~\MeV$ will then be
fully relativistic and contribute
\bea 
(\DNueff)_{\axino}^{\mathrm{therm}}
=
\frac{T_{\axino}^4}{T^4}
\simeq
\left(\frac{10.75}{g_{*S}(\Tf^{\axino})}\right)^{4/3}
\simeq
0.017 
\, ,
\label{eq:DNueffAxinoTherm}
\eea
which applies when the respective dark matter constraint
and the structure formation constraint~\eqref{eq:AxinoHDMConstraint}
discussed above are satisfied.
Since also thermally produced axinos have basically a thermal
spectrum~\cite{Brandenburg:2004du} and originate from the
high-temperature epoch with $g_{*S}\simeq 230$, the associated
population can again be characterized by $T_{\axino}$ given above.
Accordingly, for $\maxino\ll 0.36~\MeV$, this will be a relativistic
population that provides
\bea 
   (\DNueff)_{\axino}^{\mathrm{TP}}
   \simeq
   0.019\,g^6_{\mathrm{s}}(\TR) 
   \ln\left( \frac{1.211}{g_{\mathrm{s}}(\TR)}\right) 
   \left(\frac{10^{9}\,\GeV}{\fPQ}\right)^2 
   \left(\frac{\TR}{10^4\,\GeV}\right) 
   \ .
\label{eq:DNueffAxinoTPrelativ}
\eea
Indeed, this contribution cannot exceed
$(\DNueff)_{\axino}^{\mathrm{therm}}$ since thermal processes cannot
lead to an axino yield that is larger than the thermal relic yield
$Y_{\axino}^{\mathrm{therm}}$~\cite{Covi:2001nw,Brandenburg:2004du}.
Non-thermally produced axinos from stau NLSP decays with 
$\tau_\stau < 1~\seconds$
can also constitute a relativistic population at $T\simeq 1~\MeV$, as
discussed for the neutralino NLSP case in~\cite{Covi:2001nw}. With the
2-body decay being the dominant decay channel,
$|\vec{p}_{\axino}(t_i)|$ given by \eqref{Eq:paxino}, the LL
approximation of $\tau_{\stau}$ and the sudden decay approximation, we
find the associated contribution
\begin{align}
   (\DNueff)_{\axino}^{\mathrm{NTP}}
   \simeq \,\,
   & 
   6.3\times 10^{-9}
  \left(\frac{30}{\log[y^2 f_a^2/(2\mstau^2)]}\right)
  \left(\frac{1/3}{e_Q}\right)^2 \!
  \left(\frac{f_a}{10^{9}\,\GeV}\right)\! 
\nonumber\\ 
   & \times
   \left(\frac{\mstau}{100~\GeV}\right)^{1/2} \!
   \left(\frac{100~\GeV}{\mbino}\right) \! 
   \left(\frac{Y_\stau}{10^{-13}}\right) \,
   \frac{g_{*}(\tau_{\stau})^{1/4}}{g_{*S}(\tau_{\stau})^{1/3}}
   \ .
\label{eq:DNueffAxinoNTPrelativ}
\end{align}
This expression applies to cases in which non-thermally produced
axinos become non-relativistic at a temperature $\TNR\ll 1~\MeV$,
where $\TNR$ refers to temperature where the axino momentum satisfies
$p_{\axino}(\TNR)=\maxino$.
 Using again the LL and sudden decay approximations,
\begin{align}
   \frac{\TNR}{\maxino}
   \simeq \,\,
   & 
   1.8\times 10^{-4}
   \left(\frac{\log[y^2 f_a^2/(2\mstau^2)]}{30}\right)
   \left(\frac{e_Q}{1/3}\right)^2 \!
   \left(\frac{10^{9}\,\GeV}{f_a}\right)\! 
\nonumber\\ 
   & \times
   \left(\frac{100~\GeV}{\mstau}\right)^{1/2} \!
   \left(\frac{\mbino}{100~\GeV}\right)
   \left(\frac{3.91}{g_{*S}(\TNR)}\right)^{1/3} \,
   \frac{g_{*S}(\tau_{\stau})^{1/3}}{g_*(\tau_{\stau})^{1/4}} \,
   \ .
\label{eq:TNRAxinoNTP}
\end{align}
Note that the total energy density at $T\simeq 1~\MeV$ receives also
contributions in the other cases with a non-relativistic non-thermally
produced axino population, $\TNR\gtrsim 1~\MeV$, or a non-relativistic
thermally produced one, $\maxino>0.36~\MeV$.
However, if the dark matter constraint~\eqref{Eq:OmegaDM} is
respected, the resulting effect on the Hubble expansion rate and the
$\Hefour$ abundance is negligible, i.e., orders of magnitude smaller
than the one of~\eqref{eq:DNueffAxinoTherm}.

Considering $(\DNueff)_{\axino}^{\mathrm{therm/TP}}$ and
$(\DNueff)_{\axino}^{\mathrm{NTP}}$ for characteristic values of the
thermal relic stau yield~\eqref{Eq:Ystau}, we find that these
contributions cannot provide a sizeable non-standard $\DNueff\sim 1$.
In fact, they are clearly compatible with current BBN constraints on
$N_{\nu}$ such as the ones summarised above. In turn, there are
presently no associated constraints.
On the other hand, considering $(\DNueff)_{\axino}^{\mathrm{NTP}}$ for
the stau yield~\eqref{Eq:YSWIMP}, we find that an upper limit
$(\DNueff)^{\max}$ translates into the following lower limit on the
axino mass:
\begin{align}
  \maxino
  > \,\,
  & 
  23.8~\keV
  \left[\frac{1}{(\DNueff)^{\max}}\right]
  \left(\frac{30}{\log[y^2 f_a^2/(2\mstau^2)]}\right)
  \left(\frac{1/3}{e_Q}\right)^2 \!
  \left(\frac{f_a}{10^{9}\,\GeV}\right)\! 
\nonumber\\ 
   & \times
   \left(\frac{\mstau}{100~\GeV}\right)^{1/2} \!
   \left(\frac{100~\GeV}{\mbino}\right) \! 
   \left(\frac{\Omega_{\dm}}{0.105/h^2}\right)
   \left(\frac{g_{*S}(T_{\mathrm{BBN}})}{10.75}\right)^{4/3}
   \frac{g_{*}(\tau_{\stau})^{1/4}}{g_{*S}(\tau_{\stau})^{1/3}}
   \ .
\label{eq:maxinoLimitDNueffAxinoNTP}
\end{align}
This constraint applies to cases in which the stau NLSPs decay before
the temperature $T_{\mathrm{BBN}}\sim 1~\MeV$ is reached and is of
similar restrictiveness as~\eqref{eq:maxinoStdCosmology} for
$\fstau\sim 0.1$.
For the neutralino NLSP case, the constraint analogous
to~\eqref{eq:maxinoLimitDNueffAxinoNTP} can be found in
Ref.~\cite{Covi:2001nw}.

Here it should be mentioned that $\DNueff$ is constrained not only at
the time of BBN but also much later by observations of the CMB, galaxy
clustering and the Lyman-$\alpha$ forest; see
e.g.~\cite{Hamann:2007pi,Simha:2008zj} and references therein.
At those later times, ultra-light thermal relic or thermally produced
axinos and non-thermally produced axinos with sufficiently small
$\TNR$ will contribute to $\DNueff$.
Here again the observationally inferred $\DNueff$ limits are such that
these contributions can be accommodated, e.g., a 95\%-credible interval
of $2.2<\Nueff<5.8$ is obtained in~\cite{Hamann:2007pi}.
For non-thermally produced axinos, we find compatibility with those
limits for a thermal relic stau yield~\eqref{Eq:Ystau} and also
for~\eqref{Eq:YSWIMP} respecting~\eqref{eq:maxinoStdCosmology} with
$\fstau\sim 0.1$.

Let us now turn to the BBN constraints imposed by potential
hadrodissociation of primordial nuclei via hadrons emitted in stau
NLSP decays. 
At $\tau_{\stau}\gtrsim 100~\seconds$, the emitted hadrons remain
energetic and can affect the primordial abundances of $\Deut$,
$\Hefour$, $\Hethree/\Deut$, $\Liseven$ and
$\Lisix/\Liseven$~\cite{Sigl:1995kk,Jedamzik:1999di,Jedamzik:2004er,Kawasaki:2004qu,Jedamzik:2006xz,Cyburt:2006uv,Cyburt:2009pg
}.
However, in the region allowed by CBBN constraints for typical stau
yields~\eqref{Eq:Ystau},
$\tau_\stau\lesssim 4 \times 10^3\,\seconds$ (see later),
the effect on the $\Deut$ abundance through the hadrodissociation of
$\Hefour$ provides the most restrictive constraint. This can be seen,
e.g., in Figs.~38--41 of \cite{Kawasaki:2004qu} and in Figs.~6--8 of
\cite{Jedamzik:2006xz}.
Accordingly, we work with the $\tau_{\NLSP}$-dependent upper limits
(95\%~CL) on
\begin{align}
  \xi_{\HAD} \equiv \epsilon_{\HAD}\, Y_{\NLSP}
\label{Eq:EnergyRelease}
\end{align}
given in Fig.~9 of Ref.~\cite{Steffen:2006hw} as obtained in
Ref.~\cite{Kawasaki:2004qu} for observationally inferred primordial
$\Deut$ abundances of
\begin{eqnarray}
  \!\!\!\!\!\!\!\!\!\!\!\!
  & \Deut/\Hyd|_{\mathrm{mean}} 
  &=\, 
  (2.78^{+0.44}_{-0.38})\times 10^{-5} 
  \,\,\,\, \mathrm{(severe)}
  \ ,
\label{Eq:D_sev}\\
  \!\!\!\!\!\!\!\!\!\!\!\!
  &\Deut/\Hyd|_{\mathrm{high}} 
  &=\, 
  (3.98^{+0.59}_{-0.67})\times 10^{-5}
  \,\,\,\,
  \mathrm{(conservative)}
  \ .
\label{Eq:D_cons}
\end{eqnarray}
Note that we are not trying to give extra credence to the latter
rather high $\Deut/\Hyd$ value but consider~\eqref{Eq:D_cons} simply
as a limiting value that leads to robust conservative limits.

When using the limits on~\eqref{Eq:EnergyRelease},
$\tau_{\NLSP}=\tau_{\stau}$, $Y_{\NLSP}=Y_{\stau}$ and
$\epsilon_{\HAD}$ is the average amount of hadronic energy release per
decay.
As the mesons produced from the $\tau$'s emitted in the 2-body decay
$\stauR\to\tau\axino$ typically decay too quickly to interact
hadronically, the main contribution to the injection of hadronic
energy comes from the 4-body decay $\stauR\to\tau\axino q\qbar$
(considered in \refsec{Sec:Axino4Body}):
\bea
   \epsilon_{\had} 
   \equiv 
   \frac{1}{\Gamma_{\tot}^{\stau_\mathrm{R}}}
   \int_{m_{q\qbar}^{\mathrm{cut}}}^{\mstau-\maxino-\mtau} 
   \mathrm{d} m_{q\qbar} \, 
   m_{q\qbar}\,
   \frac{\mathrm{d}\Gamma(\stau_{\mathrm{R}}\to\tau\axino q\qbar)}
   {\mathrm{d} m_{q\qbar}}
   \ ;
\label{eq:epshad}
\eea
see also \reffig{fig:4br}.
Similarly, the mesons from this process also decay before they can
interact with the background nuclei. As already mentioned in
\refsec{Sec:Axino4Body}, we therefore place a cut on the invariant
mass of the produced $q\qbar$ pairs,
$m_{q\qbar}>m_{q\qbar}^{\mathrm{cut}}=2\,\GeV$, so as to consider only
those that hadronise into nucleons.
While~\eqref{eq:epshad} is used with this cut value to determine the
hadronic BBN constraints in this work, we note that a more precise
calculation of the constraints would require, in addition to
$\mathrm{d}\Gamma(\stauR\to\tau\axino q\qbar)/\mathrm{d}m_{ q\qbar}$,
a treatment of the fragmentation of the quarks into hadrons and of the
propagation of the resulting hadron spectra when computing the
abundances of primordial light elements;
cf.~\cite{Kawasaki:2004qu,Kohri:2005wn,Bailly:2008yy,Cyburt:2009pg}.

\reffigure{fig:EpsHadArray}
\FIGURE[ht]{
\centerline{\epsfig{figure=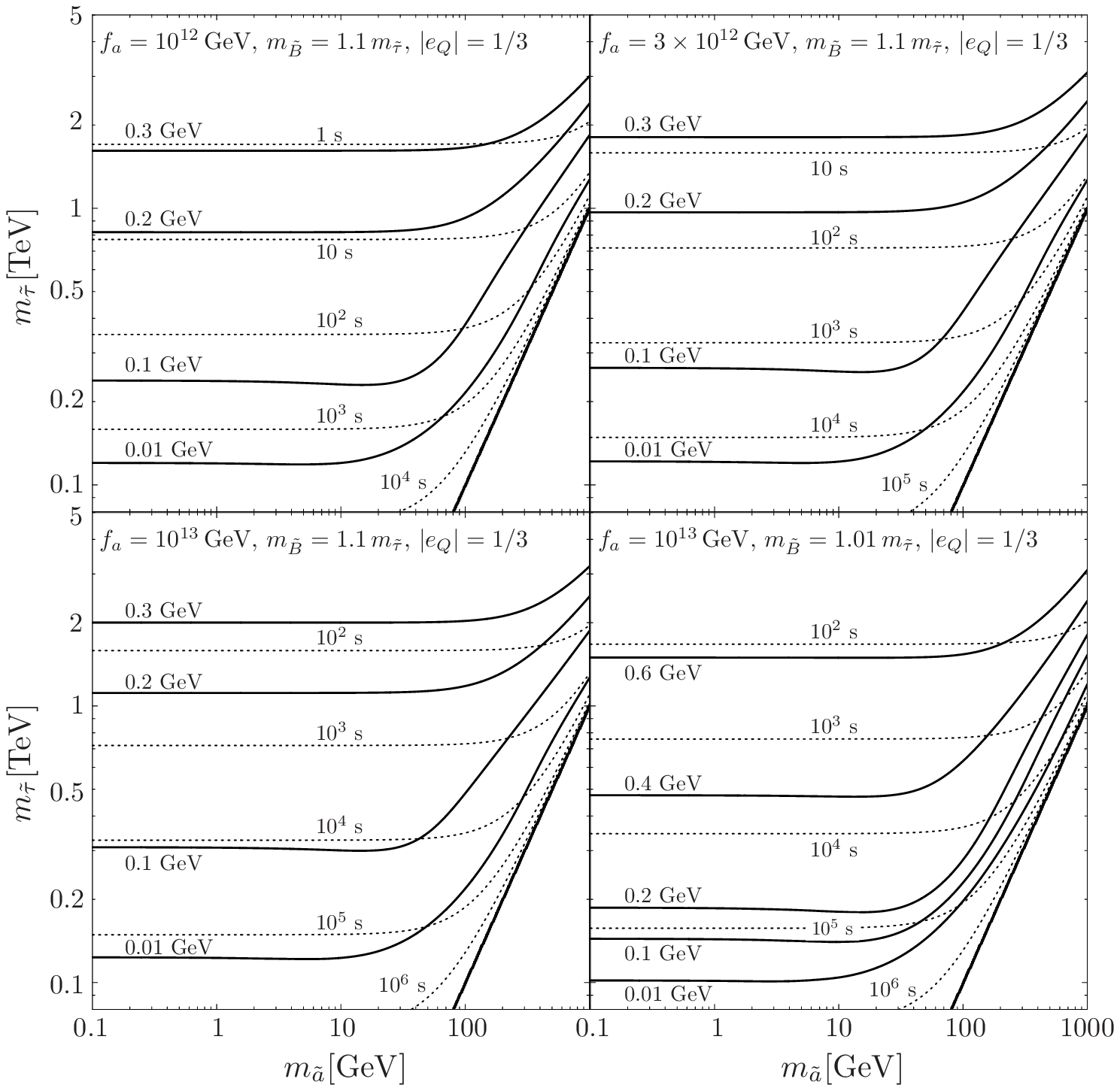, width=11cm}}
\caption{Average amount of hadronic energy release per stau NLSP decay
$\epsilon_{\had}$ (solid lines), as governed by the 4-body dec ay
$\stau_{\mathrm{R}}\rightarrow\tau\axino q\qbar$, and stau NLSP
lifetime contours (dashed lines) for $|e_Q|=1/3$ and $y=1$. Starting
from the top left panel and going row by row, we consider
  ($f_a,\,\mbino/\mstau$)=($10^{12}\,\GeV$, $1.1$),
  ($3\times 10^{12}\,\GeV$, $1.1$),
  ($10^{13}\,\GeV$, $1.1$)
  and ($10^{13}\,\GeV$, $1.01$).}
\label{fig:EpsHadArray}}
%
shows the hadronic energy release $\epsilon_{\had}$ as calculated
with~\eqref{eq:epshad} (solid lines) along with the lifetime contours
of the stau NLSP (dotted lines) for $|e_Q|=1/3$, $y=1$ and four
different combinations of $f_a$ and $\mbino/\mstau$ (as indicated).
The dependency of $\epsilon_{\had}$ on $\maxino$ is as weak as that of
$\tau_{\stau}$, until we go to large enough $\maxino$ where
phase-space suppression sets in.
As $f_a$ increases to larger values, the average hadronic energy
release $\epsilon_{\had}$ decreases, although not as drastically as
the increase in the lifetime. This is due to the fact that, after normalisation of $\epsilon_{\had}$ to $\Gamma_{\tot}^{\stauR}$, 
the leading dependence of $\epsilon_{\had}$ on $f_a$ is 
$\propto 1/\log^2\left[y^2 f_a^2/(2\mstau^2)\right]$,
rather than $\propto 1/f_a^2$.

Using $\epsilon_{\HAD}$ as calculated with~\eqref{eq:epshad} and the
upper limits on hadronic energy release $\xi_{\HAD}^{\max}$ mentioned
above, we obtain upper limits on the yield of the stau NLSP prior to
decay
\bea
    Y_{\stau\,\had}^{\max}=\xi_{\had}^{\max}/\epsilon_{\had} \ .
\eea
\reffigures{fig:YLimsevere} and \ref{fig:YLimconservative}
%
\FIGURE[ht]{
\centerline{\epsfig{figure=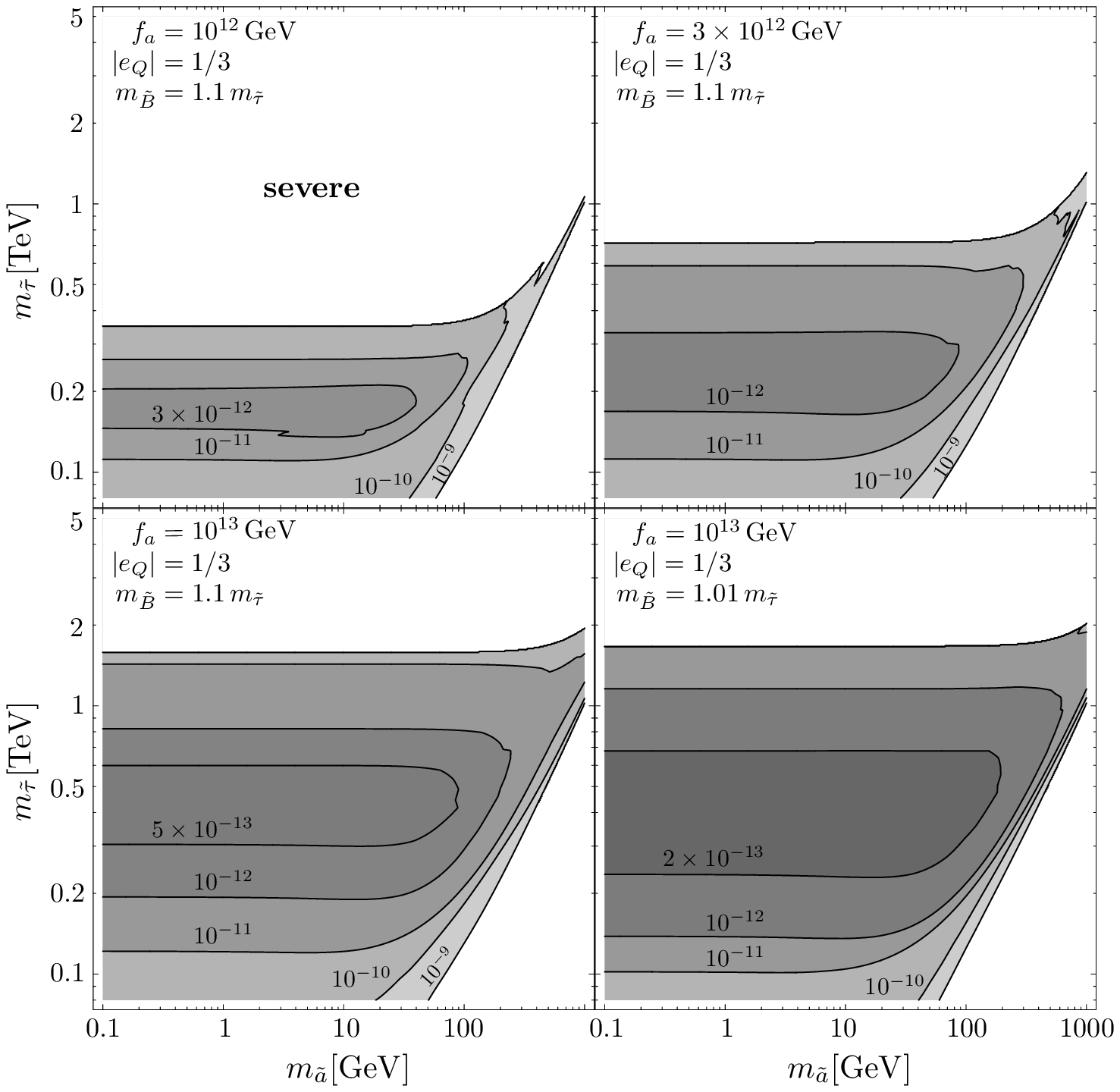, width=11cm}}
\caption{Upper limits on the yield of the stau NLSP prior to decay,
  $Y_{\stau\,\had}^{\max}$, imposed by the severe limit on late
  hadronic energy injection associated with~\eqref{Eq:D_sev}. The
  contour shading darkens with stronger limits. The choice of the
  panels and associated parameters matches the one
  in~\reffig{fig:EpsHadArray}.}
\label{fig:YLimsevere}}
%
\FIGURE[ht]{
\centerline{\epsfig{figure=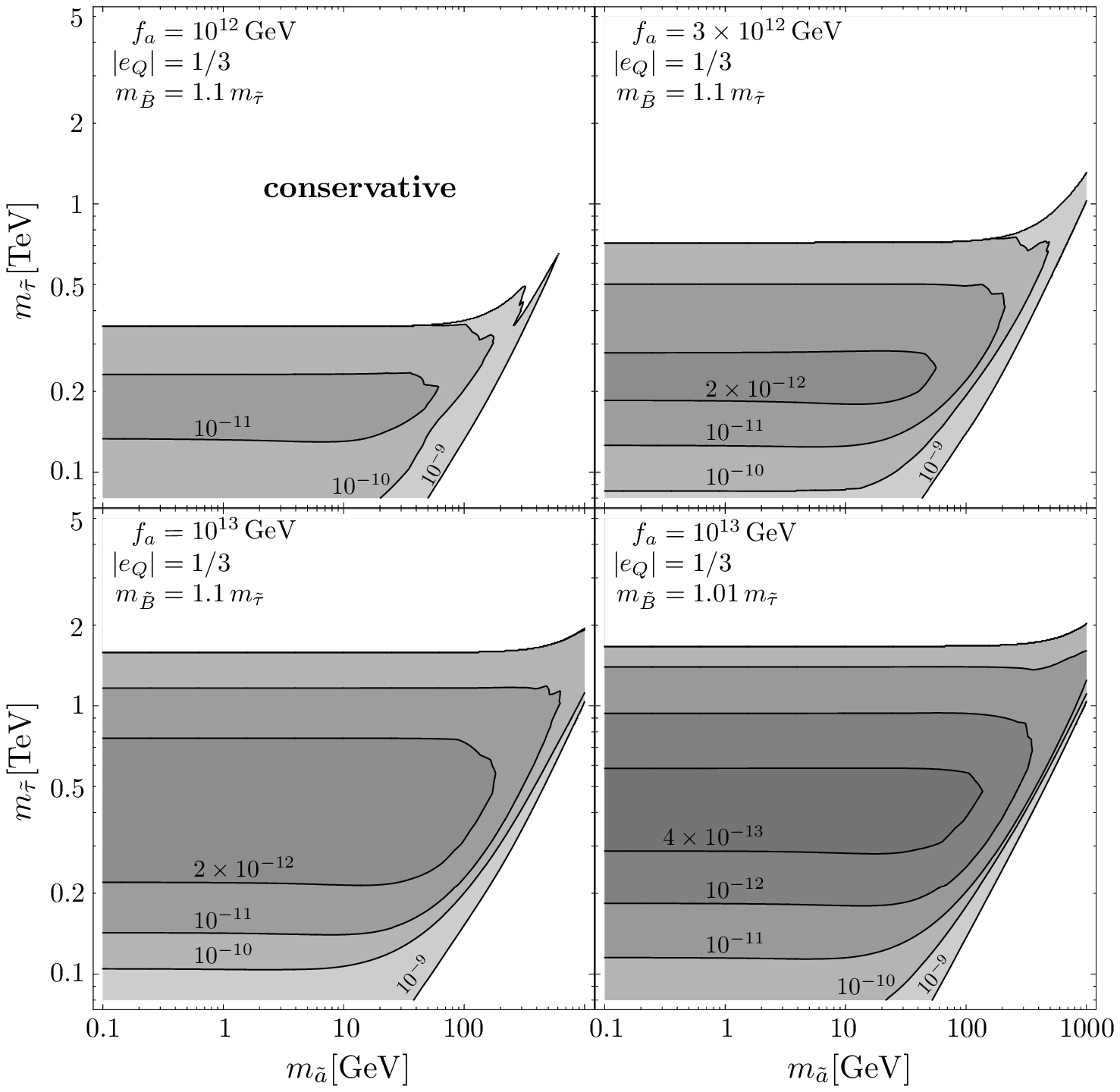, width=11cm}}
\caption{The same as \reffig{fig:YLimsevere} but for the conservative
  limit on late hadronic energy injection associated
  with~\eqref{Eq:D_cons}.}
\label{fig:YLimconservative}}
%
show those limits as imposed by the severe limit associated
with~\refeq{Eq:D_sev} and by the conservative limit associated
with~\refeq{Eq:D_cons}, respectively. 
The contours correspond to the indicated $Y_{\stau\,\had}^{\max}$
values which range between $2\times 10^{-13}$ to $10^{-9}$ with darker
shading representing more restrictive constraints.
The considered ranges of $\maxino$ and $\mstau$, the panels and the
parameters are set as in \reffig{fig:EpsHadArray}.
As seen from the first three panels of each figure, we find that
increasing values of $f_a$ are associated with increasingly tighter
constraints on $Y_{\stau}$.
Moreover, a comparison of the lower panels in each figure shows that
smaller values of $\mbino/\mstau$ imply tighter limits
$Y_{\stau\,\had}^{\max}$. This results from the enhancement of
$\epsilon_{\had}$ for $\mbino/\mstau\to 1$ described above.
In each panel, one also finds common features of the constraints.
The $Y_{\stau\,\had}^{\max}$ constraints become weaker for decreasing
$\mstau\to 80~\GeV$, which results from the decrease in
$\epsilon_{\had}$ documented in~\reffig{fig:EpsHadArray}, and one for
$\tau_{\stau}\to 100~\seconds$, which is due to a relaxation of the
$\xi_{\had}^{\max}$ limits~\cite{Kawasaki:2004qu,Steffen:2006hw}.
Furthermore, since $\epsilon_{\had}$ and $\tau_{\stau}$ become
insensitive to $\maxino$ below $10~\GeV$, the same holds also for
$Y_{\stau\,\had}^{\max}$. Accordingly, the $Y_{\stau}$ limits in
\reffigs{fig:YLimsevere} and \ref{fig:YLimconservative} shown for
$\maxino\lesssim 10~\GeV$ apply also to cases with a much lighter
axino LSP, including even those that
respect~\eqref{eq:AxinoHDMConstraint}.
Going into the other direction, $\maxino\to\mstau$, $\epsilon_{\had}$
becomes phase-space suppressed (cf.~\reffig{fig:EpsHadArray}). This
leads to the relaxation of the $Y_{\stau}$ constraints for
$\maxino\to\mstau$ that can be seen in \reffigs{fig:YLimsevere} and
\ref{fig:YLimconservative}.

Now by confronting the stau yield $Y_{\stau}$ with the maximum yield
$Y_{\stau\,\had}^{\max}$ shown, e.g., in \reffigs{fig:YLimsevere} and
\ref{fig:YLimconservative}, we obtain the regions in the parameter
space that are disfavoured by the hadronic BBN constraints associated
with the $\Deut$ abundances~\eqref{Eq:D_sev}
and~\eqref{Eq:D_cons}~\cite{Freitas:2009jb}.
For characteristic values of the thermal relic yield~\eqref{Eq:Ystau}
and for~\eqref{Eq:YSWIMP} satisfying
$\Omega_{\axino}^{\NTP}=\Omega_{\mathrm{dm}}$ by construction, those
regions are the ones enclosed by the short-dash-dotted (blue) lines in
\reffigs{fig:MB} and \ref{fig:SWIMPMB}, respectively.
As a guidance, dotted lines are shown to indicate $\tau_{\stau}=10^2$,
$5\times 10^3$ and $10^5\,\seconds$ in \reffig{fig:MB} and
$\tau_{\stau}=10^2$, $10^3$ and $10^4\,\seconds$ in
\reffig{fig:SWIMPMB}.
In fact, we limit our consideration of the hadronic BBN constraints to
parameter regions with $\tau_{\stau}\geq 10^2~\seconds$.
%
\FIGURE[ht]{
\centerline{\epsfig{figure=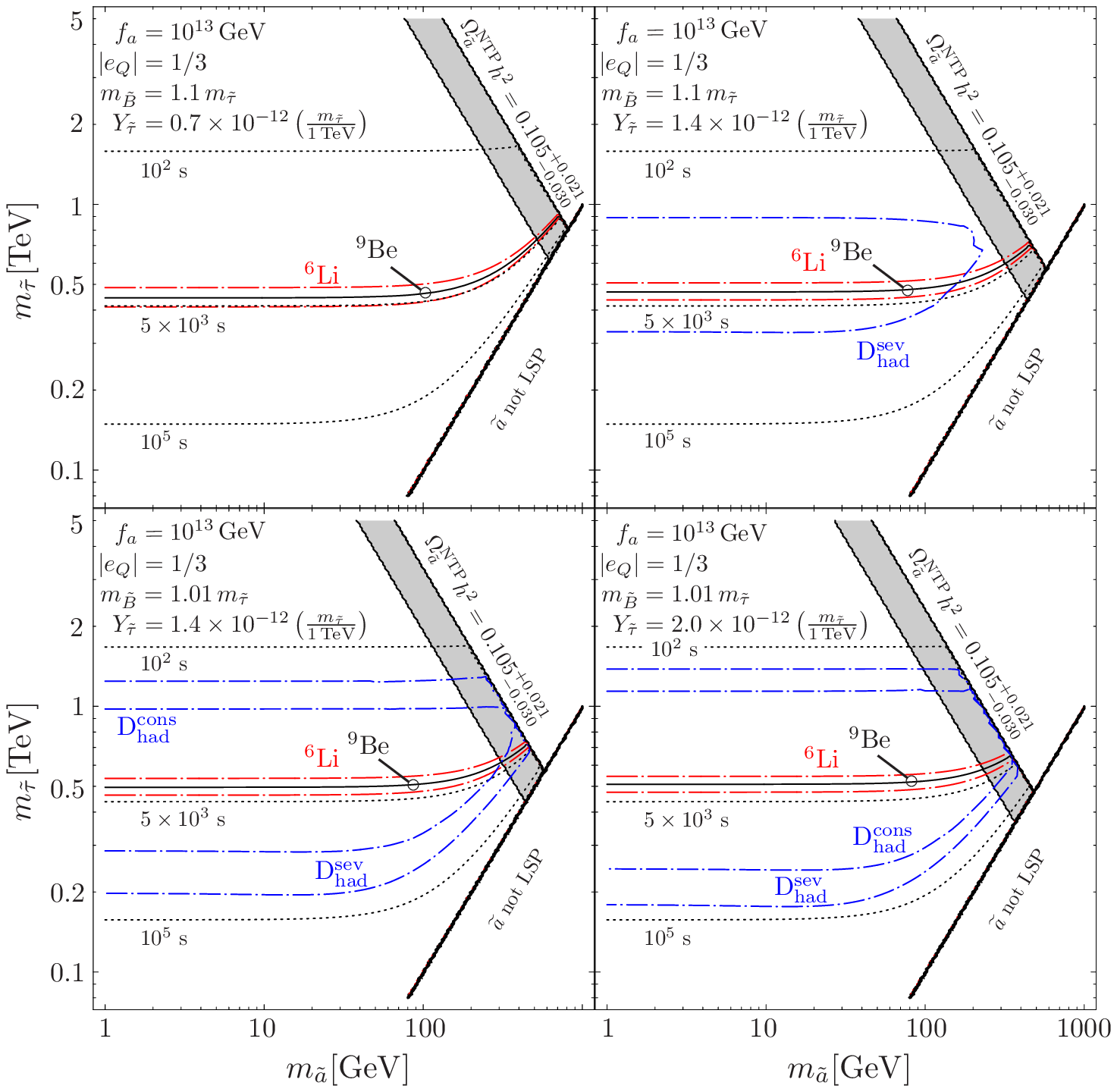, width=11cm}}
\caption{Cosmological constraints on the masses of the axino LSP and
  stau NLSP for $f_a=10^{13}\,\GeV$, $|e_Q|=1/3$, $y=1$ and $Y_\stau$
  given by representative thermal relic abundances~\eqref{Eq:Ystau}.
  Starting from the top left panel and going row by row, we have
  $(\mbino/\mstau,\,\kappa)=(1.1,\,0.7)$, $(1.1,\,1.4)$,
  $(1.01,\,1.4)$ and $(1.01,\,2.0)$. Contours of $\tau_{\stau}=10^2$,
  $5\times 10^3$ and $10^5\,\seconds$ are shown by the dotted lines.
  On (above) the grey band,
  $\Omega_{\axino}^{\NTP}\in\Omega_{\mathrm{dm}}^{3\sigma}$
  ($\Omega_{\axino}^{\NTP}h^2>0.126$). Hadronic BBN constraints
  imposed by the $\Deut$ abundances~\eqref{Eq:D_sev}
  and~\eqref{Eq:D_cons} disfavour the regions enclosed by the
  respective short-dash-dotted (blue) lines. Electromagnetic BBN
  constraints do not appear in the shown parameter space. The CBBN
  constraints imposed by the $\Lisix$ range~\eqref{Eq:LiSix} and the
  $\Benine$ abundance~\eqref{Eq:BeNine} disfavour the regions below
  the long-dash-dotted (red) and the solid lines, respectively. Not
  shown are the constraints from structure formation discussed
  in~\refsec{Sec:SmallScaleStructure} which apply only within the grey
  bands. For the upper (lower) panels, it can be seen
  in~\reffig{fig:FSvel} (\reffig{fig:SWIMPMB}, bottom right panel)
  that only the most restrictive velocity
  limit~\eqref{eq:vFSrestrictive} disfavours parts of the grey bands
  for $\mstau\gtrsim 3~\TeV$ ($\mstau> 1~\TeV$).}
\label{fig:MB}}
%
\FIGURE[ht]{
\centerline{\epsfig{figure=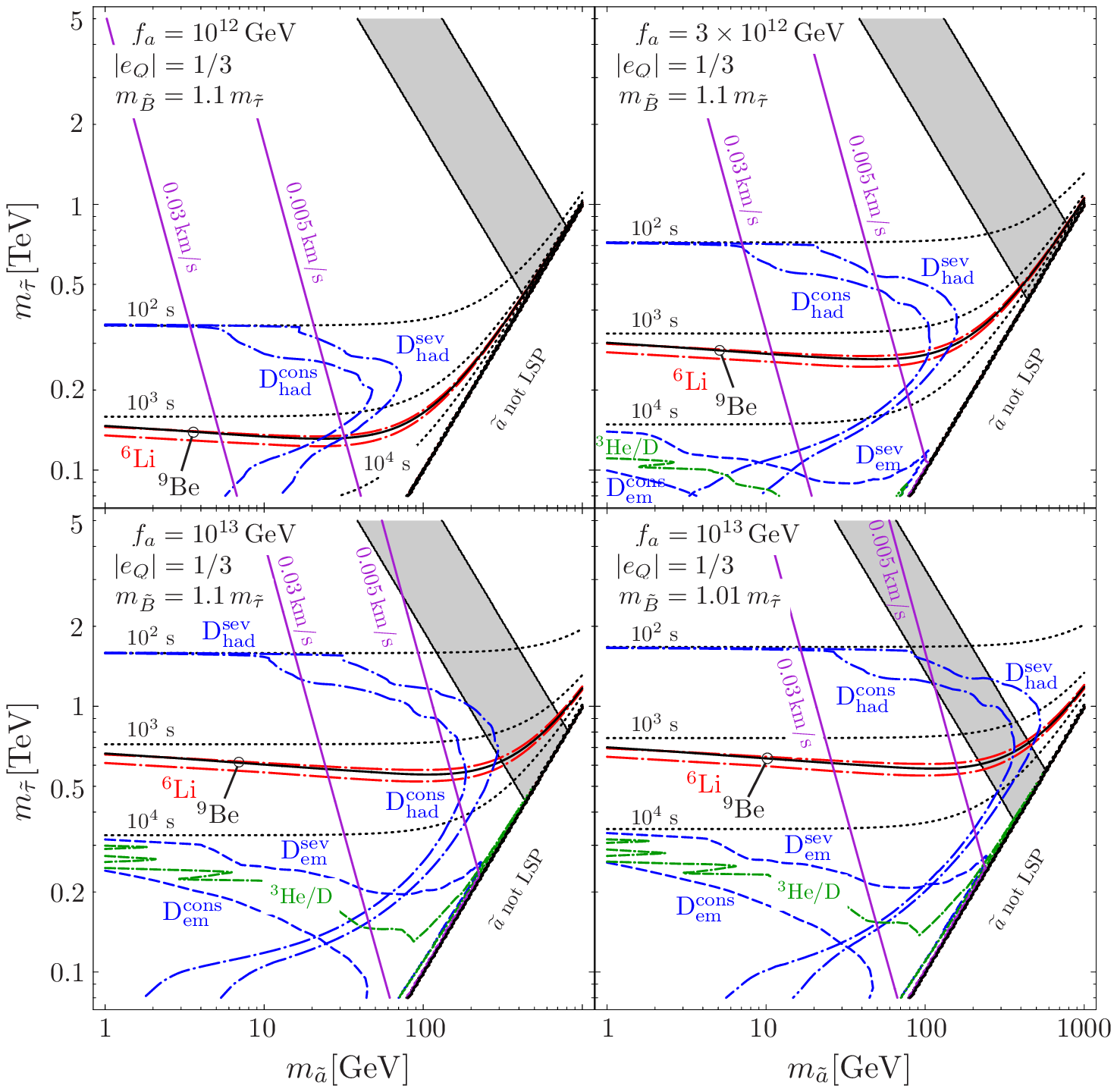, width=11cm}}
\caption{Cosmological constraints on the masses of the axino LSP and
  stau NLSP for the stau yield $Y_\stau$ given by~\eqref{Eq:YSWIMP}
  that satisfies $\Omega_{\axino}^{\NTP}=\Omega_{\mathrm{dm}}$ by
  construction everywhere in the shown parameter space.  Starting from
  the top left panel and going row by row, we have
  $(f_a,\,\mbino/\mstau)=(10^{12}\,\GeV,\,1.1)$, $(3\times
  10^{12}\,\GeV,\,1.1)$, $(10^{13}\,\GeV,\,1.1)$ and
  $(10^{13}\,\GeV,\,1.01)$, with the other parameters set to
  $|e_Q|=1/3$ and $y=1$, as in~\reffigs{fig:EpsHadArray}--\ref{fig:MB}
  above.  Contours of $\tau_{\stau}=10^2$, $10^3$ and
  $10^4\,\seconds$ are shown by the dotted lines. The constraints from
  structure formation associated with~\eqref{eq:vFSconservative}
  and~\eqref{eq:vFSrestrictive} disfavour the regions to the left of
  the respective solid (purple) diagonal lines. Here the grey band is
  only shown to indicate the region with
  $\Omega_{\axino}^{\NTP}\!\!=\Omega_{\CDM}$ for the thermal relic
  yield \eqref{Eq:Ystau} with $\kappa=0.7$ or $1.4$ in the first three
  panels and with $\kappa=1.4$ or $2.0$ in the bottom right panel.
  Hadronic BBN constraints imposed by the $\Deut$
  abundances~\eqref{Eq:D_sev} and~\eqref{Eq:D_cons} disfavour the
  regions enclosed by the respective short-dash-dotted (blue) lines.
  The corresponding $\Deut$-imposed electromagnetic BBN constraints
  disfavour the regions enclosed by the respective short-dashed (blue)
  lines and the $^3$He/D-imposed ones the regions enclosed by the
  double-dash-dotted (green) lines. The CBBN constraints imposed by
  the $\Lisix$ range~\eqref{Eq:LiSix} and the $\Benine$
  abundance~\eqref{Eq:BeNine} disfavour the regions below the
  long-dash-dotted (red) and the solid lines, respectively.}
\label{fig:SWIMPMB}}

In \reffig{fig:MB} we have chosen the parameters such that the
constraints shown in the panels become progressively more limiting.
For the generic case of $\mbino=1.1\,\mstau$ and a thermal relic
yield~\eqref{Eq:Ystau} characterised by $\kappa=0.7$, both the severe
and conservative hadronic BBN limits are still absent for
$f_a=10^{13}\,\GeV$ (top left panel).
For a larger thermal relic yield given by $\kappa=1.4$, accounting for
a possible stau--slepton coannihilation, more severe constraints are
obtained. In fact, the severe one associated with~\eqref{Eq:D_sev}
disfavours a stau mass between 300 and 900~GeV for $\maxino\lesssim
100\,\GeV$ (top right panel).
Decreasing $\mbino$ disfavours also larger regions of parameter space
due to the increase in $\epsilon_{\had}$ illustrated in the lower
panels of~\reffig{fig:EpsHadArray}. Accordingly, for the same yield
characterised by $\kappa=1.4$ but now accounting for a possible
stau--bino coannihilation with $\mbino=1.01\,\mstau$, the conservative
constraints associated with~\eqref{Eq:D_cons} appear for the first
time (bottom left panel).
In the extreme case of having simultaneous bino--stau--slepton
coannihilation with $\mbino=1.01\,\mstau$ and $\kappa=2.0$, one
encounters the most restrictive case shown in that figure.  There the
severe D limit disfavours almost the entire parameter space with
$10^2\,\seconds<\tau_{\stau}<10^5\,\seconds$ (bottom right panel).

In \reffig{fig:SWIMPMB} both the severe and the conservative hadronic
BBN constraints already show up at $f_a=10^{12}\,\GeV$ and disfavour
increasingly larger areas of parameter space in the panels for
$f_a=3\times 10^{12}$ and $10^{13}\,\GeV$. This is due to the rather
large stau yield~\eqref{Eq:YSWIMP} that must be assumed at small
$\maxino$ to fulfil
$\Omega_{\dm}^{\mathrm{obs}}=\Omega_{\tilde{a}}^{\NTP}$.
The $\maxino$ dependence of the yield~\eqref{Eq:YSWIMP} explains also
the significant variation of the hadronic BBN constraints with
$\maxino$. In contrast, those constraints do not change for
$\maxino\lesssim 10~\GeV$
when varying $\maxino$ in case of a thermal relic
yield~\eqref{Eq:Ystau} as can be seen in \reffig{fig:MB}.

Electromagnetic BBN constraints have been explored in our study in
close analogy to the hadronic ones.
However, we condense the discussion since those constraints become
relevant only for $\tau_{\stau}\gtrsim 10^4~\seconds$, i.e., in a
region already disfavoured by the CBBN constraints discussed below.
To determine the electromagnetic BBN constraints, we work with the
$\tau_{\stau}$-dependent upper limits on $\xi_{\EM} \equiv
\epsilon_{\EM}\, Y_{\stau}$ imposed by the primordial abundances of
$\Deut$ and by $^3$He/D.
In particular, we consider severe $\Deut_{\EM}^{\mathrm{sev}}$ and
$^3$He/D constraints inferred from the respective limits given in
Fig.~42 of Ref.~\cite{Kawasaki:2004qu} and a conservative
$\Deut_{\EM}^{\mathrm{cons}}$ constraint inferred from the respective
limit given in Fig.~6 of Ref.~\cite{Cyburt:2002uv}.
The electromagnetic energy release $\epsilon_{\EM}$ is governed by the
tau emitted in the 2-body decay with an energy 
$E_{\tau} = (\mstau^2 - \maxino^2 + \mtau^2)/(2\mstau)$ 
in the rest frame of the $\stauR$. As each $\tau$ decays into at least
one $\nu$, which does not interact electromagnetically, only a
fraction of $E_{\tau}$ contributes to the total electromagnetic energy
injection~\cite{Feng:2003uy,Steffen:2006hw}. We use the conservative
estimate
\bea
  \epsilon_{\EM} 
  = 0.3\, E_\tau 
  = 0.3\, \frac{\mstau^2-\maxino^2+m_{\tau}^2}{2\mstau} \ ,
\label{Eq:epsEM}
\eea
to avoid that the electromagnetic BBN constraints presented are overly
restrictive.
Proceeding as outlined for the hadronic BBN constraints above, we then
obtain upper limits $Y_{\stau\,\EM}^{\max}$ on the stau yield
$Y_{\stau}$ prior to decay.
For the settings considered in \reffig{fig:MB}, we find that the
thermal relic yields~\eqref{Eq:Ystau} respect the obtained
$Y_{\stau\,\EM}^{\max}$ limits in the shown parameter space such that
no electromagnetic BBN constraints appear in that figure; see also
Fig.~5 in Ref.~\cite{Freitas:2009jb}.
This is different in \reffig{fig:SWIMPMB} where we encounter
electromagnetic BBN constraints in the panels with 
$f_a\geq 3\times 10^{12}\,\GeV$.
There the $\Deut$-imposed constraints disfavour the regions enclosed by
the short-dashed (blue) lines and the $^3$He/D-imposed ones the
regions enclosed by the double-dash-dotted (green) lines.
However, we see in \reffig{fig:SWIMPMB} that these constraints only
show up in the regions well within those already excluded by the CBBN
constraints, which will be discussed next, and are absent altogether
for $f_a<3\times10^{12}\,\GeV$.

For the alternative $\selectronR$ and $\smuonR$ NLSP cases, the
electromagnetic BBN constraints will look somewhat different. This is
expected from the different properties of the respective emitted
lepton; cf.\ Sec.~2.2 of Ref.~\cite{Steffen:2006hw}.
For example, in the $\selectronR$ NLSP case, the electromagnetic
energy release is
$\epsilon_{\EM}=E_\mathrm{e}=(m_{\selectronR}^2-\maxino^2+m_{\mathrm{e}}^2)/(2
m_{\selectronR})$
instead of \eqref{Eq:epsEM} since the electron does not decay.
However, we refrain from a further detailed assessment since the
resulting electromagetic BBN constraints will still be less
restrictive than the CBBN constraints for scenarios with the
$\selectronR$ NLSP or the $\smuonR$ NLSP.

Let us now address CBBN constraints on axino LSP scenarios with a
long-lived stau NLSP~\cite{Freitas:2009fb}.
At cosmic times $t>10^3\,\seconds$, the presence of negatively-charged
$\stauR^-$'s can lead to the formation of $(\Hefour\stauR^-)$ and
$(\Beeight\stauR^-)$ bound states once the rate of photodissociation
of those bound states drops below the Hubble expansion rate.
Then, depending on the $(\Hefour\stauR^-)$ and $(\Beeight\stauR^-)$
abundances at the relevant times, which are governed by $\tau_{\stau}$
and $Y_{\stau}$, primordial $\Lisix$ and $\Benine$ production can
become efficient via the following CBBN
reactions~\cite{Pospelov:2006sc,Pospelov:2007js,Pospelov:2008ta}%
\footnote{The large $\Benine$-production cross section reported and
  used in~\cite{Pospelov:2007js,Pospelov:2008ta} has been questioned
  by~\cite{Kamimura:2008fx}, in which a 4-body-model study is
  announced as work in progress to clarify the efficiency of $\Benine$
  production.}
\begin{align}
    &(\Hefour\,\stauR^-)+\Deut \, \rightarrow \, \Lisix + \stauR^- \ ,
\label{Eq:CBBNLiSix} \\
    &\Hefour + (\Hefour\,\stauR^-) \, \rightarrow \, (\Beeight\,\stauR^-)+\gamma \ ,
\label{Eq:RadFusion} \\
    &(\Beeight\,\stauR^-)+n \, \rightarrow \, \Benine+\stauR^- \ .
\label{Eq:ResNCapture}
\end{align}
Here we work with the observationally-inferred limits on the
respective primordial fractions of
$\Lisix$~\cite{Cyburt:2002uv,Asplund:2005yt,Jedamzik:2007qk} and
$\Benine$~\cite{Pospelov:2008ta},
\bea
  \Lisix/\mathrm{H}|_{\mathrm{obs}} 
  & \leq & 
  10^{-11}\!-\!10^{-10},
\label{Eq:LiSix}\\
  \Benine/\mathrm{H}|_{\mathrm{obs}} 
  & \leq &
  2.1\times 10^{-13},
\label{Eq:BeNine}
\eea
and obtain the CBBN constraints by confronting $Y_{\stau}$ with the
limits in Fig.~5 of Ref.~\cite{Pospelov:2008ta}.

In \reffigs{fig:MB} and \ref{fig:SWIMPMB}, the CBBN constraints
associated with the range \eqref{Eq:LiSix} are indicated by pairs of
long-dash-dotted ($\Lisix$, red) lines and the ones associated
with~\eqref{Eq:BeNine} by solid ($\Benine$) lines.
The regions below those lines are disfavoured by an excess of $\Lisix$
and $\Benine$ above the respective limits.
As seen in \reffig{fig:MB}, the CBBN constraints on scenarios with a
thermal relic yield~\eqref{Eq:Ystau} follow contours of constant
$\tau_{\stau}$ and become independent of $\maxino$ for
$\maxino^2/\mstau^2\ll 1$. In fact, we find that the CBBN constraints
disfavour $\tau_{\stau} \gtrsim 4\times 10^3\,\seconds$ and that they
show only a mild dependence on $Y_{\stau}$ in contrast to the hadronic
BBN constraints.
For the yield~\eqref{Eq:YSWIMP}, the CBBN constraints can disfavour
even smaller
$\tau_{\stau}$ of about $10^3\,\seconds$ (cf.~\reffig{fig:SWIMPMB})
as they become more restrictive towards smaller $\maxino$ due to the
significant increase of $Y_{\stau}$.

Here one should also comment on the potential interplay between late
energy injection and CBBN. We use the CBBN limits from
Ref.~\cite{Pospelov:2008ta} which have been derived for a $\Deut$
abundance obtained with SBBN. For an increased $\Deut$ abundance
resulting from hadrodissociation of $\Hefour$, CBBN of $\Lisix$ and
$\Benine$ becomes even more efficient.
For $\Lisix$, this is evident since its catalysis proceeds
via~\eqref{Eq:CBBNLiSix} and since the primordial abundance of $\Deut$
stays significantly below the one of $\Hefour$ at the relevant times
(even for a maximum of observationally tolerable hadrodissociation of
$\Hefour$); cf.\ Fig.~2.4 in Ref.~\cite{Pradler:2009mt}.
For $\Benine$, an increased output results from the final step of its
catalysis~\eqref{Eq:ResNCapture}, which becomes more efficient since
an enhanced abundance of $\Deut$ increases the number of neutrons $n$
at the relevant times~\cite{Mukhanov:2003xs}; cf.\ Fig.~4 in
Ref.~\cite{Pospelov:2008ta}.
Moreover, the debris of hadrodissociated $\Hefour$ can hit ambient
$\Hefour$ and thereby fuse additional
$\Lisix$~\cite{Jedamzik:2004er,Kawasaki:2004qu,Jedamzik:2006xz}.
The interplay of late energy injection and CBBN will thus lead to
constraints that can only be stronger than the ones presented in this
work.
Aiming at conservative limits, this allows us to neglect those
intricacies which will have to be faced in future refinements of the
presented constraints.

\subsection{Comparison and Discussion of the Cosmological Constraints}
\label{Sec:DiscussionConstraints}

Let us close this section with a brief comparative discussion of the
cosmological constraints addressed in this work. We consider again
cosmological settings with a thermal relic stau yield~\eqref{Eq:Ystau}
and such with~\eqref{Eq:YSWIMP} satisfying
$\Omega_{\axino}^{\NTP}=\Omega_{\mathrm{dm}}$ by construction.

When presenting \reffig{fig:MB}, negligible
$\Omega_{\axino}^{\TP}+\Omega_a$ was assumed and thereby
$\TR \ll 10^4\,\GeV$ and $\Theta_i\ll 1$.
While the shown BBN constraints are not affected by this assumption,
significant additional parts of the parameter space can be excluded by
the dark matter constraint~\eqref{Eq:OmegaConstraint} for $\TR\gtrsim
10^4\,\GeV$, as explicitly illustrated in \reffig{Fig:DMConstraint}.
With thermally produced axinos dominating $\Omega_{\dm}$, which are
compatible with cosmic structure formation down to $\maxino$ as low as
$12~\keV$ or even smaller (cf.\ \refsec{Sec:SmallScaleStructure}), the
structure formation constraints shown in \reffig{fig:FSvel} do not
apply.
In other cases with $\TR \ll 10^4\,\GeV$ and a thermal relic stau
yield~\eqref{Eq:Ystau}, those constraints apply within the grey bands
shown in \reffig{fig:MB}.
However, they are still relatively mild. For the upper (lower) panels,
it can be seen in~\reffig{fig:FSvel} (\reffig{fig:SWIMPMB}, bottom
right panel) that only the most restrictive velocity
limit~\eqref{eq:vFSrestrictive} disfavours parts of the grey bands for
$\mstau\gtrsim 3\,\TeV$ ($\mstau> 1~\TeV$).
Moreover, the contributions to $\DNueff$ at $T\simeq 1~\MeV$ are
typically very small and clearly compatible with current limits for
thermally produced axinos and for those from decays of thermal relic
staus with $Y_\stau$ described by~\eqref{Eq:Ystau}.

In \reffig{fig:SWIMPMB}, in which $Y_\stau$ given by~\eqref{Eq:YSWIMP}
is used, $\Omega_{\axino}^{\NTP}=\Omega_{\mathrm{dm}}$ is considered
everywhere in the shown parameter space. This also implies negligible
$\Omega_{\axino}^{\TP}+\Omega_a$ and thereby $\TR \ll 10^4\,\GeV$ and
$\Theta_i\ll 1$. Moreover, here the constraints from structure
formation associated with~\eqref{eq:vFSconservative}
and~\eqref{eq:vFSrestrictive} apply not only in the grey bands but
disfavour the regions to the left of the respective solid (purple)
diagonal lines. Accordingly, even the conservative structure formation
constraint associated with~\eqref{eq:vFSconservative} is significantly
more restrictive than both the
constraint~\eqref{eq:maxinoStdCosmology} with $\fstau\sim 0.1$ imposed
by our working hypothesis of a standard thermal history and the BBN
constraint~\eqref{eq:maxinoLimitDNueffAxinoNTP} with
$(\DNueff)^{\max}=\Order(1)$ at
$T\simeq 1~\MeV$.
In the regions with larger $\maxino$ that are allowed by the
respective structure formation constraint, the hadronic BBN
constraints are often the most restrictive ones. While those become
weaker for $\maxino\to\mstau$, the CBBN constraints become
increasingly more restrictive in that region which is associated with
larger $\tau_{\stau}$ values due to phase-space suppression of the
stau decays. As already mentioned, the electromagnetic BBN constraints
show only up for $\tau_{\stau}>10^4\,\seconds$ and only in parts of
the parameter space already disfavoured by the CBBN constraints.
We also show grey bands in~\reffig{fig:SWIMPMB} but merely as a guide to
show where the stau yield~\eqref{Eq:YSWIMP} coincides with that of the
stau thermal relic yield~\eqref{Eq:Ystau} with $\kappa=0.7$ or $1.4$
(first three panels) and with $\kappa=1.4$ or $2.0$ (bottom right
panel).
Those are the regions in which one might consider~\eqref{Eq:YSWIMP} to
take on its most natural values, i.e., values motivated by
thermal-freeze-out considerations.

Looking once more at \reffig{fig:MB}, one finds that the hadronic BBN
constraints are much more sensitive to the stau yield $Y_{\stau}$ and
the $\mbino/\mstau$ ratio than the CBBN constraints.
While the CBBN constraint disfavours basically 
$\mstau\lesssim 500~\GeV$ in all four panels,
the hadronic BBN constraint may not show up at all or may disfavour
even significantly larger $\mstau$ values of $1~\TeV$ or above,
depending on $Y_{\stau}$ and the $\mbino/\mstau$ ratio.
Indeed, this sensitivity motivated us to show those four combinations
of $\kappa$ and $\mbino/\mstau$ accounting systematically for the
possibilities of no coannihilation, stau--slepton coannihilation,
stau--bino coannihilation, and simultaneous stau--bino--slepton
coannihilation.
While $f_a$ is fixed to $10^{13}\,\GeV$ in \reffig{fig:MB}, an
illustration of the $f_a$ dependence of the CBBN constraints can be
found in Fig.~4 of Ref.~\cite{Freitas:2009fb} and in Fig.~5 of
Ref.~\cite{Freitas:2009jb} which also includes hadronic and
electromagnetic BBN constraints.
For $|e_Q|=1/3$ and $y=1$, those figures demonstrate that BBN
constraints start to become an issue for $f_a\gtrsim 10^{12}\,\GeV$
while settings with smaller $f_a$ are free from those constraints when
considering thermal relic stau yields and the limit $\mstau\gtrsim
80\,\GeV$ from searches at LEP~\cite{Nakamura:2010zzi}.
With $\epsilon_{\HAD/\EM}$ being independent of $e_Q$ and
$\tau_{\stau}\propto 1/e_Q^4$, even settings with $f_a$ up to
$10^{13}\,\GeV$ will be free from BBN constraints if $|e_Q|=1$ and
$y=1$.

Finally, there is the possibility that very light thermal relic axinos
may exist as HDM while CDM is provided by
axions~\cite{Brandenburg:2004du}.  Here the obtained limit
$\maxino\lesssim 37~\eV$ from structure formation constraints is
significantly more restrictive than the previously known limit
$\maxino\lesssim 200~\eV$ imposed by the dark matter
constraint~\eqref{Eq:OmegaConstraint}.
Interestingly, also for very light axinos the hadronic,
electromagnetic and CBBN constraints apply since they are sensitive to
the yield of the stau NLSP, its SM decay products and its lifetime,
which is independent of $\maxino$ in that regime.
As mentioned, for $Y_\stau$ given by~\eqref{Eq:Ystau}, the location of
those constraints can then be seen directly in \reffig{fig:MB} as they
agree with the ones shown for $\maxino\lesssim 10~\GeV$.

\section{Collider Phenomenology of Axino Dark Matter Scenarios}
\label{Sec:Collider}

Cosmological considerations can significantly constrain the parameter
space of SUSY axion models. Nevertheless, in order to investigate and
to determine the properties of axinos, one must look towards
laboratory experiments. Due to the axino's highly suppressed
interactions, its production cross sections at various colliders are
very small. Instead, one would look for the NLSPs, the staus, either
from direct pair production or from cascade decays of heavier
sparticles. In fact, a first hint towards axino dark matter would come
from the appearance of quasi-stable staus at the Tevatron, the LHC or
future colliders.%
\footnote{Collider studies of long-lived charged massive particles
  (CHAMPs), a category to which the staus belong, can be found for
  example in
  Refs.~\cite{Drees:1990yw,Nisati:1997gb,Ambrosanio:1997rv,Feng:1997zr,Fairbairn:2006gg,Martin:1998vb}.}

The most straightforward property of the stau to determine would be
its mass $\mstau$, which can be measured using time-of-flight (TOF)
data from the muon chambers. At the LHC, it is estimated that this can
be done with $<1\%$ accuracy \cite{Ambrosanio:2000ik,Ellis:2006vu}. In
order to measure other properties, such as the stau lifetime
$\tau_{\stau}$ and the axino mass $\maxino$, it is important to be
able to study the stau decays. The approach that one should take to
analyse these decays depends on the decay length $c\tau_{\stau}$ and
on the size of the detector $L$. Here it is natural to distinguish the
following three cases,
which we now relate to the considered axino dark matter scenarios:

\begin{itemize}

\item If the lifetime is very short such that $c\tau_{\stau} \ll
  \mathcal{O}({\rm cm})$ then the staus would effectively decay
  immediately after production. This decay length corresponds to
  $\tau_{\stau} \ll 10^{-10}\,\seconds$. For the SUSY axion models
  considered in this paper, a stau with such a short lifetime can only
  be achieved with a very low $f_a$ and a very heavy stau, as well as
  $|e_Q|$ larger than 1/3. For example, with the lowest $f_a$ value
  respecting~\eqref{Eq:f_a_axion}, $f_a\simeq 6\times 10^8\,\GeV$, one
  would need $\mstau\gtrsim 2~\TeV$, along with $\mbino/\mstau=2$,
  $\maxino/\mstau\ll 1$, $y=1$ and $|e_Q|=1$, to obtain
  $\tau_{\stau}\lesssim 2 \times 10^{-9}\,\seconds$. This very high
  stau mass is out of the reach of current and near future colliders.
  If one considers a stau mass that is more likely to be probed at
  colliders, $\mstau=300\,\GeV$ while keeping $f_a\simeq 6 \times
  10^8\,\GeV$, along with the selection of $\mbino/\mstau=1.1$,
  $\maxino/\mstau\ll 1$, $y=1$ and $|e_Q|=1/3$, one obtains a lifetime
  of $\tau_{\stau}\approx 10^{-4}\,\seconds$ and $c\tau_{\stau}\approx
  30~\km$.

\item If the lifetime is such that $\mathcal{O}({\rm cm}) \lesssim
  c\tau_{\stau} \lesssim L$ then the staus will decay with a displaced
  vertex within the beam pipe or the detectors. With a typical
  detector size, $L< 0.1~\km$, it is evident from the arguments given
  above that this case still requires a very low $f_a$ and a very
  heavy stau to be relevant for the considered axino LSP scenarios.

\item If the decay length is $c\tau_{\stau}\gtrsim L$, there would be
  only few in-flight decays within the detector and most of the staus
  would escape the detector. This is the case on which we focus in
  this section. It applies to axino LSP scenarios
  respecting~\eqref{Eq:f_a_axion} and with $\mstau$ such that the stau
  NLSP can be probed at current or near future colliders.

\end{itemize}

There are different ways to access stau decays in the considered third
case.
For $\tau_{\stau} \sim 10^{-(3-5)}\,\seconds$ (still corresponding to
decay lengths well over $L$), it has been argued that a substantial
number of in-flight decays can still be observed within the
detectors~\cite{Ishiwata:2008tp,Kaneko:2008re}.
Also for larger $\tau_{\stau}$, decays could be accessible since the
produced staus will be slowed down (due to ionisation energy loss) so
that some of them are expected to be stopped within the main
detectors~\cite{Martyn:2006as,Asai:2009ka,Pinfold:2010aq} or in an
additional dedicated
stopping-detector~\cite{Goity:1993ih,Hamaguchi:2004df,Hamaguchi:2006vu}.%
\footnote{Note that this method is not limited to staus with long
  lifetimes. It can also be used to study staus with very short
  lifetimes but highly boosted momenta.}
By recording the stopping point/time with those devices, it seems
possible to study the subsequent decays of the trapped staus and to
determine their lifetime.
For very long lifetimes, there was also a suggestion to place a water
tank around the detector to trap the staus, such that the water can be
moved to a different location at a later date for further
analyses~\cite{Feng:2004yi}.
Even the surrounding rock has been suggested to be used for the study
of stau decays~\cite{DeRoeck:2005bw}.

In the coming parts of this section we first explore the prospects of
stopping staus and studying their 2-body decays with the main
detectors at the LHC and the ILC,
i.e., without additional stopping-detectors. Four benchmark scenarios
are proposed and studied for this purpose. Thereafter, we turn to the
3-body stau decays into the LSP, a tau and a photon, whose study might
require additional stopping-detectors. Based on those decays, we
address the possibility to distinguish between the axino LSP and the
gravitino LSP at colliders.

\subsection{Probing 2-Body Decays of Trapped Staus at the LHC and the ILC}
\label{sec:trappedstaus}

To illustrate the potential of stopping staus and analyzing their
2-body decay into the axino LSP at the LHC and the ILC, we consider
the four SUSY benchmark points specified in
\reftable{tab:CMSSMpoints}. To simplify matters and minimise the
number of parameters that have to be specified, we work in the
framework of CMSSM. Here the gaugino masses, the scalar masses and the
trilinear scalar interactions are assumed to take on the respective
universal values $m_{1/2}$, $m_0$ and $A_0$ at the scale of grand
unification $M_{\mathrm{GUT}}\simeq 2\times 10^{16}\,\GeV$. Specifying
also the mixing angle $\tan\beta$ in the Higgs sector and the sign of
the higgsino mass parameter $\mu$, one can calculate associated mass
spectra at the weak scale and stau yields $Y_{\stau}$ prior
to decay.  For the considered benchmark points, we obtain those
quantities by employing the spectrum calculator
SuSpect~\cite{Djouadi:2002ze} and
MicrOMEGAs~2.4~\cite{Belanger:2001fz,Belanger:2004yn} for $m_t =
172.5\,\GeV$. Those tools are also used to verify that the accelerator
constraints are respected for our chosen points.
%
\TABLE[t]{
\begin{tabular}{c c c c c c c c}
\toprule
Scenario 
& $m_0$\,[GeV] 
& $m_{1/2}$\,[GeV] 
& $\tan\beta$ 
& $\mstau$\,[GeV] 
& $\mbino$\,[GeV] 
& $m_{\tilde{e}_{{\rm R}}}$\,[GeV] 
& $\kappa$ \\
\hline 
A 
& 170      
& 560                 
& 30                            
& 210.3                 
& 232.4                 
& 270.1         
& 0.73 \\
B         
& 50        
& 550                        
& 8                             
& 207.6                 
& 226.8                 
& 212.4         
& 1.34 \\
C         
& 138      
& 435                 
& 25                            
& 174.3                 
& 177.7                 
& 215.6         
& 1.42 \\
D         
& 135      
& 700                 
& 6                             
& 290.2                 
& 292.6                 
& 292.8         
& 1.94 \\
\bottomrule
\caption{Benchmark points in the CMSSM that emulate some of the scenarios
explored in \refsec{Sec:AxinoConstraints}. For all four point s,
$A_0=0$ and ${\rm sgn}(\mu)=1$. The weak scale masses of the lighter
stau, the bino-like neutralino and the lighter selectron and the stau
yield $Y_{\stau}$ given by the respective $\kappa$ with
\eqref{Eq:Ystau} are calculated with the tools SuSpect
\cite{Djouadi:2002ze} and micrOMEGAs 2.4
\cite{Belanger:2001fz,Belanger:2004yn} for $m_t = 172.5\,\GeV$.}
\label{tab:CMSSMpoints}
\end{tabular}}

We have chosen the benchmark points such that they fall into the
classes for which we studied cosmological constraints in
\refsec{Sec:AxinoConstraints}. This can be seen in
\reftable{tab:CMSSMpoints}, where $\kappa$ values specify the
associated stau yield $Y_{\stau}$ given by \eqref{Eq:Ystau} with the
respective $\mstau$ values. Scenario~A shows no coannihilation, B
stau--slepton coannihilation, C stau--bino coannihilation and D
simultaneous stau--slepton-bino coannihilation; cf.\ text
below~\eqref{Eq:Ystau}.
These four scenarios come with a bino-like lightest neutralino and
with lighter staus that have $\lesssim 10\%$ left-handed component. 
This left-admixture can be reduced further by fine-tuning the parameters
or considering other SUSY scenarios that offer greater flexibility in
choosing parameters.  In fact, we use our result for the stau lifetime
derived for right-handed staus in \refsec{Sec:NLSPDecays} also
for the four benchmark points even if they show a non-negligible
left-right mixing.

The SUSY hadronic axion model is not part of the CMSSM and the
associated parameters have to be specified in addition to those in
\reftable{tab:CMSSMpoints}. For concreteness, we choose
$\maxino=10\,\GeV$, $f_a=10^{11}\,\GeV$, $y=1$ and $|e_Q|=1/3$.
While this scenario respects all constraints from structure formation
and BBN, the dark matter constraint limits the maximum reheating
temperature to a relatively low value of about $10^3\,\GeV$;
cf.~\reffig{Fig:DMConstraint} and associated explanations given in
\refsec{Sec:OmegaDMConstraints}.
For smaller $\maxino$ and/or larger $f_a$, this $\TR$ limit would be
more relaxed.
Note also that a change in $\maxino$ will not noticeably affect the
stau lifetime and the associated error derived below, as long as
$\maxino^2/\mstau^2\ll 1$. However, the same is not true for a change
in $f_a$, on which the stau lifetime strongly depends; see
\refeq{eq:2lamp_Gamma}.

Let us now turn to the LHC. As mentioned above, we concentrate on the
fraction of staus that are produced with small enough values of
$\beta\gamma=|\vec{p_{\stau}}|/\mstau$ such that they lose enough
kinetic energy through ionisation to be stopped within the main
detectors. For this setting, the authors of \cite{Asai:2009ka} suggest
the possibilities of having a trigger for a beam dump should a stau be
detected to have stopped, such that its subsequent decay can be
observed, or of studying the decays of the staus during beam
downtime.%
\footnote{Another recent LHC proposal
\cite{Pinfold:2010aq} suggested to measure the lifetime of staus
trapped in the ATLAS detector during scheduled shutdown periods. It
is thus more suitable for studying staus with a very long lifetime 
of several days or more. However, in the scenarios that we are considering,
a lifetime longer than about three hours is disfavoured by the 
CBBN constraints discussed in \refsec{Sec:NucleosynthesisConstraints}.}

Our LHC study follows closely the approach of Ref.~\cite{Asai:2009ka}.
Using Pythia~6.4 \cite{Sjostrand:2006za} and the cuts 1--4 outlined in
\cite{Asai:2009ka}, we have generated proton--proton collision 
events  with a centre of mass energy
$\sqrt{s}=14~\TeV$ and an integrated luminosity
$\mathcal{L}=10~\fb^{-1}$
for each CMSSM point
given in \reftable{tab:CMSSMpoints}. 
The resulting $|\vec{p}_{\stau}|/m_{\stau}$ spectra are shown in
\reffig{fig:LHCmomspectra}.
%
\FIGURE[t]{$\begin{array}{c@{\hspace{7mm}}c}
    \epsfig{figure=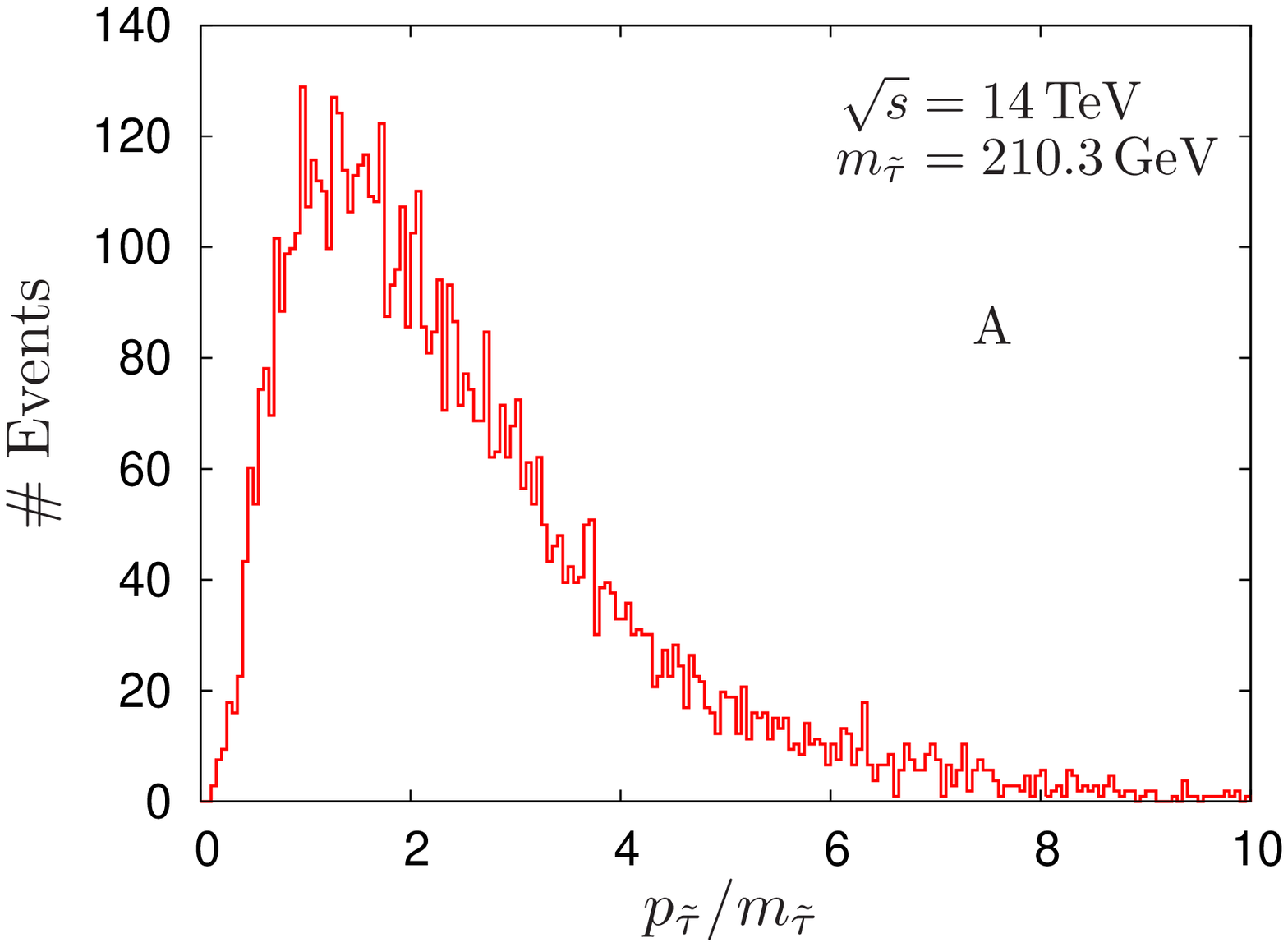, width=0.47\textwidth} &
    \epsfig{figure=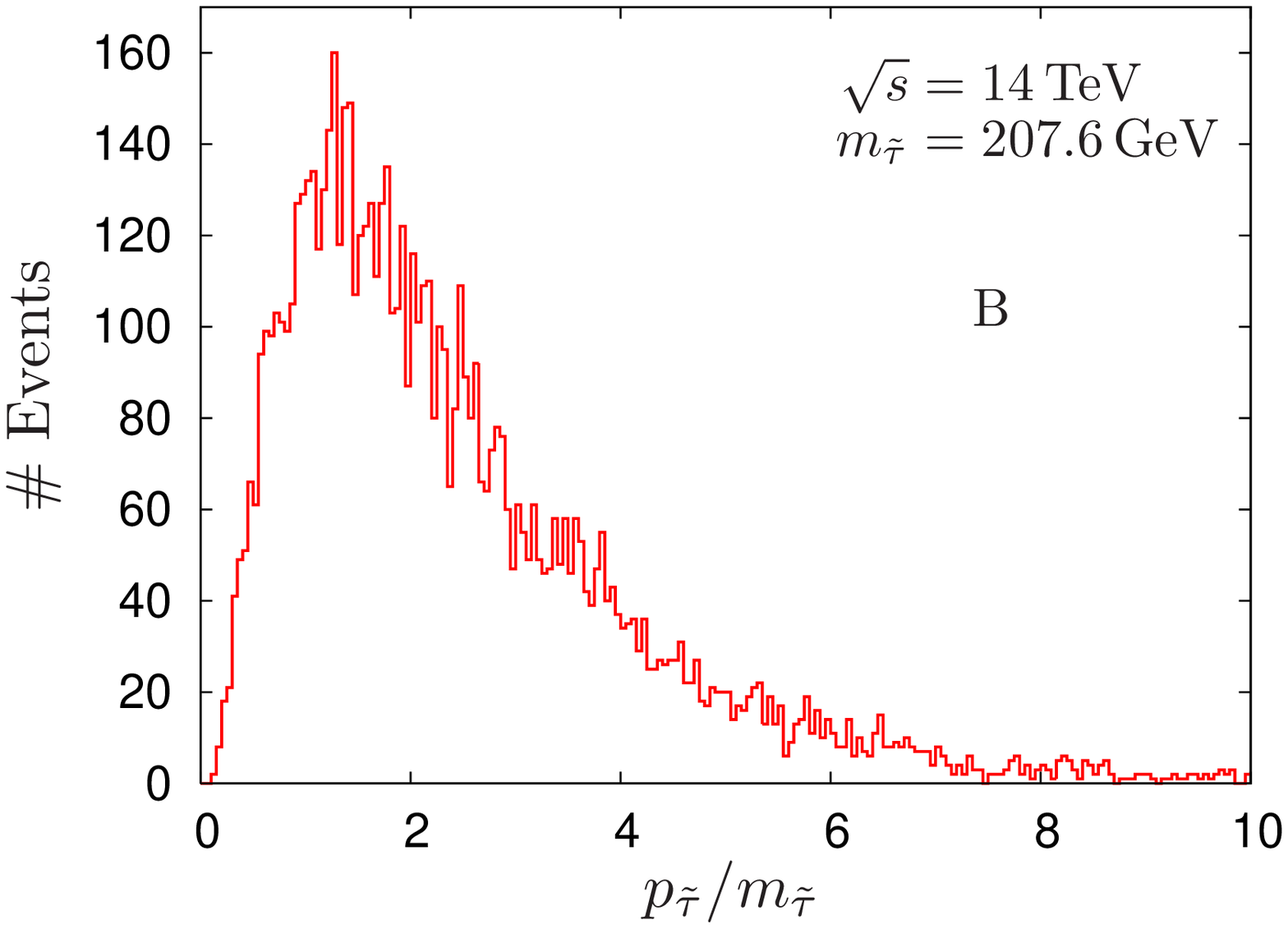, width=0.47\textwidth} \\ [0.2cm]
    \epsfig{figure=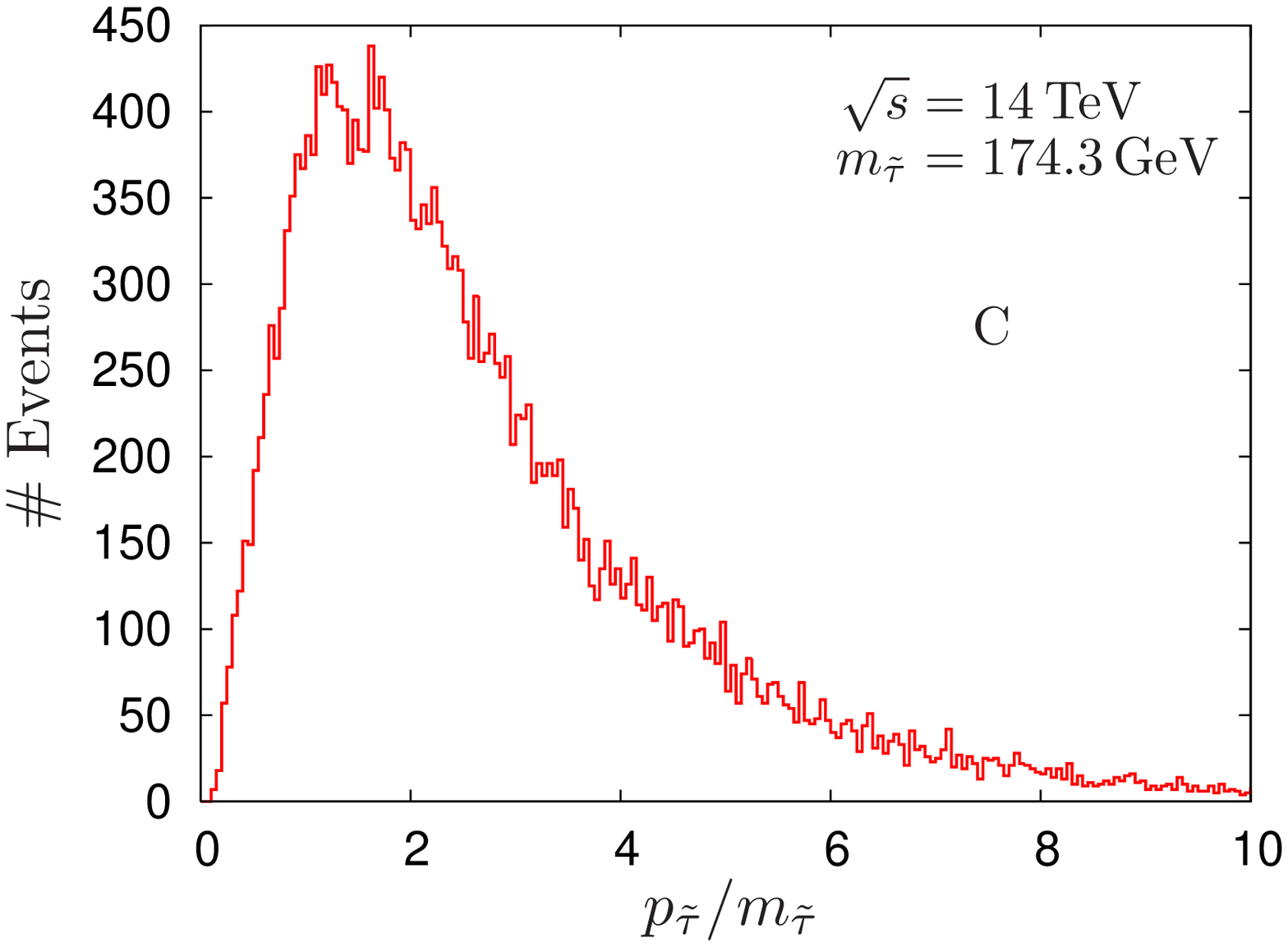, width=0.47\textwidth} &
    \epsfig{figure=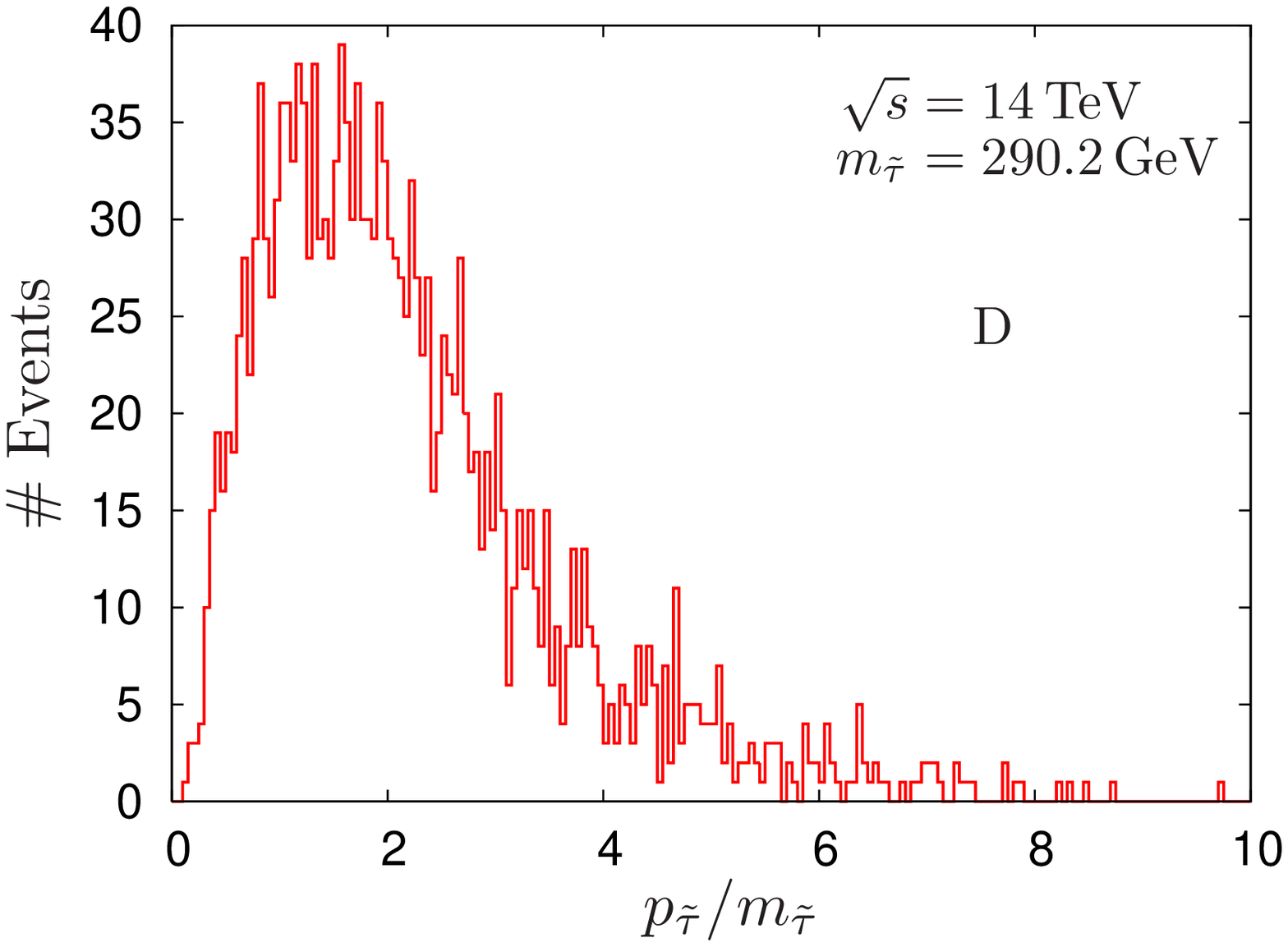, width=0.47\textwidth}
  \end{array}$
  \caption{Expected $|\vec{p_{\stau}}|/\mstau$ distributions of
    lighter staus at the LHC with $\sqrt{s}=14~\TeV$ and an
    integrated luminosity of $\mathcal{L}=10~\fb^{-1}$ for the CMSSM
    points A--D specified in \reftable{tab:CMSSMpoints}. The events
    have been generated with Pythia~6.4 \cite{Sjostrand:2006za} and
    the cuts 1--4 outlined in \cite{Asai:2009ka} have been applied.}
\label{fig:LHCmomspectra}}
%
Following the strategy outlined in \cite{Asai:2009ka}, we consider
only the staus produced with $\beta\gamma\leq0.45$ as those that get
trapped in the calorimeters of the ATLAS detector. 
For the considered benchmark scenarios, we list the corresponding
numbers of staus expected to be stopped in the ATLAS calorimeters,
$N_{\stau}^{{\rm stopped}}$,
together with the right-handed up squark mass $m_{\tilde{u}_R}$, the
gluino mass $m_{\tilde{g}}$ and the total SUSY production cross
section $\sigma_{\rm SUSY}$ in \reftable{tab:LHCruns}.
%
\TABLE[ht]{
\centerline{\begin{tabular}{c c c c c c}
\toprule
Scenario 
& $m_{\tilde{u}_R}$ [GeV]
& $m_{\tilde{g}}$ [GeV] 
& $\sigma_{\rm SUSY}$ [pb]
& $N_{\stau}^{{\rm stopped}}$ 
& $\tau_{\stau}\,[\seconds]$ 
\\ \hline 
A                  
& 1127.2                        
& 1272.8                
& 0.66          
& 119           
& 5.4\,$\pm$\,0.8 
\\
B        
& 1097.5                  
& 1250.3                
& 0.76          
& 182           
& 5.8\,$\pm$\,0.6 
\\
C        
& 896.4                  
& 1008.3                
& 2.60          
& 522           
& 10.8\,$\pm$\,0.7 
\\
D        
& 1372.1                  
& 1563.6                
& 0.185         
& 58                    
& 2.5\,$\pm$\,0.5 
\\
\bottomrule
\caption{Expected number $N_{\stau}^{{\rm stopped}}$ of lighter staus to be
stopped in the calorimeters of the ATLAS detector at the LHC with
$\sqrt{s}=14~\TeV$ and an integrated luminosity of
$\mathcal{L}=10~\fb^{-1}$ for the CMSSM points A--D specified in
\reftable{tab:CMSSMpoints}, and the resulting expected statistical
68\%~CL accuracy of the determination of $\tau_{\stau}$ for
$\maxino=10~\GeV$, $f_a=10^{11}\,\GeV$, $y=1$ and $|e_Q|=1/3$. For
each CMSSM point, we also list the weak scale values of
$m_{\tilde{u}_R}$, $m_{\tilde{g}}$ and $\sigma_{\rm SUSY}$, as
obtained with the spectrum calculator SuSpect \cite{Djouadi:2002ze}
for $m_t = 172.5~\GeV$ and with Pythia~6.4 \cite{Sjostrand:2006za}.}
\label{tab:LHCruns}
\end{tabular}}}
%
Clearly $N_{\stau}^{{\rm stopped}}$ is larger for larger 
$\sigma_{\rm SUSY}$,%
\footnote{\reftable{tab:LHCruns} lists leading order results for $\sigma_{\rm SUSY}$. The values increase by factors of 1.2--1.4 when including the next-to-leading-order SUSY-QCD corrections so that $N_{\stau}^{{\rm stopped}}$ may be above the given estimates.}
and thus governed by $m_{\tilde{u}_R}$ and $m_{\tilde{g}}$ at the LHC.

The lifetime of the trapped staus is determined by measuring the time
between when the stau is stopped and when its subsequent decay is
detected, where we refer to \cite{Asai:2009ka} for details on the
different techniques one would use for different $\tau_{\stau}$.
By extrapolating the results of \cite{Asai:2009ka} to our parameter
points, we obtain the expected statistical 68\%~CL precision of the
$\tau_{\stau}$ determination listed in \reftable{tab:LHCruns}.
Clearly the statistical accuracy depends on the number of events that
can be observed,
and at the LHC, this is very much dictated by the mass spectrum of the
SUSY scenario that is realised. With the conditions assumed for the
generation of the above events, light squarks and gluinos with masses
$m_{\tilde{q},\tilde{g}} \lesssim 1~\TeV$ would be preferred for a
statistical 68\%~CL accuracy of the $\tau_{\stau}$ determination of 
$\lesssim 10\%$. 

A similar analysis can be done for a future linear collider.  
In Ref.~\cite{Martyn:2006as} it has been proposed to use the Large
Detector Concept (LDC)~\cite{Brau:2007zza} at the ILC to study the
decays of the trapped staus. The staus produced in the collisions can
become trapped either in the hadronic calorimeter or the yoke of the
LDC. We follow very closely the strategy of \cite{Martyn:2006as} to
determine the statistical accuracy with which the properties of the
stau and the axino can be determined.  

Again using Pythia~6.4~\cite{Sjostrand:2006za}, we now generate 
electron--positron collision events
with a centre of mass energy of
$\sqrt{s}=600~\GeV$ and an integrated luminosity of
$\mathcal{L}=250~\fb^{-1}$
for each CMSSM point given in \reftable{tab:CMSSMpoints} .
The obtained $|\vec{p}_{\stau}|/m_{\stau}$ distributions for the
produced lighter staus are shown in \reffig{fig:ILCmomspectra}.
%
\FIGURE[t]{$\begin{array}{c@{\hspace{7mm}}c}
    \epsfig{figure=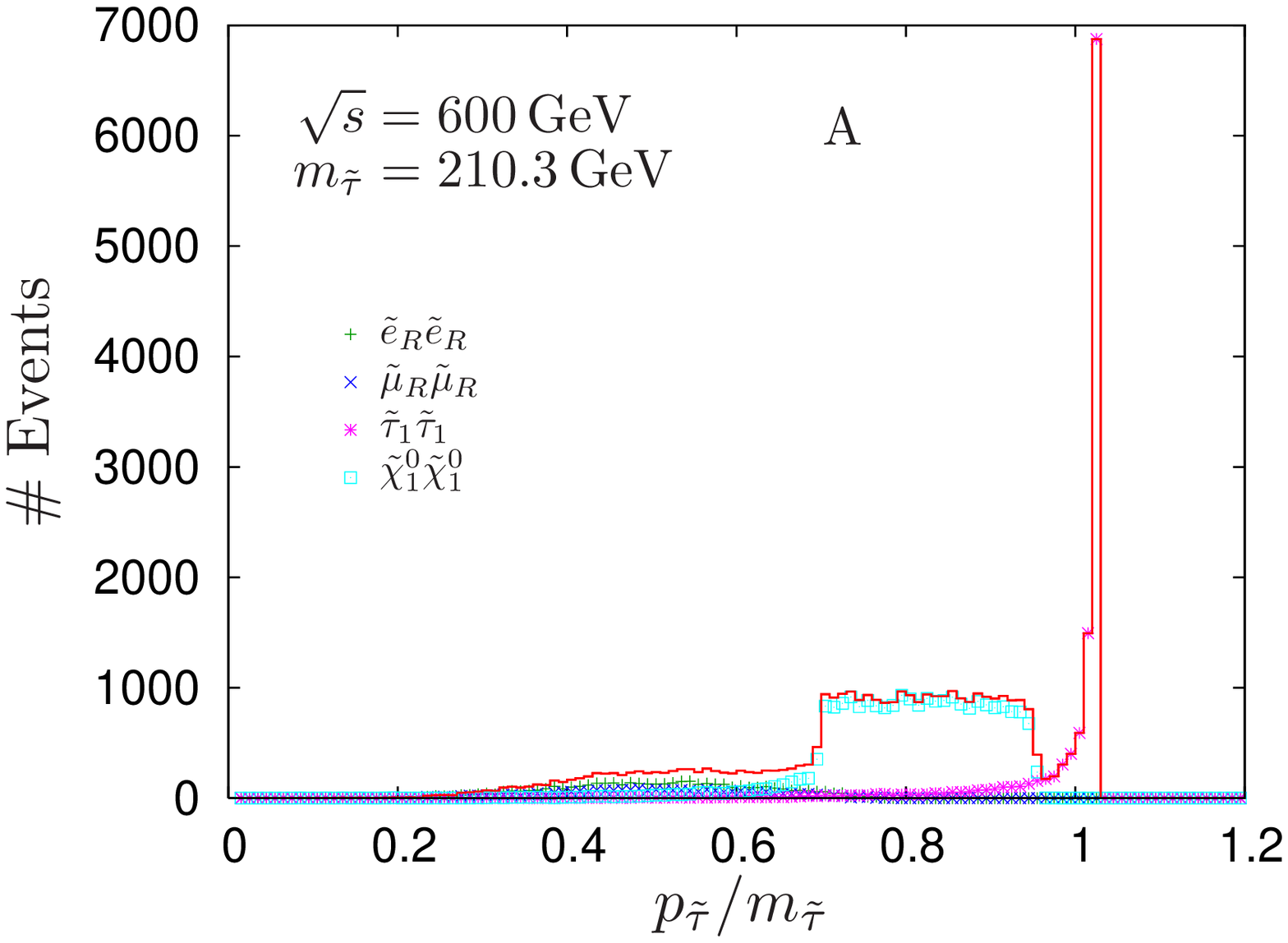, width=0.47\textwidth} & 
    \epsfig{figure=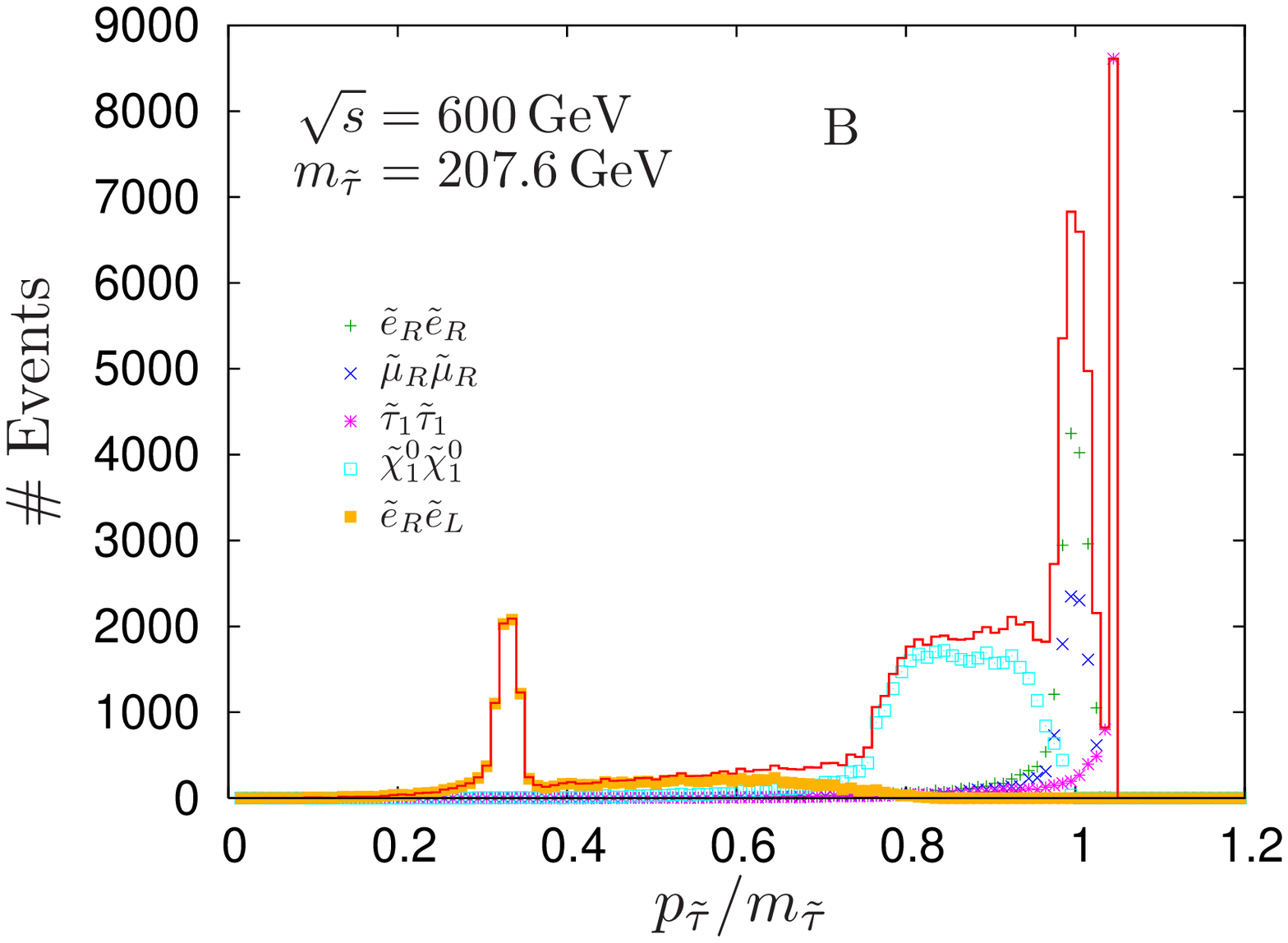, width=0.47\textwidth} \\ [0.2cm]
    \epsfig{figure=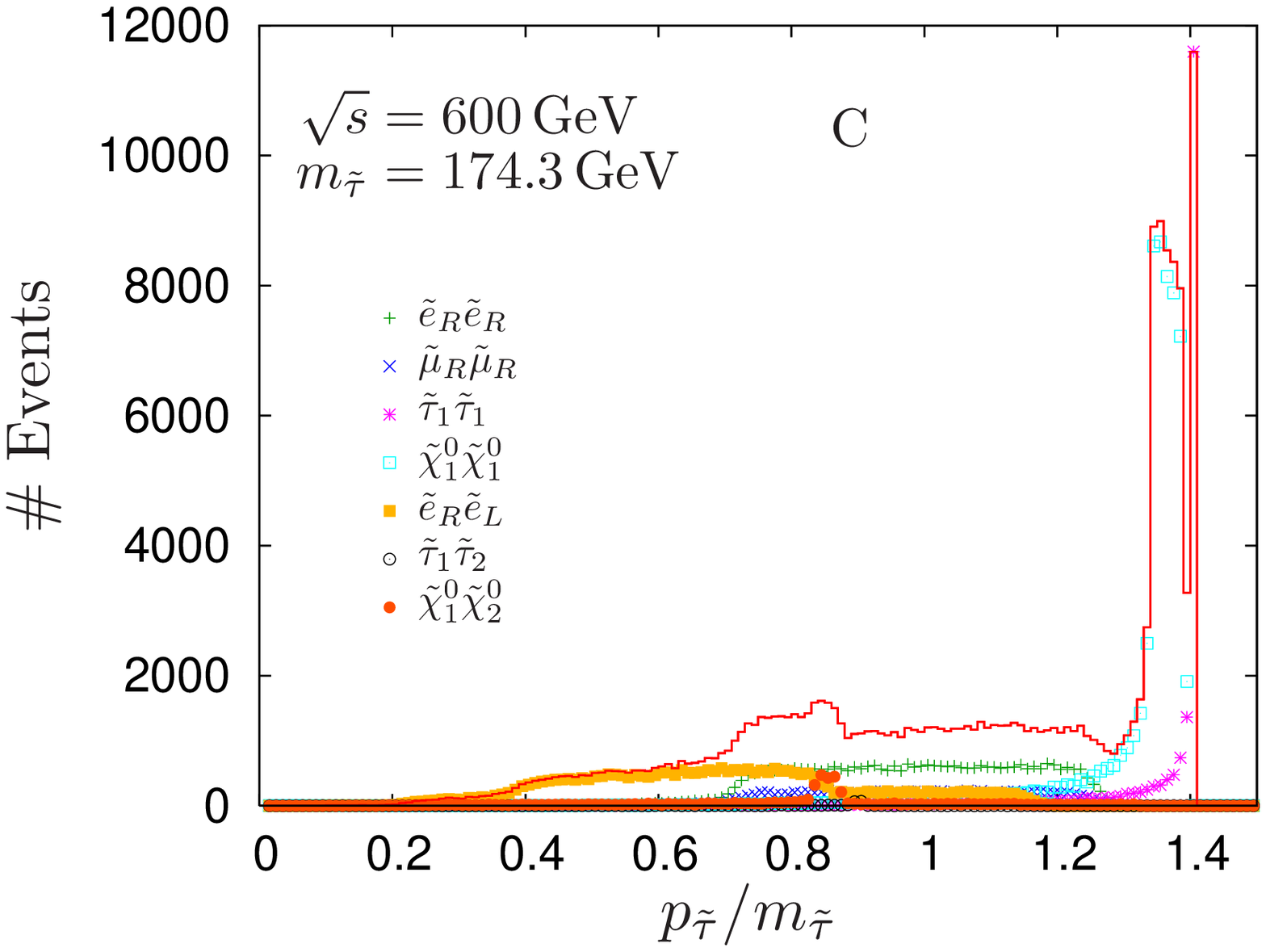, width=0.47\textwidth} &
    \epsfig{figure=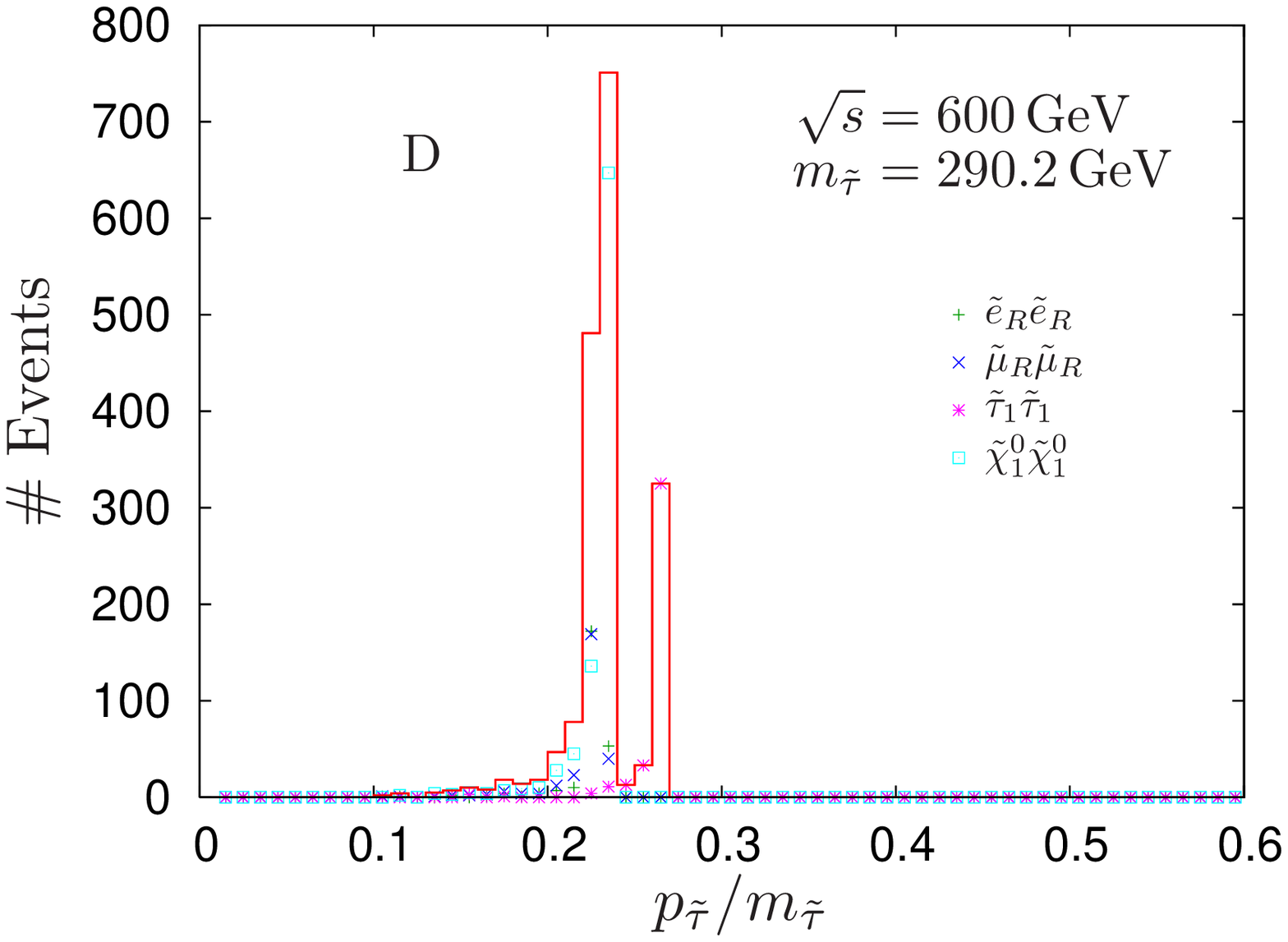, width=0.47\textwidth}
  \end{array}$
  \caption{Expected $|\vec{p_{\stau}}|/\mstau$ distributions of
    lighter staus (solid line, red) in electron--positron collisions
    at the ILC with
    $\sqrt{s}=600~\GeV$ and an integrated luminosity of
    $\mathcal{L}=250~\fb^{-1}$ for the CMSSM points A--D specified in
    \reftable{tab:CMSSMpoints}. The events have been generated with
    Pythia~6.4~\cite{Sjostrand:2006za}. There are contributions from
    $\positron\electron\to\selectronR\selectronR^*$ (plus signs,
    green), $\smuonR\smuonR^*$ (crosses, blue), $\stauone\stauone^*$
    (stars, magenta), $\tilde{\chi}_1^0\tilde{\chi}_1^0$ (boxes,
    cyan), $\selectronR\selectronL^*$ and $\selectronR^*\selectronL$
    (circles with crosses, yellow), $\stauone\stautwo^*$ and
    $\stauone^*\stautwo$ (open circles, black) and
    $\tilde{\chi}_1^0\tilde{\chi}_2^0$ (filled circles, orange).}
\label{fig:ILCmomspectra}}
%
In addition to the total distribution (solid line, red), we illustrate
the contributions of the following processes to the production of
lighter staus: $\positron\electron\to\selectronR\selectronR^*$ (plus
signs, green), $\smuonR\smuonR^*$ (crosses, blue),
$\stauone\stauone^*$ (stars, magenta),
$\tilde{\chi}_1^0\tilde{\chi}_1^0$ (boxes, cyan),
$\selectronR\selectronL^*$ and $\selectronR^*\selectronL$ (circles
with crosses, yellow), $\stauone\stautwo^*$ and $\stauone^*\stautwo$
(open circles, black) and $\tilde{\chi}_1^0\tilde{\chi}_2^0$ (filled
circles, orange).
This breakdown of the contributing processes allows us to see effects
associated with small mass differences between the lighter stau
$\stauone^{(*)}$ and the sleptons $\selectronR^{(*)}$, $\smuonR^{(*)}$
and/or the lightest neutralino $\tilde{\chi}_1^0$:
Pronounced peaks just below the ones from direct $\stauone^{(*)}$
production come from cascade decays of $\selectronR^{(*)}$ and
$\smuonR^{(*)}$ at point~B (top right panel), of $\tilde{\chi}_1^0$ at
point~C (bottom left panel) and of $\selectronR^{(*)}$,
$\smuonR^{(*)}$ and $\tilde{\chi}_1^0$ at point~D (bottom right
panel).
In addition, one finds also for points A and B that a significant
fraction of the staus is produced in decays of the lightest neutralino
$\tilde{\chi}_1^0$.
This indicates that a lightest neutralino with a mass close to the one
of the lighter stau would be beneficial for phenomenological studies.

Following Ref.~\cite{Martyn:2006as}, 
we have calculated the numbers of staus expected
to be stopped in the hadronic calorimeter, $N_{\stau}^{{\rm hcal}}$,
and in the yoke, $N_{\stau}^{{\rm yoke}}$, of the LDC. 
For the CMSSM
points A--D, the results are tabulated together with the total SUSY
production cross section $\sigma_{\rm SUSY}$ in
\reftable{tab:ILCruns}.
The respective $\beta\gamma$ ranges for which the staus are expected 
to get stopped 
in those detector parts depend on $\mstau$ and are typically
$0.3$--$0.4$ (hcal) and $0.4$--$0.5$ (yoke) for our benchmark points;
see also Table~2 in Ref.~\cite{Martyn:2006as}.
%
\TABLE[t]{
\begin{tabular}{c c c c c c c}
\toprule
Scenario 
& $\sigma_{\rm SUSY}$ [fb] 
& $N_{\stau}^{{\rm hcal}}$ 
& $N_{\stau}^{{\rm yoke}}$ 
& $\mstau$ [GeV] 
& $\tau_{\stau}$ [s] 
& $\maxino$ [GeV] \\
\hline 
A     
& 84         
& 508           
& 1362          
& 210.3\,$\pm$\,0.2       
& 5.4\,$\pm$\,0.2                     
& 10\,$\pm$\,64 
\\
B     
& 190                
& 681           
& 1311          
& 207.6\,$\pm$\,0.2       
& 5.8\,$\pm$\,0.2             
& 10\,$\pm$\,58 
\\
C     
& 310                
& 1569          
& 2933          
& 174.3\,$\pm$\,0.1     
& 10.8\,$\pm$\,0.3            
& 10\,$\pm$\,35 
\\
D     
& 3.6                
& 0                     
& 0                     
& 290.2\,$\pm$\,0.2       
& 2.5\,$\pm$\,0.1             
& 10\,$\pm$\,53 
\\
\bottomrule
\caption{
Expected number of lighter staus to be stopped in the hadronic
calorimeter ($N_{\stau}^{{\rm hcal}}$) and the yoke ( $N_{\stau}^{{\rm yoke}}$) 
of the LDC detector at the ILC with
$\sqrt{s}=600~\GeV$ and an integrated luminosity of
$\mathcal{L}=250~\fb^{-1}$ for the CMSSM points A--D specified in
\reftable{tab:CMSSMpoints}, and expected 68\%~CL accuracies for the
determination of $\mstau$, $\tau_{\stau}$ and $\maxino$ for
$\maxino=10~\GeV$, $f_a=10^{11}\,\GeV$, $y=1$ and $|e_Q|=1/3$. For
each CMSSM point, we also list the value of $\sigma_{\rm SUSY}$, as
obtained with Pythia~6.4 \cite{Sjostrand:2006za}.
}
\label{tab:ILCruns}
\end{tabular}}
%

Note that $N_{\stau}^{{\rm hcal}}$ and $N_{\stau}^{{\rm yoke}}$ depend
not only on $\sigma_{\rm SUSY}$ but are also very sensitive to the
specific shape of the $|\vec{p}_{\stau}|/m_{\stau}$ distribution,
which is governed by the mass spectrum and the considered $\sqrt{s}$.
In particular, for scenario D and $\sqrt{s}=600~\GeV$, almost all the
events are clustered around the
$|\vec{p_{\stau}}|/\mstau=0.2$--$0.3$ region,
as can be seen in the bottom right plot of \reffig{fig:ILCmomspectra}.
The momenta would thus be so low that the staus do not have enough
energy to reach the LDC hadronic calorimeter/yoke.
Nevertheless, they are energetic enough to travel out of the beam pipe
and become trapped in the inner detectors. We thus treat all the staus
produced, 1814, as stopped to derive the expected $\tau_{\stau}$ and
$\maxino$ accuracies listed for scenario~D. This is different for the
scenarios A--C where we use
$N_{\stau}^{{\rm hcal}}$ and $N_{\stau}^{{\rm yoke}}$ 
to infer those accuracies.
Here we should note that a future linear collider will have a tunable
centre of mass energy $\sqrt{s}$. This will allow us to maximize the
number of staus stopped in those detector parts most suitable to study
the decays. However, such optimisations for scenarios A--D are beyond
the scope of this work and left for future studies.

Independent of the stopping of staus, the kinematics from the direct
production of stau pairs will allow for a determination the stau mass
with an accuracy listed in \reftable{tab:ILCruns}.
As shown, we expect that the stau mass can be measured very precisely
at the ILC, with an error below $1\%$ for all four CMSSM points. This
can perhaps be further improved on if combined with a mass extraction
using TOF data. 
The $\tau_\stau$ measurements are not expected to be that accurate,
but greatly improve with increasing numbers of trapped staus as can be
seen in \reftable{tab:ILCruns}.
Accordingly, these measurements would benefit from the $\sqrt{s}$
optimisations mentioned above. 
However, the statistical accuracy of the $\tau_\stau$ determination
expected to be achievable at the ILC is already very promising at this
stage.

For scenarios A--C, the explored measurement of the stau lifetime
utilises the staus trapped both in the hadronic calorimeter and the
yoke, while only those trapped in the hadronic calorimeter are
considered when probing the axino mass.
This restriction results from the fact that the axino mass exploration
relies on the kinematics of the 2-body decay $\stauR\to\tau\axino$,
\bee
\maxino=\sqrt{\mstau^2+m_{\tau}^2-2\mstau E_{\tau}},
\label{Eq:maxino}
\eee 
and thereby on the determination of the energy of the emitted tau
lepton $E_{\tau}$~\cite{Brandenburg:2005he}. With $E_{\tau}$ being the
maximum energy of the observable $\tau$ decay products, the latter
requires to study the endpoints of the energy spectrum of the emitted
`$\tau$ jets' and thus a hadronic calorimeter~\cite{Martyn:2006as}.
As shown in \reftable{tab:ILCruns}, only rough upper limits on
$\maxino$ are expected for $\maxino=10~\GeV$. This is due to the
difficulties to determine $E_{\tau}$ with sufficiently high precision.
While the situation will improve for $\maxino/\mstau>0.1$, it will be
even worse for smaller $\maxino/\mstau$ values because of the demand
for an even more precise $E_{\tau}$ determination.%
\footnote{In a more general setting than the CMSSM, 
it may also be the $\selectronR$ that is the NLSP. Then it will be
easier to achieve the precision required to probe smaller $\maxino$ 
values since this relies now on a measurement of the energy 
$E_\mathrm{e}$ of the electron that is emitted (instead of the unstable tau) 
in the decay $\selectronR\to\mathrm{e}\,\axino$.}

Measurements of the values of $\tau_{\stau}$, $\mstau$, $\mbino$ and
$\maxino$ can be used to approximately constrain the PQ scale
$f_a$~\cite{Brandenburg:2005he}. Here the $\maxino$ dependence becomes
only relevant in the $\maxino/\mstau>0.1$ region, where one should be
able to infer $\maxino$ with a better accuracy than reported in
\reftable{tab:ILCruns}.
Using the LL expression \eqref{eq:2bLL}, we find that only the
following combination can be probed by measuring $\tau_{\stau}$,
$\mstau$, $\mbino$ and $\maxino$ at colliders
\bee
  \frac{f_a^2}{e_Q^4} 
  \log^{-2}\left(\frac{y^2 f_a^2}{2\mstau^2}\right)
  \approx 
  \frac{81\alpha^4}{512\pi^5\,c^8_{\rm W}}
  \left(1-\frac{\maxino^2}{\mstau^2}\right)
  \tau_{\stau}\,\mstau\,\mbino^2,
\label{eqn:fa}
\eee
where one sees explicitly that the $\maxino$ dependence becomes
negligible for $\maxino/\mstau<0.1$.
Note that it will not be possible to constrain $f_a$ alone from
collider data. For the considered SUSY axion models, $f_a$ always
appears as the ratio $e_Q^4/f_a^2$ in the prefactor of the decay
widths and as the product $y f_a$ in the logarithm; see
\eqref{eq:2lamp_A} and \eqref{eq:3amp}.
However, ongoing and future axion searches might yield valuable
additional insights into the axion model possibly realised in nature
as detailed, e.g., in Sec.~6 of Ref.~\cite{Freitas:2009fb}.
In fact, an axion discovery via the axion--photon
interaction~\eqref{Eq:Laxionphoton} may point to $e_Q$ and $f_a$. 
To some extent, ambitious precise future measurements of 
$\tau_{\stau}$, $\mstau$, $\mbino$ (and $\maxino$) at
colliders could then probe $y$ via~\eqref{eqn:fa} and thereby -- as
advertised in the last paragraph of \refsec{Sec:Axino2Body} -- the
hierarchy between $f_a$ and the heavy (s)quark masses
$M_{Q/\tilde{Q}_{1,2}}$.

\subsection{Distinguishing between Axino LSP and Gravitino LSP}
\label{Sec:DistAxGrav}

Should a long-lived charged massive particle be detected at a
collider, it is not guaranteed that its decay products include an
axino. Despite being able to accurately determine the NLSP's lifetime
and other properties, there can remain ambiguities over the identities
of its non-SM decay products, i.e., the LSP. 
In particular, as mentioned at the end of \refsec{Sec:HadronicAxionModels}, 
the gravitino $\gravitino$ 
is also a well-motivated LSP candidate in SUSY
scenarios that include gravity.  Similarly to the axino, it is very
weakly interacting so that a long-lived stau NLSP can also occur
naturally in gravitino LSP scenarios. Here the stau NLSP lifetime is
governed by a 2-body decay which proceeds via a supergravity
coupling~\cite{Cremmer:1982en}
\bee
\tau_\stau^{\gravitino\,{\rm LSP}}
\simeq
1/\Gamma(\stauR\to\tau\gravitino) 
= 
\frac{48\pi\mgravitino^2 \MPl^2}{\mstau^5}
\left(1-\frac{\mgravitino^2}{\mstau^2}\right)^{-4} \ ,
\label{Eq:StauLifetimeGravitinoLSP}
\eee
where the gravitino mass $\mgravitino$ does not only enter via the
kinematics but also governs the interaction strengths via the spin-1/2
goldstino components. 

Comparing \eqref{Eq:StauLifetimeGravitinoLSP} with 
$\tau_\stau^{\axino\,{\rm LSP}}\simeq 1/\Gamma(\stauR\to\tau\axino)$ 
given by~\eqref{eq:2lamp_Gamma} in the axino LSP case, agreement
between the stau NLSP lifetimes in both scenarios is found on the
lines shown in \reffig{Fig:AxGrAmbiguity}.
%
%
\FIGURE[t]{ 
\centerline{\epsfig{figure=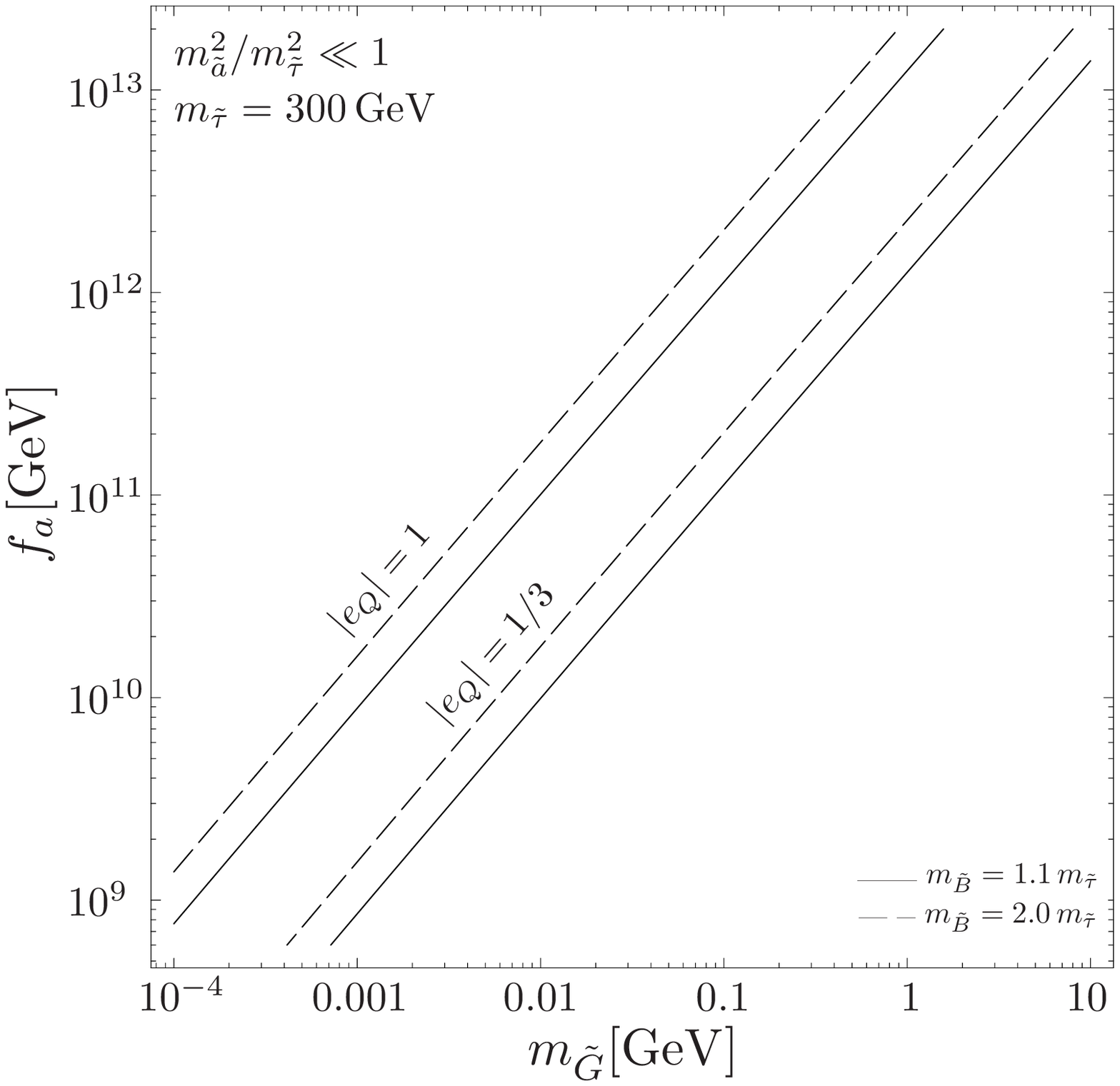, width=8cm}}
\caption{Agreement between the stau NLSP lifetime in the axino LSP
  case and the one in the gravitino LSP case,
  $\tau_\stau^{\axino\,{\rm LSP}}=\tau_\stau^{\gravitino\,{\rm LSP}}$,
  is indicated by the solid (dashed) lines for $\mstau=300~\GeV$ and
  $\mbino=1.1\,\mstau$ ($\mbino=2.0\,\mstau$). The considered SUSY
  axion model parameters are $|e_Q|=1/3$ (lower lines) and $|e_Q|=1$
  (upper lines) with $\maxino^2/\mstau^2\ll 1$, $y=1$, and $f_a$ above
  $6\times 10^8\,\GeV$ in line with the lower limit from astrophysical
  and cosmological axion studies~\eqref{Eq:f_a_axion}.}
\label{Fig:AxGrAmbiguity}}
%
For concreteness, we consider $\mstau=300~\GeV$ and
$\mbino=1.1\,\mstau$ (solid lines) or $2.0\,\mstau$ (dashed lines).
Those masses may be known quite precisely from collider studies when
$\tau_\stau$ will be measured. Indeed, the LSP mass $m_{\axino}$ or
$m_{\gravitino}$ will be much harder to probe (cf.\ 
\reftable{tab:ILCruns}) and also $f_a$ might still not be more
strongly constrained than summarised by~\eqref{Eq:f_a_axion}. To
account for the potential additional $e_Q$ dependence, the
$\tau_\stau$ agreement is indicated for the two representative cases
$|e_Q|=1/3$ (lower lines) and $|e_Q|=1$ (upper lines), whereas $y=1$
is kept fixed.
In the limit $m_{\axino,\gravitino}^2\ll \mstau^2$, we find that
\bee
\mgravitino 
= 
\sqrt{\frac{32}{3}}\,
\frac{\pi^2 c^4_{\rm W}}{9\,\alpha^2 e_Q^2}\,
\frac{f_a}{\MPl}\,
\frac{\mstau^2}{\mbino}
\log^{-1}\left(\frac{y^2 f_a^2}{2\mstau^2}\right).
\label{Eq:taustauAx_eq_taustauGr}
\eee
in \eqref{Eq:StauLifetimeGravitinoLSP} gives the same stau NLSP
lifetime as the LL result~\eqref{eq:2bLL} in the axino LSP case.

The smaller $\mgravitino$ or $f_a$, the shorter is the stau NLSP
lifetime. Accordingly, one finds a lower $\tau_\stau$ limit imposed
by~\eqref{Eq:f_a_axion} in the axino LSP case. In
\reffig{Fig:AxGrAmbiguity} ($\mstau=300~\GeV$, $\maxino^2/\mstau^2\ll
1$), this lower limit ranges between $1.2\times 10^{-4}~\seconds$ (for
$\mbino=330~\GeV$ and $e_Q=1/3$) and $5.1\times 10^{-7}~\seconds$ (for
$\mbino=600~\GeV$ and $e_Q=1$). Much smaller $\tau_\stau$ are possible
with a light gravitino LSP, $\mgravitino<40~\keV$, and could as such
provide hints against the axino LSP. In Ref.~\cite{Brandenburg:2005he}
it had been pointed out that $\tau_\stau\lesssim {\rm ms}$ and
$\tau_\stau\gtrsim {\rm days}$ will point to a gravitino LSP scenario.
While we agree on the short lifetime being a potential indicator for
the gravitino LSP, the mentioned long lifetime cannot exclude the
axino LSP since $f_a$ may well be larger than $5\times 10^{12}\,\GeV$
considered in~\cite{Brandenburg:2005he}.  Instead, we find that the
upper limit $\tau_\stau\lesssim (4\!-\!6)\times
10^3\,\seconds=67\!-\!100~{\rm min}$ imposed by CBBN
for typical yields~\eqref{Eq:Ystau} applies for both LSP candidates.
The additional constraints from hadronic energy injection might
tighten this upper $\tau_\stau$ limit but differently for the
gravitino LSP (see~\cite{Steffen:2006wx,Pospelov:2008ta}) and the
axino LSP (see~\cite{Freitas:2009jb} and
\refsec{Sec:NucleosynthesisConstraints} above).

Another way in which one can try to distinguish between an axino LSP
and a gravitino LSP is by studying the 3-body decay of the stau
$\stauR\to\tau+\gamma+{\not}E$~\cite{Brandenburg:2005he,Hamaguchi:2006vu}.
In Ref.~\cite{Brandenburg:2005he} the authors focused on the branching
ratio of the integrated 3-body decay rate and on the double
differential distribution of the 3-body decays.
For the axino LSP case, our result of the latter quantity is given
in~\eqref{eq:3amp} above and also used to calculate that branching
ratio~\eqref{eq:3bodyBR} with results illustrated in \reffig{fig:3br}.
Also Ref.~\cite{Hamaguchi:2006vu} considered the branching ratios of
the 3-body decays in the following form and with the given choice of
cuts,
\bea
\frac{\Gamma(\stauR \to \tau\, \gamma\, X;
x_\gamma^{{\rm cut}}=\frac{20~\GeV}{\mstau}, 
x_\theta^{{\rm cut}}=0.134)}{\Gamma(\stauR \to\tau\, X)}, 
\qquad 
X=\axino , \gravitino \ .
\label{Eq:3bodyBRcomparison}
\eea
Here we present an updated comparison of both the double differential
distributions
\bea
\frac{1}
{\Gamma(\stauR \to \tau\, \gamma\, X ; 
x_\gamma^{{\rm cut}}=0.2,
x_\theta^{{\rm cut}}=0.134 )} 
\frac{{\rm d}^2\Gamma(\stauR\to\tau\, \gamma \, X)}
{{\rm d}x_\gamma\,{\rm d}\!\cos\theta} , 
\qquad 
X=\axino,\gravitino \ ,
\label{Eq:doublediff3body}
\eea
and of \eqref{Eq:3bodyBRcomparison} using the results of our
calculations of the 2- and 3-body decays of the stau NLSP into the
axino LSP presented in \refsec{Sec:NLSPDecays}. For the 3-body decay
of the stau NLSP into the gravitino LSP, we use the result obtained
in~\cite{Brandenburg:2005he}, which is
applicable for finite $\mbino$.%
\footnote{In Ref.~\cite{Buchmuller:2004rq} the decay
  $\stauR\to\tau\,\gamma\,\gravitino$ was also considered but in the
  limit $\mbino\to\infty$.}

\reffigure{fig:ddiff3body} shows \eqref{Eq:doublediff3body} as a
function of $x_\gamma$ and $\cos\theta$ for $\mstau=100~\GeV$ and the
given choice of cuts: $x_\gamma^{{\rm cut}}=0.2$ and $x_\theta^{{\rm
    cut}}=0.134$.
%
\FIGURE[t]{
\centerline{\epsfig{figure=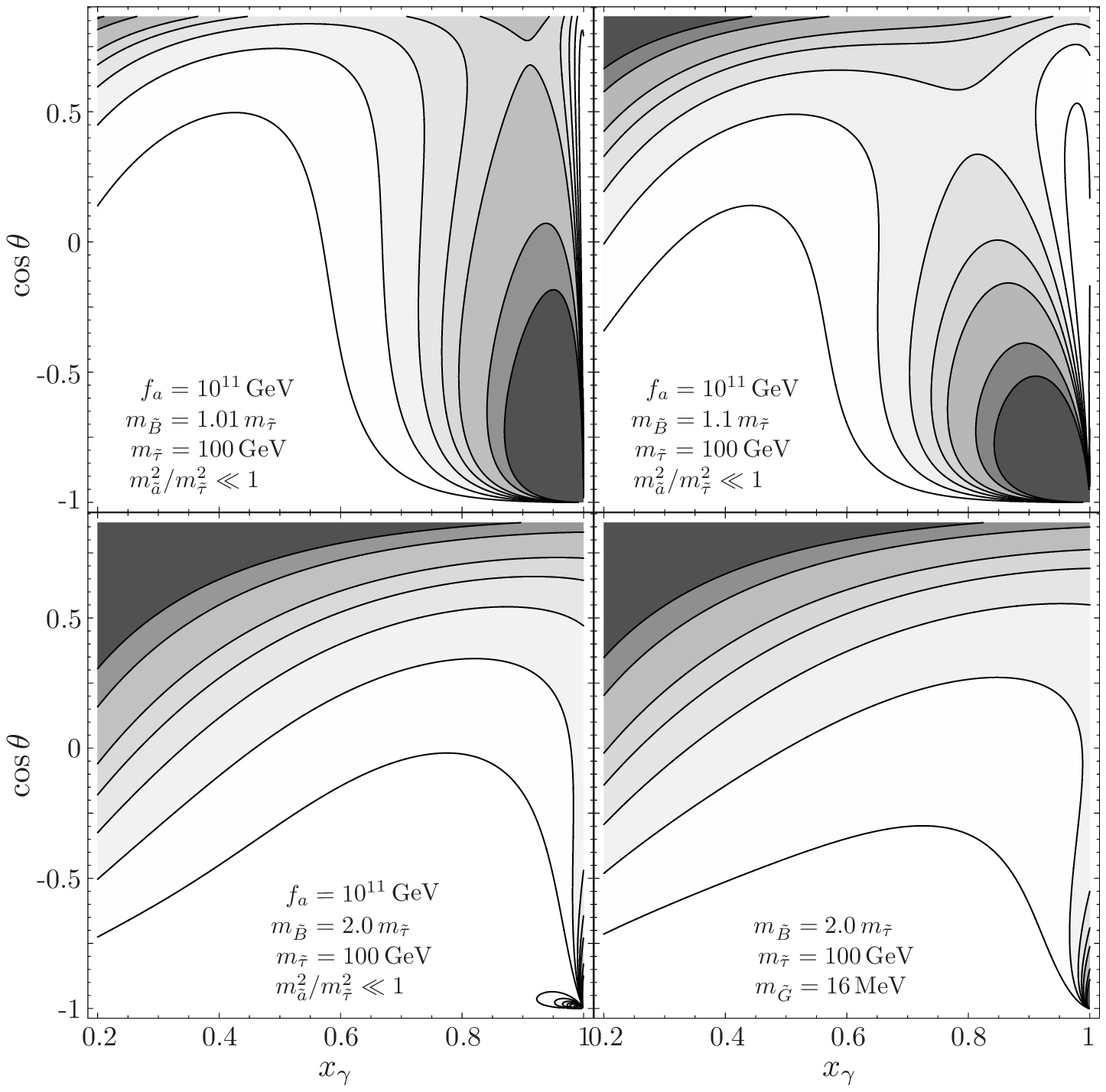, width=11cm}}
\caption{The normalised double differential
  distribution~\eqref{Eq:doublediff3body} of the 3-body decays
  $\stauR\to\tau\,\gamma\,X$ with $X=\axino$ (first three panels) and
  $X=\gravitino$ (bottom right panel) for $\mstau=100~\GeV$.  The
  contours shown in the plots represent the values 0.2, 0.4, 0.6, 0.8,
  1.0, 1.5 and 2.0, with darker shadings corresponding to higher
  number of events. The cut parameters are set to $x_\gamma^{{\rm
      cut}}=0.2$ and $x_\theta^{{\rm cut}}=0.134$. For the axino LSP,
  $f_a=10^{11}\,\GeV$, $\maxino^2/\mstau^2\ll 1$, $|e_Q|=1/3$ and
  $y=1$ in all three panels while $\mbino/\mstau$ varies:
  $1.01\,\mstau$ (top left), $1.1$ (top right) and $2.0$ (bottom
  left). For the gravitino LSP, $\mbino/\mstau=2.0$ and
  $\mgravitino=16~\MeV$, for which a $\tau_\stau$ agreement is
  encountered in the two scenarios considered in the bottom panels;
  cf.~\eqref{Eq:taustauAx_eq_taustauGr}.}
\label{fig:ddiff3body}}%
%
The darker shadings correspond to higher number of events, with the
contours representing the values 0.2, 0.4, 0.6, 0.8, 1.0, 1.5 and 2.0.
The first three panels show the axino LSP case for
$f_a=10^{11}\,\GeV$, $\maxino^2/\mstau^2\ll 1$, $|e_Q|=1/3$, $y=1$ and
for different values of $\mbino/\mstau=1.01$ (top left), $1.1$ (top
right) and $2.0$ (bottom left). The lower right panel shows the
gravitino LSP case for $\mbino/\mstau=2.0$ and $\mgravitino=16~\MeV$.
This value of $\mgravitino$ is chosen such that one finds an agreement
of the stau lifetime in the two scenarios considered in the bottom
panels; cf.~\eqref{Eq:taustauAx_eq_taustauGr}.
We however note that changes in $\mgravitino$, while significantly
altering the stau lifetime, do not greatly affect the shown
differential distribution as long as
$\mgravitino^2/\mstau^2\ll 1$.
Likewise, changes in the values of $f_a$ and $\mstau$ cause only
slight changes of the distributions.

To varying degrees, all the panels show a preference for producing
soft and collinear photons.
The top right panel is the scenario most closely resembling the plot
shown for the axino LSP case in Fig.~5 of
Ref.~\cite{Brandenburg:2005he}. 
Here we find an excess of events for highly energetic photons produced
back-to-back to the emitted taus in comparison to what is obtained in
the corresponding gravitino LSP case (cf.\ Fig.~5
of~\cite{Brandenburg:2005he}). This is in agreement with the result
found in \cite{Brandenburg:2005he}, and we now note how this excess
region changes for different values of the bino--stau--mass ratio
$\mbino/\mstau$.
For smaller $\mbino/\mstau$, the excess is more pronounced with events
covering a phase space that extends towards larger values of
$\cos\theta$ as can be seen in the top left panel of
\reffig{fig:ddiff3body}.
This is due to the one-loop decay mode, shown in
\reffig{fig:3body}~(a), being more dominant over the two-loop decay
mode in that region as the bino becomes more and more on-shell.
In turn, towards larger values of $\mbino/\mstau$, the excess region
shrinks such that for $\mbino/\mstau=2.0$ (bottom left panel) the
excess region is basically missing altogether.
In the gravitino LSP case, \eqref{Eq:doublediff3body} is less
sensitive to $\mbino/\mstau$ with more events expected in the very
hard photon region for larger $\mbino/\mstau$. It must be noted
however that this increase is very marginal and would require high
statistics to be noticeable.
Thus, the larger the $\mbino/\mstau$ ratio, the more difficult it
becomes to differentiate between the axino LSP and the gravitino LSP.
This is shown explicitly in the bottom panels of
\reffig{fig:ddiff3body}. Here we have chosen parameters for the
gravitino LSP case to most closely resemble phenomenologically the
axino LSP case to its left to test a particularly difficult situation:
$\mbino/\mstau=2.0$ and, as already mentioned, 
$\mgravitino$ such that $\tau_{\stau}$ agrees in the lower panels.
If we look closely at those two distributions, we find small
differences, with the axino LSP scenario having a larger phase space
area with small values of \eqref{Eq:doublediff3body}. However, these
differences are small and we would need large statistics and precise
measurements to rely on them for a distinction between the two
scenarios.

To give an idea of the numbers that can be expected, we assume $5000$
trapped staus whose decays can be observed in a detector component
(such as a hadronic calorimeter) that allows for the analysis of the
3-body decays.
This may seem somewhat optimistic in light of our $N_{\stau}^{{\rm
    hcal}}$ estimates (obtained without exploring ILC-parameter
optimisations) given in \reftable{tab:ILCruns}. Nevertheless, even
$N_{\stau}^{{\rm hcal}}=10^4$ could be feasible at the ILC already
without additional stopping-detectors~\cite{Martyn:2006as}.  Also at
the LHC one might be able to stop the assumed number staus but only
when invoking additional
stopping-detectors~\cite{Hamaguchi:2004df,Feng:2004yi,Hamaguchi:2006vu}.
For example, with massive stoppers set up 
next to the Compact Muon Solenoid (CMS) detector at the LHC, 
it seems possible to stop staus
with $\beta\gamma$ up to $0.6$ and 
to achieve $N_{\stau}^{{\rm stopped}}$ values 
that are more than an order of magnitude above
the ones reported in \reftable{tab:LHCruns}
above~\cite{Hamaguchi:2006vu}.
Now, for 5000 trapped staus, we expect to see 149$\pm12$(stat),
58$\pm8$, 26$\pm5$ and 28$\pm5$ 3-body events with
$0.2 \leq x_\gamma \leq 1$ 
and 
$-1 \leq \cos\theta \leq 0.866$
when considering the scenarios in \reffig{fig:ddiff3body} in the
sequence of left to right and top to bottom, respectively.
Looking at the region with $0.8 \leq x_\gamma \leq 1$ and 
$-1 \leq \cos\theta \leq -0.5$, in which an excess of events may be
found,
we would expect 53$\pm7$(stat) (36\% of the above number of 3-body
events), 19$\pm 4$ (33\%), 0.3$\pm 0.6$ (1.4\%), 0.4$\pm 0.6$ (1.4\%)
events, respectively.
Clearly, this looks very pessimistic for distinguishing between the
axino LSP and gravitino LSP scenario with $\mbino/\mstau=2.0$. If we
enlarge the phase space region to $-1\leq\cos\theta\leq 0$ while
keeping $0.8\leq x_\gamma \leq 1$, we would obtain 0.7$\pm0.8$ (3.5\%)
and 1.5$\pm1.2$ (5.2\%), respectively, for these two cases.
Admittedly the improvement is not large, but in lieu of other signals,
this discrepancy could be indicative.

In the axino LSP case, the markedness of the potential excess of
highly energetic photons produced back-to-back to the emitted taus is
not only sensitive to $\mbino/\mstau$ but also to the size of the
logarithm
$\log[y^2 f_a^2/(2 \mstau^2)]$.
As can be seen in~\eqref{eq:3amp}, the smaller this logarithm, the
larger is the relative importance of the one-loop 3-body decay mode
shown in \reffig{fig:3body}~(a) and thereby the number of excess
events.
This can affect the success of the distinction between the axino LSP
and the gravitino LSP in a way that we now illustrate by establishing
a connection to the results of Ref.~\cite{Hamaguchi:2006vu}.

In Ref.~\cite{Hamaguchi:2006vu} the branching
ratios~\eqref{Eq:3bodyBRcomparison} and $E_\gamma$ distributions in
$\theta$ intervals were explored based on the 2- and 3-body decay
widths calculated in~\cite{Brandenburg:2005he}.
Accordingly, the results for the axino case did rely on a particular
choice of $\xi\log(f_a/m)$, where the factor $\xi$ and the mass scale
$m$ were introduced to parametrise the uncertainty associated with the
cutoff procedure applied in~\cite{Brandenburg:2005he} instead of the
two-loop calculations presented by us in this work; see also
\refsec{Sec:Axino3Body}. In particular, in the comparisons
of~\eqref{Eq:3bodyBRcomparison}
in~\cite{Brandenburg:2005he,Hamaguchi:2006vu}, $\log(f_a/m)=20.7$ was
fixed and different choices of $\xi$ including $C_{\rm
  aYY}^{-1}$~\cite{Brandenburg:2005he}, 0.5, 0.75, 1 and
2~\cite{Hamaguchi:2006vu} were considered. Relying on our 3-body
result~\eqref{eq:3amp} and on the 2-body rate in the LL
approximation~\eqref{eq:2bLL}, we can now relate those choices to the
parameters of the underlying model:%
\bee
\xi 
\equiv 
\frac{1}{20.7}\,
\log\left(\frac{y f_a}{\sqrt{2}\mstau}\right)
\ .
\label{Eq:XiMapping}
\eee
The different $\xi$ scenarios considered in
\cite{Brandenburg:2005he,Hamaguchi:2006vu} can thus be related to
combinations of $y$ and $f_a$ for a given $\mstau$.
For example, with $\mstau\simeq 100~\GeV$, $\xi\simeq 0.5$ ($1$) is
found for $y=0.01$ and $f_a=6\times 10^8\,\GeV$ ($y=1$ and
$f_a=10^{11}\,\GeV$), while $\xi\simeq 2$ would require $y\gg 1$
and/or $f_a\gg 10^{16}\,\GeV$.
With such relations in mind, the results for $\xi=0.5$--$1$ shown in
Figs.~16--20 of Ref.~\cite{Hamaguchi:2006vu} apply also to our
considerations. 
There it has been estimated, e.g., that $N_{\stau}^{{\rm
    stopped}}\simeq 300$ ($3000$) would be sufficient for $\xi=0.5$
($1$) to establish a $3\sigma$ difference between the branching
ratios~\eqref{Eq:3bodyBRcomparison} when $\mstau\simeq 100~\GeV$ and
$\mbino=1.1\,\mstau$, while the required $N_{\stau}^{{\rm stopped}}$
increases to about $700$ ($10^4$) for $\mbino=1.2\,\mstau$.
For further results and details, we refer to
Ref.~\cite{Hamaguchi:2006vu} and to relation~\eqref{Eq:XiMapping} for
the mapping to parameters of the SUSY axion models considered in this
work.
Note however that relation~\eqref{Eq:XiMapping} is only relevant for
making contact with the phenomenological investigations
of~\cite{Brandenburg:2005he,Hamaguchi:2006vu}.  When the logarithm is
not fixed to a number such as $\log(f_a/m)=20.7$, agreement of our
3-body result~\eqref{eq:3amp} and of the 2-body rate in the LL
approximation~\eqref{eq:2bLL} with~\cite{Brandenburg:2005he} is found
for $\xi=1$, $m=\sqrt{2}\mstau$ and $C_{\rm aYY}=6 e_Q^2$, as already
discussed in \refsec{Sec:NLSPDecays}.
Still the size of $\log[y^2 f_a^2/(2 \mstau^2)]$ will be relevant for
the viability of the axino--gravitino LSP distinction as illustrated
by the results of Ref.~\cite{Hamaguchi:2006vu} as mentioned above.

\section{Scenarios with \boldmath$\maxino, \mgravitino<\mstau$}
\label{Sec:AxGrSt}

In the above sections, we mainly considered the axino LSP scenario
with a stau NLSP, where $\maxino<\mstau<\mgravitino$.  In
\refsec{Sec:DistAxGrav} we also considered the hierarchy
$\mgravitino<\mstau<\maxino$ (or equivalently, a SUSY scenario with no
PQ extension) in the context of differentiating the experimental
signatures of an axino LSP and a gravitino LSP. Now limiting ourselves
to the hierarchy $\maxino<\mstau$, yet another two orderings of the
masses are possible, $\maxino<\mgravitino<\mstau$ and
$\mgravitino<\maxino<\mstau$, as already mentioned in
\refsec{Sec:HadronicAxionModels}.  We will explore the consequences of
both of these in this section. In particular, we consider effects on
decays and the lifetime of the lighter stau, on cosmological
constraints and on collider phenomenology.

Scenarios in which both the axino and the gravitino are lighter 
than the other sparticles (cf.~\cite{Asaka:1998ns,Asaka:1998xa}) 
have been considered first in studies 
aiming at the understanding of structure formation 
and of the matter power spectrum~\cite{Olive:1984bi,Chun:1993vz,Kim:1994ub}.
Moreover, it had been realised that such a spectrum may provide a
solution to the gravitino problem of thermal
leptogenesis~\cite{Asaka:2000ew}; see
also~\cite{Tajuddin:2010dr,Kawasaki:2007mk,Hasenkamp:2010if,Baer:2010gr}.
Such a setting may also allow for a late non-standard increase of
$\DNueff$ after BBN but before the beginning of structure
formation~\cite{Ichikawa:2007jv}.
Brief comments on $\maxino, \mgravitino<\mstau$ scenarios were already
given in Refs.~\cite{Brandenburg:2005he,Freitas:2009fb}.
We now elaborate on those and provide several new insights, where our
focus is on the behaviour of the lighter stau and on
associated constraints and phenomenology, see also
Ref.~\cite{Tajuddin:2010dr}.%
\footnote{As this work was being finalised, a study of 
$\TR$ constraints in scenarios with the gravitino LSP and 
the axino NLSP appeared~\cite{Cheung:2011mg}, which is 
complementary to our considerations.}

\subsection{Lifetime and Branching Ratios}
\label{Sec:LifetimeBRAxGrSt}

Let us first take a look at how the lifetime of the stau is affected.
In addition to the 2-body decay $\stauR\to\tau\axino$, the 2-body
decay $\stauR\to\tau\gravitino$ is also kinematically viable.  The
lifetime of the stau is then governed by these 2-body decays:
\bea
\tau_{\stau} 
= 
\frac{1}
{\Gamma_{\tot}^{\stau}}
\approx 
\frac{1}
{\Gamma_{\tot}^{\stau,{\rm 2b}}} 
\equiv
\frac{1}
{\Gamma(\stauR\to\tau\axino)+\Gamma(\stauR\to\tau\gravitino)} 
\label{eq:comblifetime}
\eea
with the partial decay widths \eqref{eq:2lamp_Gamma} and
\eqref{Eq:StauLifetimeGravitinoLSP}.
For given $\mstau$ and $\mbino$, the relative importance of the
different 2-body decays depends on $f_a$ and $\mgravitino$, which
govern $\Gamma(\stauR\to\tau\axino)$ and
$\Gamma(\stauR\to\tau\gravitino)$, respectively.
This is illustrated in \reffig{fig:combltwidthdiff} for
$\mstau=300~\GeV$, $\mbino=1.1\,\mstau$, $|e_Q|=1/3$, $y=1$ and
$\maxino^2/\mstau^2\ll 1$, where the latter limit minimises the
dependency of $\Gamma(\stauR\to\tau\axino)$ on $\maxino$.
%
\FIGURE[t]{
\centerline{\epsfig{figure=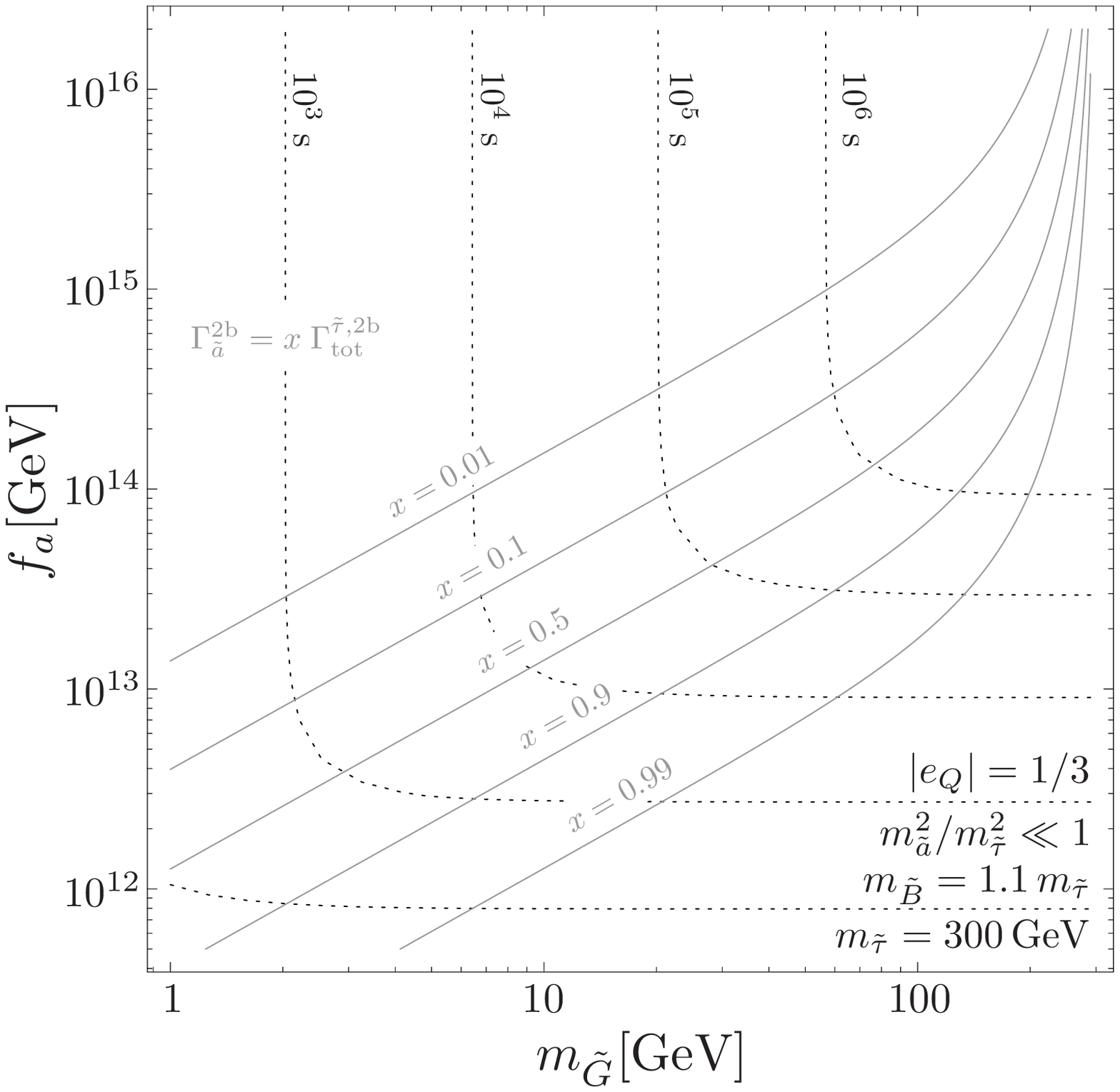, width=8cm}}
\caption{The lifetime of the lighter stau (dotted lines) when it can decay into
both the axino and the gravitino ($\maxino,\mgravitino<\mstau$) as a
function of $\mgravitino$ and $f_a$, for $\mstau=300\,\GeV$,
$\mbino=1.1\,\mstau$, $|e_Q|=1/3$, $y=1$ and $\maxino^2/\mstau^2\ll
1$.  The solid grey lines indicate the relative importance of
$\Gamma(\stauR\to\tau\axino)$ as a fraction $x$ of the total 2-body
decay width $\Gamma_{\tot}^{\stau,{\rm 2b}}$. Above (below) the
$x=0.5$ contour, the stau decays dominantly into the gravitino
(axino).}
\label{fig:combltwidthdiff}}
%
The dotted lines indicate contours of the stau lifetime $\tau_{\stau}$
and the solid grey lines the fractional contribution of
$\Gamma_\axino^{\rm 2b}\equiv\Gamma(\stauR\to\tau\axino)$ 
to the total 2-body decay width of the stau: 
$x=\Gamma_\axino^{\rm 2b}/\Gamma_{\tot}^{\stau,{\rm 2b}}
\simeq{\mathrm{BR}}(\stauR\to\tau\axino)$.
Note that the upward bending of the $x$ contours
towards larger $\mgravitino$ results from the factor
$(1-\mgravitino^2/\mstau^2)^{4}$
in $\Gamma(\stauR\to\tau\gravitino)$,
cf.~\eqref{Eq:StauLifetimeGravitinoLSP}, which also accounts for the
phase space suppression of the $\stauR\to\tau\gravitino$ decay for
$\mgravitino\to\mstau$.
Above (below) this contour, the stau decays dominantly into the
gravitino (axino), which explains the clear turning points in the
shown lifetime contours.

For $\maxino,\mgravitino<\mstau$, both 3-body decays
$\stauR\to\tau\gamma\axino/\gravitino$ contribute simultaneously to
$\stauR\to\tau+\gamma+{\not}E$. Accordingly, the corresponding
branching ratio with cuts is given by
\bea
{\rm BR}({\mbox{3-body}};x_\gamma^{\rm cut},x_\theta^{\rm cut})
\approx
\frac{\Gamma_{\tot}^{\stau,{\rm 3b}}}
{\Gamma_{\rm tot}^{\stau,{\rm 2b}}}
\equiv
\frac{\Gamma(\stauR\!\to\!\tau\axino\gamma;
x_\gamma^{\rm cut},x_\theta^{\rm cut}) 
+ 
\Gamma(\stauR\!\to\!\tau\gravitino\gamma;
x_\gamma^{\rm cut},x_\theta^{\rm cut})}
{\Gamma_{\rm tot}^{\stau,{\rm 2b}}} 
\eea
with~\eqref{Eq:3BodyWidthwCuts} and the corresponding $\gravitino$
result of Ref.~\cite{Brandenburg:2005he}.
In \reffig{fig:comb3and4bodyBR}~(a) this branching ratio is
illustrated by the dotted lines for the same parameters as in
\reffig{fig:combltwidthdiff} and for $x_\gamma^{\rm
  cut}=20~\GeV/\mstau$ and $x_\theta^{\rm cut}=0.134$, as in
\refsec{Sec:Collider}. The shown contours vary only between 0.8\% and
1.5\% and confirm that the total stau decay width
$\Gamma_{\tot}^{\stau}$ is governed by the 2-body decays.
The solid grey lines show the fractional contribution of
$\Gamma_\axino^{\rm 3b} 
\equiv
\Gamma(\stauR\to\tau\gamma\axino;
x_\gamma^{\rm cut},x_\theta^{\rm
  cut})$
to the total 3-body decay width of the stau: 
$x=\Gamma_\axino^{\rm 3b}/\Gamma_{\tot}^{\stau,{\rm 3b}}$.
One finds that $\Gamma_\axino^{\rm 3b}$ dominates in a
$(\mgravitino,f_a)$ range that is slightly larger than the one in
which $\Gamma_\axino^{\rm 2b}$ dominates in the 2-body case (see
\reffig{fig:combltwidthdiff}).
%
\FIGURE[t]{
\epsfig{figure=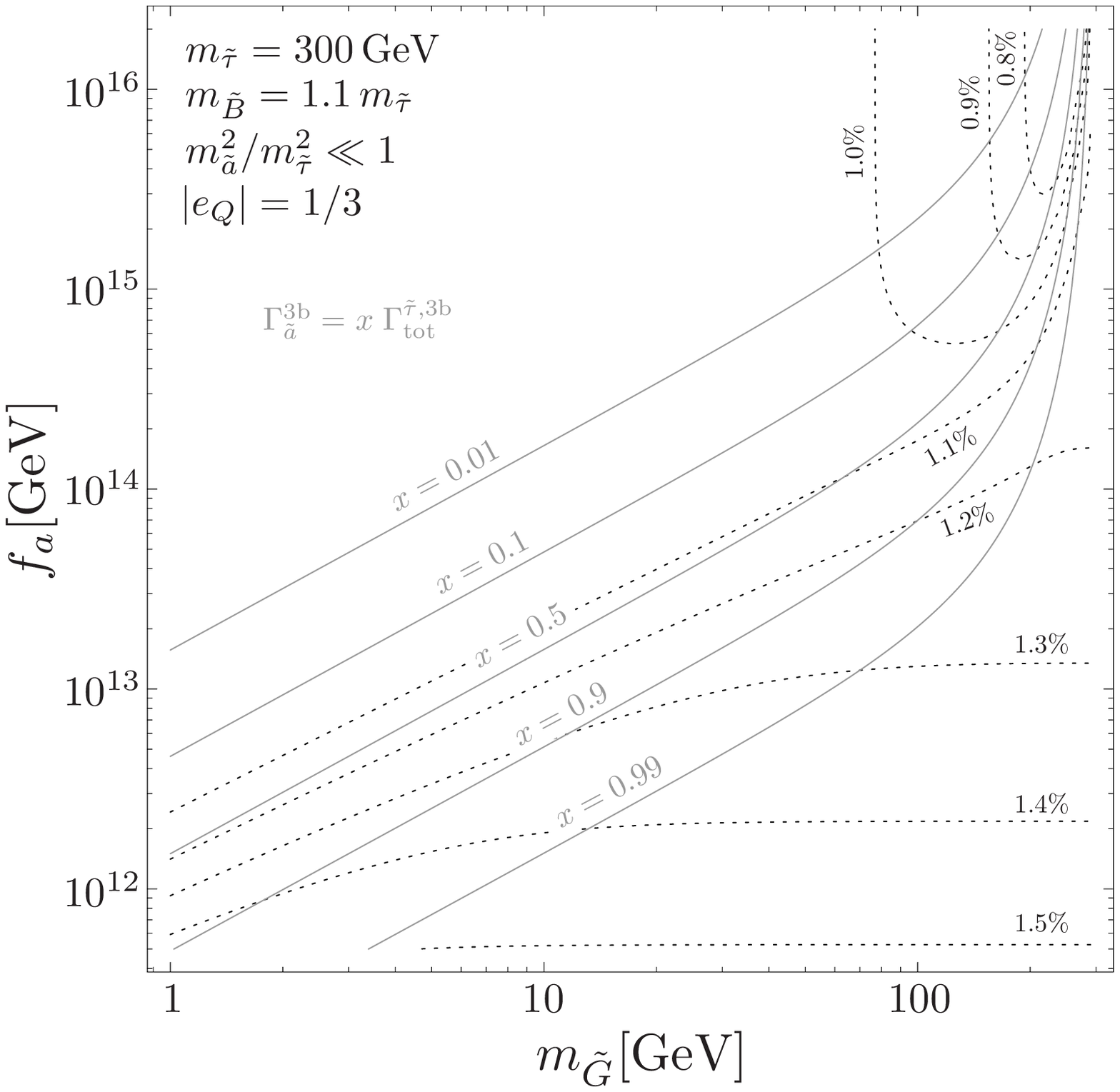, height=6.9cm}
\hfill
\epsfig{figure=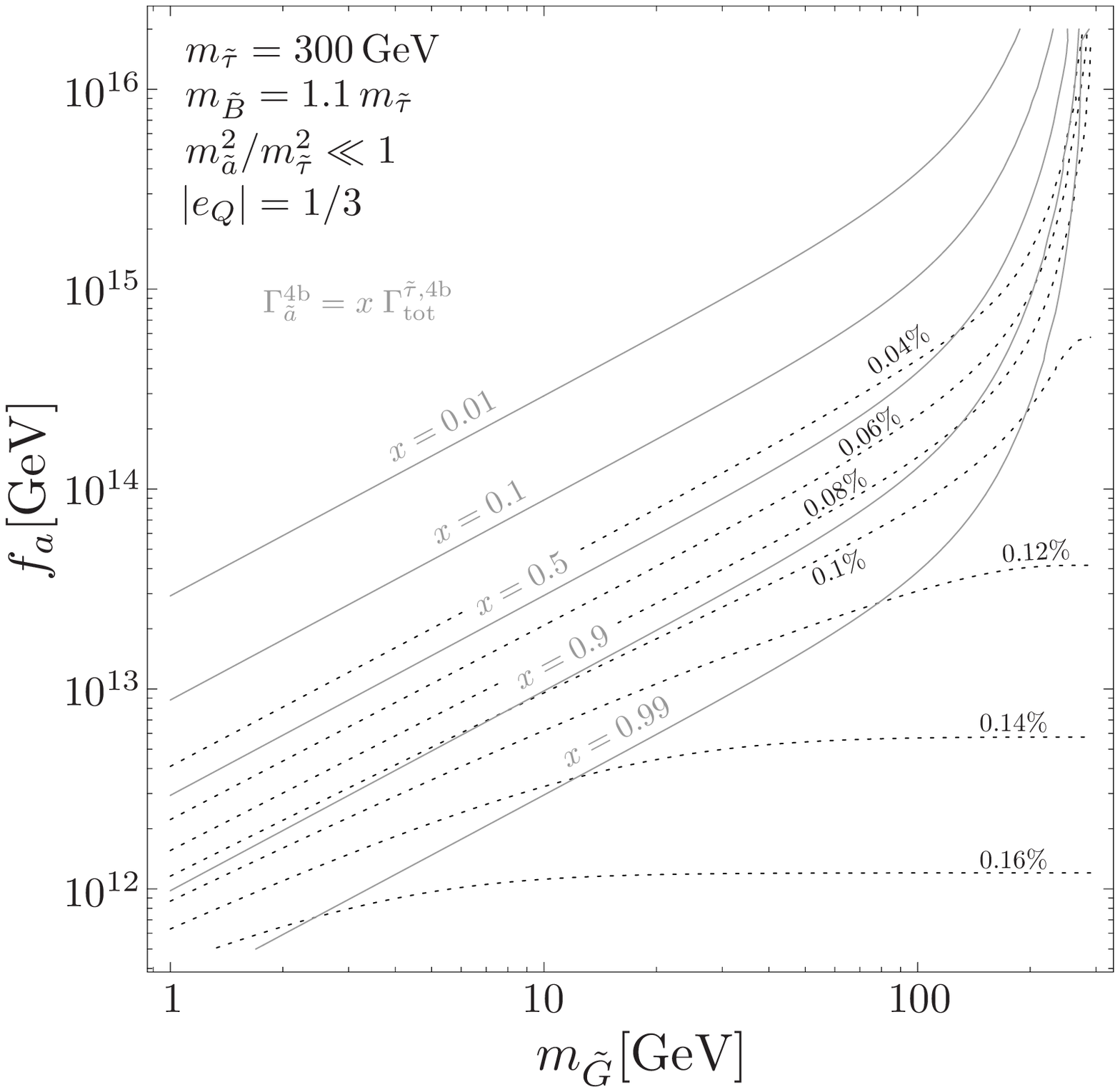, height=6.9cm}
\\
\hspace*{3cm}
\makebox[2cm][c]{(a)}\hfill
\hspace*{4cm}
\makebox[2cm][c]{(b)}\hfill
\caption{Branching ratios (dotted lines) of (a)~the 3-body decays
  $\stauR\to\tau\gamma\axino/\gravitino$ and (b)~the 4-body decays
  $\stauR\to\tau q\bar{q}\,\axino/\gravitino$ as a function of
  $\mgravitino$ and $f_a$, for the same parameters as in
  \reffig{fig:combltwidthdiff}. The cut parameters in~(a) are set to
  $x_\gamma^{\rm cut}=20~\GeV/\mstau \simeq 0.07$ and $x_\theta^{\rm
    cut}=0.134$ and in~(b) to $m_{q\bar{q}}^{\rm cut}=2~\GeV$. 
   The solid grey lines indicate the fractional
  contributions of the respective stau decays into the axino
  (a)~$x\equiv\Gamma_\axino^{\rm 3b}/\Gamma_{\tot}^{\stau,{\rm 3b}}$
  and (b)~$x\equiv\Gamma_\axino^{\rm 4b}/\Gamma_{\tot}^{\stau,{\rm
      4b}}$.}
\label{fig:comb3and4bodyBR}}

\reffigure{fig:comb3and4bodyBR}~(b) shows the corresponding 4-body
branching ratio (dotted lines)
\bea 
{\rm BR}({\mbox{4-body}}; m_{q\bar{q}}^{\rm cut})
=
\frac{\Gamma_{\tot}^{\stau,{\rm 4b}}}
{\Gamma_{\rm tot}^{\stau,{\rm 2b}}}
\equiv
\frac{
\Gamma(\stauR\to\tau\axino q\qbar; m_{q\bar{q}}^{\rm cut}) 
+
\Gamma(\stauR\to\tau\gravitino q\qbar; m_{q\bar{q}}^{\rm cut})
}{\Gamma_{\rm tot}^{\stauR,{\rm 2b}}} , 
\eea 
where
\bee
\Gamma(\stauR\to\tau\axino/\gravitino q\qbar; 
m_{q\bar{q}}^{\rm cut})
\equiv
\int_{m_{q\bar{q}}^{\rm cut}}^{\mstau-\maxino-m_\tau}
{\rm d}m_{q\qbar}\,
\frac{{\rm d}\Gamma(\stau_{\mathrm R}\to\tau\axino/\gravitino q\bar{q})}
{{\rm d}m_{q\qbar}}
\label{Eq:BR4body}
\eee
with \eqref{Eq:dGammadmqqbar} and the corresponding $\gravitino$
result of Ref.~\cite{Steffen:2006hw}.
Here we set $m_{q\bar{q}}^{\rm cut}=2~\GeV$, as in
\refsec{Sec:NucleosynthesisConstraints}, and the other parameters
again as in \reffigs{fig:combltwidthdiff} and
\ref{fig:comb3and4bodyBR}~(a). Now the solid grey lines show the
fractional contribution of
$\Gamma_\axino^{\rm 4b} 
\equiv
\Gamma(\stauR\to\tau\axino q\qbar; m_{q\bar{q}}^{\rm cut})$
to the total 4-body decay width of the stau: 
$x=\Gamma_\axino^{\rm 4b}/\Gamma_{\tot}^{\stau,{\rm 4b}}$.
One can see that $\Gamma_\axino^{\rm 4b}$ dominates in a
($\mgravitino, f_a$) region that is even a bit larger than the one in
which $\Gamma_\axino^{\rm 3b}$ dominates in the 3-body case.
Moreover, the shown
${\rm BR}({\mbox{4-body}}; m_{q\bar{q}}^{\rm cut}=2~\GeV)$
contours range between 0.04\% and 0.16\%. This demonstrates that the
3-body decays considered in panel~(a) are about an order of magnitude
more abundant than the 4-body decays considered in panel~(b).

\subsection{Cosmological Constraints}
\label{Sec:CosmologicalConstraintsAxGrSt}

Let us now turn to the cosmological constraints from primordial
nucleosynthesis for characteristic freeze-out yields of the lighter
stau $Y_{\stau}$ described by~\eqref{Eq:Ystau}. Indeed, the freeze-out
yield of the lighter stau prior to decay is not affected by the
presence of a light gravitino. Also decays such as
$\gravitino\to\axino a$ or, alternatively, $\axino\to\gravitino a$,
even if taking place after freeze out and prior to decay of the stau,
will not dilute $Y_{\stau}$ as the emitted extremely weakly
interacting particles will not thermalise and not affect the entropy
density.
However, we have just seen that $\tau_{\stau}$ can be very different
for $\maxino,\mgravitino<\mstau$, and this will affect the CBBN
constraints and also the BBN constraints associated with late hadronic
and electromagnetic energy injection.
Moreover, there are now the additional decays
$\stauR\to\tau\gravitino$ and $\stau_{\mathrm{R}}\to\tau\gravitino
q\qbar$ contributing to $\epsilon_{\EM/\HAD}$.
Accordingly, we find 
\begin{align}
  \epsilon_{\had}^{\mathrm{tot}} 
  = \frac{1}{\Gamma_{\tot}^{\stau}} 
  & 
  \left[ 
    \int_{m_{q\qbar}}^{\mstau-\maxino-\mtau} \mathrm{d} m_{q\qbar} \, 
    m_{q\qbar}\,
    \frac{\mathrm{d}\Gamma(\stau_{\mathrm{R}}\to\tau\axino q\qbar)}
    {\mathrm{d} m_{q\qbar}} \right.
  \nn \\
  & \qquad 
  + \left. 
    \int_{m_{q\qbar}}^{\mstau-\mgravitino-\mtau} \mathrm{d} m_{q\qbar} \, 
    m_{q\qbar}\,
    \frac{\mathrm{d}\Gamma(\stau_{\mathrm{R}}\to\tau\gravitino q\qbar)}
    {\mathrm{d} m_{q\qbar}} \right] 
  \ ,
\label{eq:combepshad}
\end{align}
with $\Gamma_{\tot}^{\stau}$ given in~\eqref{eq:comblifetime}, and
that the conservative estimate \eqref{Eq:epsEM} changes to
\bea
\epsilon_{\EM}
&=& 
0.3\, E_\tau^{\axino} \,\, {\mathrm{BR}}(\stauR\to\tau\axino)
+
0.3\, E_\tau^{\gravitino} \,\, {\mathrm{BR}}(\stauR\to\tau\gravitino)
\ ,
\label{Eq:epsEMAxGrSt}
\eea
where the energy of the tau emitted in $\stauR\to\tau\axino$ and
$\stauR\to\tau\gravitino$ is given respectively by
\bee
E_\tau^{\axino}=\frac{\mstau^2-\maxino^2+m_{\tau}^2}{2\mstau}
\quad
\mbox{and}
\quad
E_\tau^{\gravitino}=\frac{\mstau^2-\mgravitino^2+m_{\tau}^2}{2\mstau}
\ .
\label{Eq:EtauAxGRSt}
\eee

The resulting constraints are illustrated in the $\mgravitino$--$f_a$
plane in \reffig{fig:famaxgravaxNoCoeQ13} for (a)~$\mstau=300~\GeV$
and (b)~$\mstau=1~\TeV$, with the other parameters set to
$\maxino^2/\mstau^2\ll 1$, $|e_Q|=1/3$, $y=1$, $\mbino=1.1\,\mstau$
and the yield $Y_{\stau}=0.7\times 10^{-12}\,(\mstau/1~\TeV)$.
%
\FIGURE[t]{
\epsfig{figure=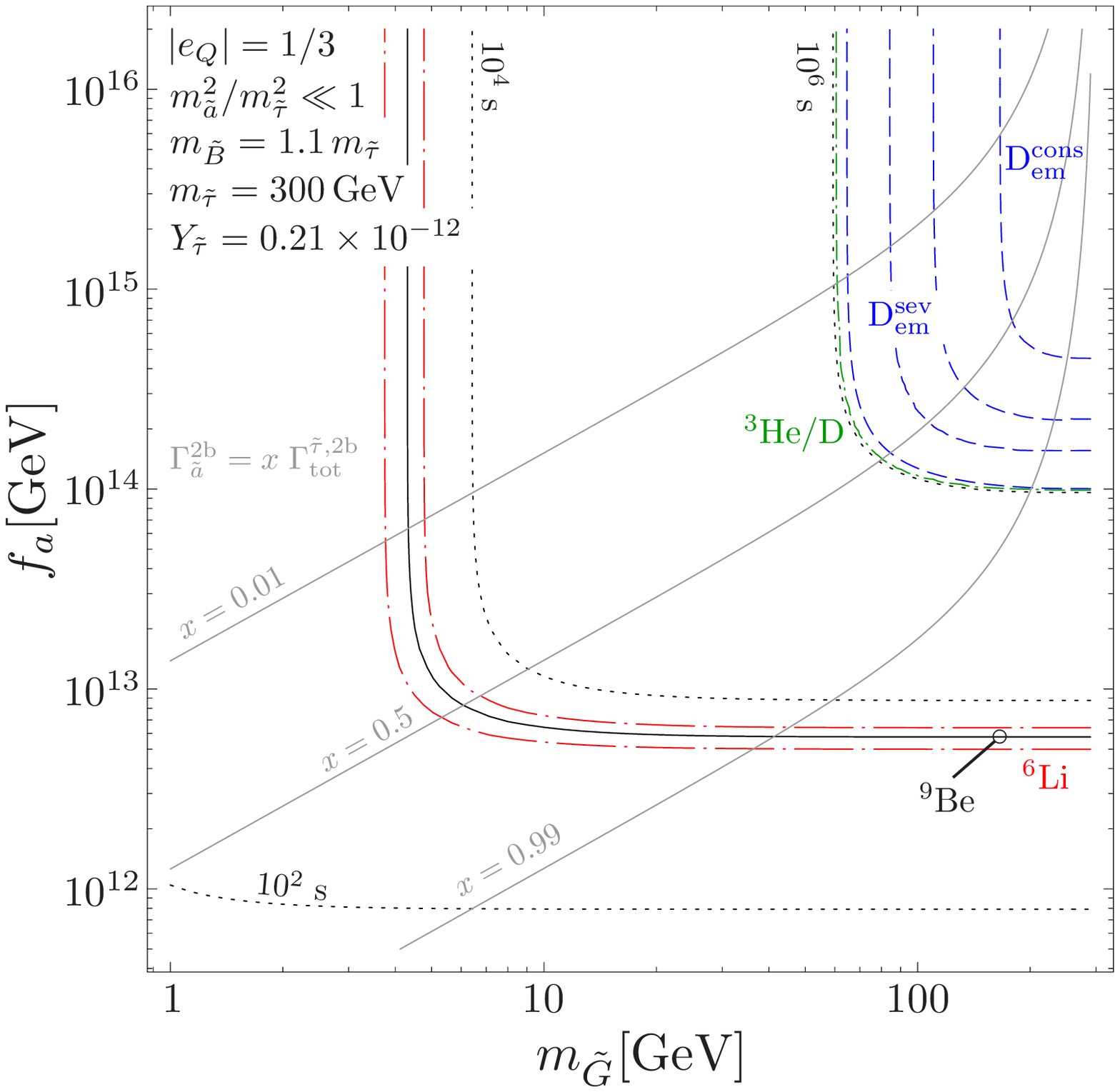, height=6.9cm}
\hfill
\epsfig{figure=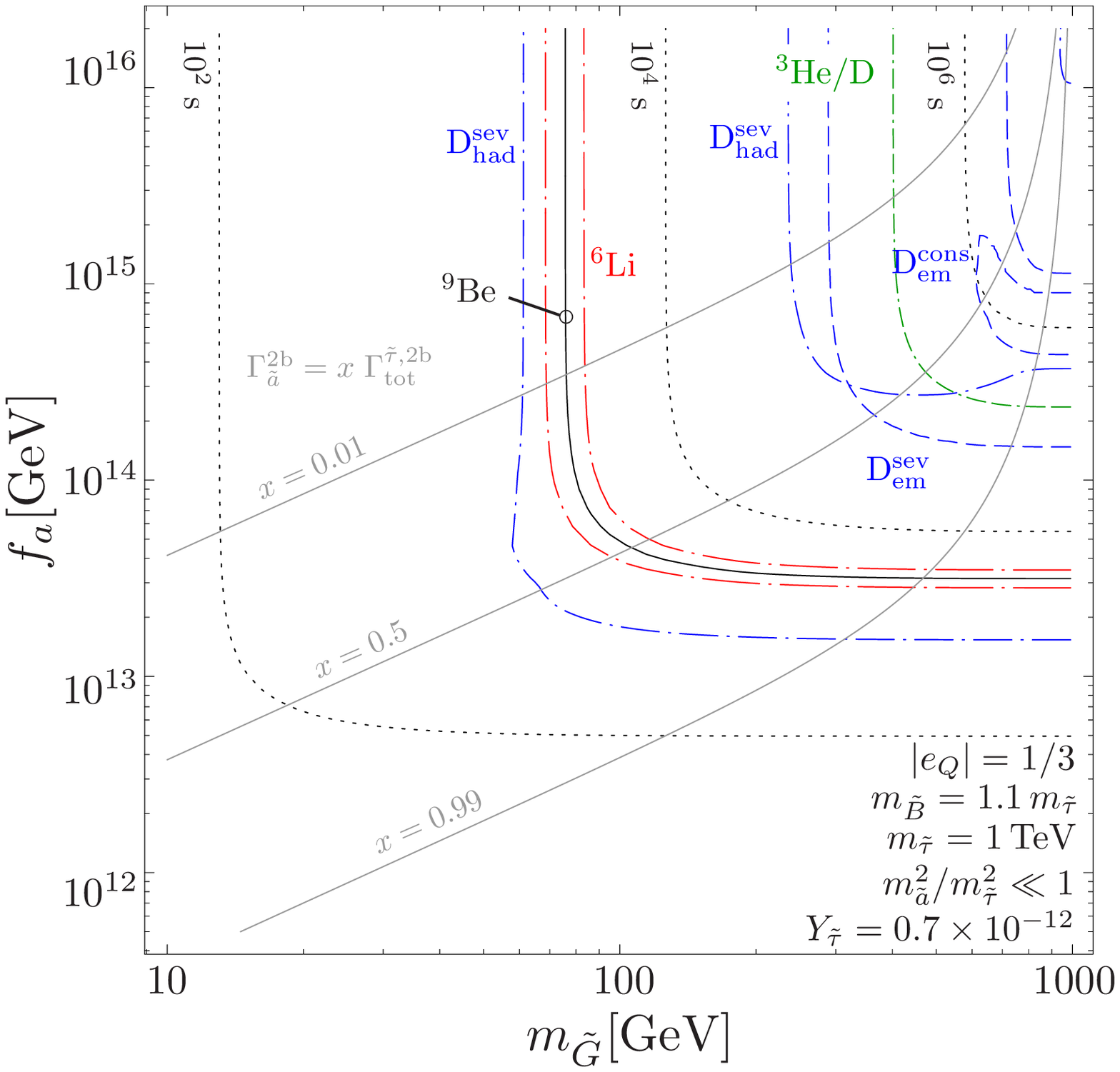, height=6.9cm}
\\
\hspace*{3cm}
\makebox[2cm][c]{(a)}\hfill
\hspace*{4cm}
\makebox[2cm][c]{(b)}\hfill
\caption{Cosmological constraints on $f_a$ and $\mgravitino$ for
  (a)~$\mstau=300~\GeV$ and (b)~$\mstau=1~\TeV$ and
  $\maxino^2/\mstau^2\ll 1$, $|e_Q|=1/3$, $y=1$, $\mbino=1.1\,\mstau$
  and $Y_{\stau}=0.7\times 10^{-12}\,(\mstau/1~\TeV)$. The regions
  above the long-dash-dotted (red) and the solid lines
  ($\Lisix,\,\Benine$) are disfavoured by the CBBN constraints
  associated with \refeq{Eq:LiSix} and \refeq{Eq:BeNine}. No hadronic
  BBN constraints appear in panel~(a). In panel~(b) the ones
  associated with \refeq{Eq:D_sev} disfavour the region between the
  two short-dash-dotted (blue, $\Deut_{\HAD}^{\mathrm{sev}}$) lines.
  Electromagnetic BBN constraints associated with
  $^3\mathrm{He}/\mathrm{D}$ disfavour the regions above the
  double-dash-dotted (green) line. In panel~(a) the electromagnetic
  BBN constraints associated with $\mathrm{D}$ disfavour the regions
  between the two leftmost dashed (blue, $\Deut_{\EM}^{\mathrm{sev}}$)
  lines and to the right of the corresponding 3rd line from the
  left-hand side or the region above the rightmost of those lines
  ($\Deut_{\EM}^{\mathrm{cons}}$). In panel~(b) the corresponding
  disfavoured regions are the ones to the right of the leftmost dashed
  (blue, $\Deut_{\EM}^{\mathrm{sev}}$) lines and the ones to the right
  of the corresponding rightmost (broken,
  $\Deut_{\EM}^{\mathrm{cons}}$) line. Contours of
  $\tau_{\stau}=10^2$, $10^4$, and $10^6\,\seconds$ are shown by the
  dotted lines, and contours of $x\equiv\Gamma_\axino^{\rm
    2b}/\Gamma_{\tot}^{\stau,{\rm
      2b}}[\simeq{\mathrm{BR}}(\stauR\to\tau\axino)]=0.01$, $0.5$ and
  $0.99$ by the solid grey lines, as in \reffig{fig:combltwidthdiff}.}
\label{fig:famaxgravaxNoCoeQ13}}
%
For orientation, we show contours of $\tau_{\stau}=10^2$, $10^4$, and
$10^6\,\seconds$ (dotted lines) and of $x\equiv\Gamma_\axino^{\rm
  2b}/\Gamma_{\tot}^{\stau,{\rm
    2b}}[\simeq{\mathrm{BR}}(\stauR\to\tau\axino)]=0.01$, $0.5$ and
$0.99$ (solid grey lines).
In the region with $x\gtrsim 0.99$, the (C)BBN constraints are
insensitive to $\mgravitino$ and basically as in the
$\maxino<\mstau<\mgravitino$ case considered in 
\refsec{Sec:NucleosynthesisConstraints}; 
see also Fig.~5~(a) of Ref.~\cite{Freitas:2009jb}.
In turn, in the region with $x\lesssim 0.01$, the shown constraints
are insensitive to $f_a$ and basically as in the $\gravitino$ LSP
cases with $\mgravitino<\mstau<\maxino$ or no PQ extension; cf.\ 
Fig.~6 of Ref.~\cite{Pospelov:2008ta}.

Let us now describe the (C)BBN constraints shown in
\reffig{fig:famaxgravaxNoCoeQ13}.
The regions to the right of and above the long-dash-dotted (red) and
the solid lines are disfavoured by the CBBN constraints associated
with \refeq{Eq:LiSix} and \refeq{Eq:BeNine}. Again, these lines are
located close to the $\tau_{\stau} \approx 5\times 10^3\,\seconds$
contour (not shown).
The hadronic BBN constraints associated with \refeq{Eq:D_sev} show up
in panel~(b), where they disfavour the region enclosed in the L-shaped
strip bounded by the two short-dash-dotted (blue) lines.  These severe
constraints do not appear in panel~(a), and also the conservative
constraints associated with \refeq{Eq:D_cons} are absent in both
panels, which is consistent with Fig.~5~(a) of
Ref.~\cite{Freitas:2009jb} and Fig.~6 of Ref.~\cite{Pospelov:2008ta}.
The electromagnetic BBN constraints lie well within the region already
disfavoured by CBBN in our panels~(a) and~(b). Here the
$^3\mathrm{He}/\mathrm{D}$ constraint disfavours the regions above the
double-dash-dotted (green) line. Moreover, in panel~(a) the
$\mathrm{D}$ constraint disfavours the regions between the two
leftmost dashed (blue, $\Deut_{\EM}^{\mathrm{sev}}$) lines and to the
right of the corresponding 3rd line from the left-hand side or the
region above the rightmost of those lines
($\Deut_{\EM}^{\mathrm{cons}}$), whereas in panel~(b) the regions to
the right of the leftmost dashed (blue, $\Deut_{\EM}^{\mathrm{sev}}$)
lines and the ones to the right of the corresponding rightmost
(broken, $\Deut_{\EM}^{\mathrm{cons}}$) line are disfavoured.

We find that the upper limits on $\mgravitino$, which have been
obtained for a given $\mstau$ in the gravitino LSP case and which
imply upper limits on the SUSY breaking
scale~\cite{Steffen:2006wx,Pospelov:2008ta}, hold only in the
parameter region with $x\ll 0.5$, i.e., for
$\Gamma(\stauR\to\tau\axino)\ll\Gamma(\stauR\to\tau\gravitino)$.
For $x\gg 0.5$, the gravitino mass range
$0.1\,\mstau\lesssim\mgravitino\lesssim\mstau$ 
becomes cosmologically viable with a standard thermal history,
R-parity conservation and a standard freeze-out yield $Y_{\stau}$.
Despite a then relatively small branching ratio of the
$\stauR\to\tau\gravitino$ decay, this might still be relevant for
collider phenomenology since a kinematic $\mgravitino$ determination
appears to be feasible only in that mass
range~\cite{Buchmuller:2004rq,Martyn:2006as,Hamaguchi:2006vu} which is
disfavoured otherwise~\cite{Steffen:2006wx,Pospelov:2008ta}.
On the other hand, our upper limits on $f_a$ obtained for a given
$\mstau$ and $\mbino$~\cite{Freitas:2009fb,Freitas:2009jb} hold only
in the parameter region with $x\gg 0.5$, i.e., for
$\Gamma(\stauR\to\tau\axino)\gg\Gamma(\stauR\to\tau\gravitino)$.
If $x\ll 0.5$, $f_a$ values even as large as
$M_{\mathrm{GUT}}\simeq 2\times 10^{16}\,\GeV$
will not be cosmologically disfavoured.

So we see that the $\maxino,\mgravitino<\mstau$ scenario opens up
regions of parameter space that are disfavoured 
if only the axino~\cite{Freitas:2009fb,Freitas:2009jb} or only 
the gravitino~\cite{Steffen:2006hw,Steffen:2006wx,Pospelov:2008ta}
is lighter than the stau.
Accordingly, without additional phenomenological insights into
$\mgravitino$ or $f_a$ from collider experiments or axion searches, it
will not be possible to infer rigid upper limits separately on $f_a$
or $\mgravitino$ once $\mstau$ and $\mbino$ are known, even under the
assumption of a standard thermal history.
In fact, for known $\mstau$ and $\mbino$ and with $Y_{\stau}$ inferred
for a standard thermal history with $\TR>\Tf^{\stau}$, we think that a
figure such as the ones shown in \reffig{fig:famaxgravaxNoCoeQ13} will
be the ideal way to display the possible (C)BBN constraints on $f_a$
and/or $\mgravitino$ for a given axion model.

Before proceeding we also note that some of the constraints disfavour
regions for $\maxino,\mgravitino<\mstau$ that were previously free of
constraints. For example, the electromagnetic and hadronic BBN
constraints in the $\gravitino$ LSP cases with
$\mgravitino<\mstau<\maxino$ or no PQ extension disappear for a finely
tuned $\mgravitino$--$\mstau$ degeneracy leading to
$\epsilon_{\EM/\HAD}\to 0$. For $\maxino,\mgravitino<\mstau$ and no
additional finely tuned $\maxino$--$\mstau$ degeneracy, this is no
longer the case since $\epsilon_{\EM/\HAD}$ will be governed
by~\eqref{Eq:epsEM} and~\eqref{eq:epshad} which can remain sizeable.
Accordingly, the $\mathrm{D}$ and $^3\mathrm{He}/\mathrm{D}$
constraints shown in \reffig{fig:famaxgravaxNoCoeQ13} do not disappear
for $\mgravitino\to\mstau$.

\subsection{Collider Phenomenology}
\label{Sec:ColliderPhenomenologyAxGrSt}

Not only the cosmological constraints but also collider phenomenology
is affected by the possible mass hierarchy
$\maxino,\mgravitino<\mstau$ considered in this section.
The methods to probe $\mstau$ and $\tau_{\stau}$ will remain the same.
Also our estimates of $N_{\stau}^{{\rm stopped}}$, $N_{\stau}^{{\rm
    hcal}}$ and $N_{\stau}^{{\rm yoke}}$ in \refsec{sec:trappedstaus}
will not change for the assumed $\tau_{\stau}$ values.
However, the experimental distinction between the axino LSP and the
gravitino LSP faces additional complications.
While brief comments on this issue had already been given in Secs.~1
and~4.3 of~\cite{Brandenburg:2005he}, we now address this issue more
extensively and in more detail, with a number of additional insights.

Even with the axino being the LSP, a short stau lifetime of
$\tau_{\stau}<10^{-7}~\seconds$ is now possible for
$\maxino<\mgravitino<40~\keV$ due to
$\Gamma(\stauR\to\tau\gravitino)\gg\Gamma(\stauR\to\tau\axino)$.
Then such $\tau_{\stau}$ will no longer provide a hint against the
axino LSP (cf.\ \refsec{Sec:DistAxGrav}) but point to a light gravitino, 
with the possibility of an even lighter axino.
Moreover, we have shown explicitly in \reffig{fig:comb3and4bodyBR}~(a)
that both
$\Gamma(\stauR\to\tau\axino\gamma)\gg\Gamma(\stauR\to\tau\gravitino\gamma)$ 
and 
$\Gamma(\stauR\to\tau\axino\gamma)\ll\Gamma(\stauR\to\tau\gravitino\gamma)$ 
is possible, whereas the corresponding regions (i.e., $x\gg 0.5$ and
$x\ll 0.5$) in parameter space are mainly governed by $\mgravitino$ 
and $f_a$.
Accordingly, double differential distributions such as the ones shown
in the top panels of \reffig{fig:ddiff3body} do not necessarily point
to the axino LSP if $\maxino,\mgravitino<\mstau$. Instead, they then
would point to a parameter region such as the one with $x\gg 0.5$ in
\reffig{fig:comb3and4bodyBR}~(a), on which cosmological constraints
can be imposed as shown in \reffig{fig:famaxgravaxNoCoeQ13}.
The 3-body decays can thereby still provide very valuable information.
Since the region with $x\gg 0.5$ in \reffig{fig:comb3and4bodyBR}~(a)
basically coincides with the $x\gg 0.5$ region in
\reffig{fig:combltwidthdiff}, an experimental signature for
$\Gamma(\stauR\to\tau\axino\gamma)\gg\Gamma(\stauR\to\tau\gravitino\gamma)$
implies
$\Gamma(\stauR\to\tau\axino)\gg\Gamma(\stauR\to\tau\gravitino)$
and thereby that the stau decays into the axino govern $\tau_{\stau}$.
This would mean that a $\tau_{\stau}$ measurement would probe the PQ
scale $f_a$, as outlined in~\eqref{eqn:fa} for the considered SUSY
axion models.
In turn, the opposite case with $x\ll 0.5$ would mean that a
$\tau_{\stau}$ measurement would probe the gravitino mass
$\mgravitino$ (and thereby the SUSY breaking scale), provided
$\MPl=2.4\times10^{18}\,\GeV$ as inferred from Newton's constant.

Note that a $\tau_{\stau}$ measurement alone
will point to a $\tau_{\stau}$ contour in
the $\mgravitino$--$f_a$ plane; cf.\ dotted lines in
\reffigs{fig:combltwidthdiff} or~\ref{fig:famaxgravaxNoCoeQ13}. Such a
$\tau_{\stau}$ contour can then be confronted with cosmological
constraints. However, as can be seen in
\reffig{fig:famaxgravaxNoCoeQ13}, the most restrictive and thereby
relevant constraints basically follow the shape of the $\tau_{\stau}$
contours. Thus, the cosmological constraints cannot tell us whether
$\tau_{\stau}$ is governed by stau decays into the axino or the
gravitino or by both. It will then still be unclear whether a
$\tau_{\stau}$ measurement probes $f_a$ or $\mgravitino$ or a
combination of both.

Axion searches and resulting limits on $f_a$ may help in particular
for small values of $\tau_{\stau}$. Already the current $f_a$ 
limit~\eqref{Eq:f_a_axion} will point to
$\Gamma(\stauR\to\tau\gravitino)\gg\Gamma(\stauR\to\tau\axino)$
if $\tau_{\stau}\lesssim 10^{-7}$--$10^{-4}\,\seconds$, where the
given range of the upper $\tau_{\stau}$ limit reflects an uncertainty
with respect to the axion model and here the parameters $e_Q$ and $y$;
cf.~\refsec{Sec:DistAxGrav}.
In the future, the lower $f_a$ limit may increase and thereby also the
above values of the $\tau_{\stau}$ limit.
Alternatively, a future axion discovery may point to an $f_a$ value
and also to an $e_Q$ value (cf.\ Sec.~6 of~\cite{Freitas:2009fb}).
Those values together with measured $\mstau$, $\mbino$ and
$\tau_{\stau}$ can then be tested with relation~\eqref{eqn:fa}.  For a
reasonable $y$ value, which appears only in the logarithm and which
will be difficult to probe independently of the stau decays (cf.\ end
of \refsec{sec:trappedstaus}), an agreement would then point to
$\Gamma(\stauR\to\tau\axino)\gg\Gamma(\stauR\to\tau\gravitino)$
and thereby to the $\mgravitino$-independent part of the
$\tau_{\stau}$ contour; cf.\ dotted lines for $x\gg 0.5$ in
\reffig{fig:combltwidthdiff} or~\ref{fig:famaxgravaxNoCoeQ13}.
In turn, a severe mismatch in~\eqref{eqn:fa} may originate from
$\Gamma(\stauR\to\tau\gravitino)\gg\Gamma(\stauR\to\tau\axino)$
for which $\tau_{\stau}$ is governed by $\mgravitino$ and independent
of $f_a$; cf.\ dotted lines for $x\ll 0.5$ in
\reffigs{fig:combltwidthdiff} or~\ref{fig:famaxgravaxNoCoeQ13}.

Additional information might be gleaned from the kinematics 
of the 2-body decays $\stauR\to\tau\axino/\gravitino$. 
If $\maxino$ and $\mgravitino$ are sufficiently different, 
two isolated peaks may be observed in the energy spectrum 
of the tau, as given by \eqref{Eq:EtauAxGRSt}. 
The heights of the peaks are goverend by 
$\Gamma(\stauR\to\tau\axino)$ and
$\Gamma(\stauR\to\tau\gravitino)$. 
This information, together with~\eqref{Eq:StauLifetimeGravitinoLSP}, could in principle allow us to determine which of the two peaks corresponds to the gravitino, and thus whether the axino or the gravitino is the LSP. This procedure will not work, however, if one of the decay channels is strongly suppressed.

While this may sound encouraging, one faces the complication of the
short tau lifetime and that the decay products of the tau always
include neutrinos. Thus, the two $E_\tau^{\axino/\gravitino}$ values
will not show up in the form of two basically monochromatic peaks but,
e.g., as tau jet energy distributions, which will be further smeared
out by the finite detector energy resolution as shown, 
for example, in Fig.~8 of Ref.~\cite{Hamaguchi:2006vu}.  
Because of such complications, it had
been pointed out before that kinematical determinations of
$\maxino$\cite{Brandenburg:2005he} and
$\mgravitino$~\cite{Buchmuller:2004rq} will be feasible only for
$0.1~\mstau\lesssim m_{\axino/\gravitino}<\mstau$. However, maybe it
is not the stau $\stauR$ that is the lightest MSSM superparticle but
the right-handed selectron $\selectronR$ which would then decay into
an electron instead of the tau:
$\selectronR\to\mathrm{e}\,\axino/\gravitino$.  This would facilitate
the above also for smaller $m_{\axino/\gravitino}$ since the
corresponding two discrete values of
$E_\mathrm{e}^{\axino/\gravitino}$ would be deposited fully in the
calorimeters. In the absence of a degeneracy of the axino and
gravitino masses, one may then observe the two decay modes very
distinctively and thereby identify the LSP and the decay(s) that
govern $\tau_{\stau}$.

\section{Conclusions}
\label{Sec:Conclusions}

Supersymmetric Peccei--Quinn models, where both the strong CP problem
and the hierarchy problem are solved simultaneously, are compelling
also from the cosmological point of view. They include attractive
candidates for dark matter. On the one hand, dark matter can just
reside in the form of
axions~\cite{Raffelt:2006cw,Sikivie:2006ni,Kim:2008hd,Nakamura:2010zzi,Graf:2010tv}.
On the other hand, the axino $\axino$ can be the stable LSP in an
R-parity-conserving realisation of SUSY and as such be another natural
dark matter
candidate~\cite{Bonometto:1989vh,Rajagopal:1990yx,Bonometto:1993fx,Covi:1999ty,Covi:2001nw,Covi:2004rb,Brandenburg:2004du,Baer:2008yd,Freitas:2009fb,Strumia:2010aa}.
In this work we have studied in detail the cosmological effects 
and potential collider phenomenology of such a model, 
in which the axino is the LSP and a charged slepton/stau the NLSP.

We focus on hadronic KSVZ axion
models~\cite{Kim:1979if,Shifman:1979if,Frere:1982sg,Kim:1983ia}, 
where the axion supermultiplet couples to the MSSM fields only 
via loops of very heavy (s)quarks, 
whose mass is of the order of the breaking scale $f_a$ 
of the PQ symmetry. 
In \refsec{Sec:NLSPDecays} we have calculated the 2-body decay
$\stau_{\mathrm R}\to\tau\axino$, which is induced only at the
two-loop level. Here the large mass hierarchy between the KSVZ fields
and the MSSM fields has allowed us to use the heavy mass expansion
technique. Due to a large logarithm, the rate of this decay mode is
large and dominates all other modes by a factor of 100 or so.
We have also calculated differential rates for the 3-body decay
$\stau_{\mathrm R}\to\tau\axino\gamma$,
which is relevant for collider prospects, 
and 4-body decays such as
$\stau_{\mathrm R}\to\tau\axino q\bar{q}$,
which are relevant for BBN constraints.
Explicit expressions are given for rates and distributions, and their
dependencies on the involved parameters have been illustrated.
For the natural mass range of the stau NLSP (and the other MSSM
sparticles), its lifetime can be typically of the order of
$0.01~\seconds$ to thousands of seconds.

In \refsec{Sec:AxinoConstraints} we have aimed at a detailed and
complete account of the cosmological constraints on axino dark matter
scenarios with such a long-lived charged slepton. Covering a wide
range of the axino mass $\maxino$ and of the reheating temperature
after inflation $\TR$, various possible axino populations have been
considered: thermal relic axinos, thermally produced axinos and
non-thermally produced axinos from NLSP decays. We have explored the
constraints on those populations imposed by the present dark matter
density, by structure formation 
(and associated limits on the corresponding present free-streaming
velocity)
and by bounds on non-standard contributions to the radiation density,
$\DNueff$, at the onset of BBN. At present, we find that the BBN
bounds on $\DNueff$ are such that they do not impose relevant
restrictions on the considered scenarios. This is different for the
dark matter density which leads to upper limits on $\TR$, $\maxino$
and $\mstau$. The detailed form of these limits depends significantly
on $f_a$ and also on the other parameters as explored and discussed 
in~\refsec{Sec:AxinoConstraints}.

Other BBN constraints relate to the primordial stau NLSP
abundance after freeze out and prior to decay. Here we have studied
various scenarios with and without stau--slepton coannihilation, 
stau--bino coannihilation and
simultaneous stau--bino--slepton coannihilation,
leading to different values of the stau yield $Y_{\stau}$.
For $f_a\gtrsim 10^{12}\,\GeV$, 
the stau decays so late that its decay products can
reprocess BBN or that negatively charged staus can form bound states
leading to catalysed BBN (CBBN) of $\Lisix$ and $\Benine$. We have
shown explicitly that the resulting constraints from late hadronic
energy injection and from CBBN are then often the most relevant ones
disfavouring significant regions in parameter space.

Those BBN constraints also apply to axino hot dark matter scenarios
with a charged slepton NLSP and an axion condensate providing the cold
dark matter~\cite{Brandenburg:2004du}.  For such light hot thermal
relic axinos, structure formation furthermore imposes the limit
$\maxino\lesssim 37~\eV$.
Assuming instead that (almost) all dark matter resides in
the form of non-thermally produced axinos, the constraints
imposed by structure formation considerations
can even disfavour $\maxino$ smaller than $1$--$100~\GeV$.

The precise position of the cosmological constraints depends not only
on $f_a$, $\mstau$ and the other parameters but also on the adopted
limits on the free-streaming velocity or the abundances of primordial
nuclei. Since both types of limits rely on astrophysical studies,
which are still subject to ongoing research, we have considered
conservative and restrictive limits when presenting the constraints
from structure formation and BBN.
Taking even the most restrictive limits into account, we find that
axino dark matter scenarios with a long-lived stau remain a viable
possibility that can be compatible with the cosmological constraints,
in particular, for $f_a\lesssim 10^{12}~\GeV$ and $\mstau$ within the
discovery reach of the LHC and a future linear collider such as the
ILC.

In \refsec{Sec:Collider} we have explored the prospects to study
scenarios with the axino LSP and a stau NLSP in collider experiments.
Evaluating the stau decay length in the axino LSP case, we find that
it is typically much larger than the detector size. While most staus
(if produced) will thus decay outside of the detectors, a fraction of
slow staus can get stopped in the main detectors or in
some additional dedicated stopping material.
Those stopped staus could then allow us to study their decays to determine their lifetime $\tau_{\stau}$ and the axino mass $\maxino$, 
to probe $f_a$ and to distinguish 
between axino LSP and gravitino LSP.

To quantify the collider prospects in a representative way, we have
studied four benchmark scenarios with CMSSM spectra that cover the
aforementioned coannihilation scenarios. We have calculated the stau
production rates and the number of staus expected to get stopped in a
typical detector at the LHC and the ILC. We find that with moderate
amounts of luminosity, several hundred staus may be stopped at the
LHC, which would allow one to determine their lifetime with about 10\%
uncertainty. At the ILC, these numbers improve to several thousand
stopped staus and an estimated uncertainty of about 3\%, respectively.
Higher statistics are conceivable with additional dedicated stopping
detectors at the LHC or by tuning the beam energy at the ILC.

While the stau lifetime and mass can be measured fairly precisely, the
determination of the properties of the LSP that it decays into is more
challenging. We have addressed the question whether it is possible to
experimentally distinguish an axino LSP from an alternative scenario
with a gravitino LSP. A short decay time 
$\tau_{\stau}\lesssim 10^{-7}$--$10^{-4}\,\seconds$ 
may point toward the gravitino LSP case, but in general a
$\tau_{\stau}$ measurement alone cannot clearly discriminate between
an axino LSP and a gravitino LSP.
On the other hand, the 3-body decays
$\stauR\to\tau\gamma\axino/\gravitino$ can help to distinguish between
these cases for large parts of the parameter space. However, the
distinction will also become challenging in this channel for a heavy
bino or large values of the ratio $f_a/\mstau$.

Finally, we have studied scenarios in which both the axino and the
gravitino are lighter than the charged slepton/stau that is assumed to
be the lightest MSSM sparticle. Such a mass hierarchy is well
motivated and may occur, e.g., in models with gauge-mediated SUSY
breaking~\cite{Asaka:1998ns,Asaka:1998xa}. We have shown explicitly
the way in which the simultaneous existence of a light axino and a
light gravitino affects the 2-body, 3-body and 4-body stau decays and
thereby the stau lifetime, cosmological constraints and collider
prospects. Our findings show that $\tau_{\stau}$ can now be much
smaller and the cosmological constraints can be much weaker than
expected without a second extremely weakly interacting sparticle being
lighter than the MSSM sparticles. In particular, $f_a$ values even as
large as $M_{\mathrm{GUT}}\simeq 2\times 10^{16}\,\GeV$ -- which are
disfavoured for $\maxino<\mstau<\mgravitino$~\cite{Freitas:2009jb} --
can now be allowed for $\mgravitino\lesssim 10$--$100~\GeV<\mstau$.
Alternatively, the gravitino mass range
$0.1\,\mstau\lesssim\mgravitino<\mstau$ can become cosmologically
viable with a standard thermal history and a standard freeze-out yield
$Y_{\stau}$ if $f_a\lesssim 10^{13}\,\GeV$ and $\maxino<\mstau$.

For a scenario with $\maxino,\mgravitino<\mstau$, the stau lifetime
depends on the PQ scale $f_a$ as well as on the gravitino mass
$\mgravitino$ (and the Planck mass $\MPl$). Thus a measurement of
$\tau_{\stau}$ yields only ambiguous information. We have shown that
this ambiguity may be resolved by extracting the branching ratios of
decays into axinos and into gravitinos.
Here we have proposed two complementary approaches one relying on a
study of the 3-body decays $\stauR\to\tau\gamma\axino/\gravitino$ and
the other on a study of the 2-body decay kinematics.
In fact, the outlined kinematical considerations may clarify 
also the mass hierarchy between the axino and the gravitino.
In practice this requires the
identification of a double-peak feature in the tau energy spectrum,
which however will only be feasible if $|\mgravitino-\maxino|$ is
relatively large.  Such an analysis would become much simpler with
electrons in the case that the NLSP is a selectron instead of a stau.
It remains to be seen how far detector technology can be driven in
order to study the various aspects of late decays of charged sleptons.

In the present exciting time in which the LHC experiments are taking
data, there are high hopes for nearby discoveries of new physics and
new particles. In fact, maybe a long-lived charged massive particle
will be seen soon in the detectors. Then our detailed study presented
in this work may become very relevant. This will be even more so if in
addition also the axion is discovered in ongoing or future axion
searches. It may then very well be axions and axinos that make up the
dark matter in our universe.


\section*{Acknowledgements}

We are grateful to P.~Graf, K.~Hamaguchi, J.~Lindert and S.~Schilling
for valuable discussions.
This research was partially supported by the Cluster of Excellence
`Origin and Structure of the Universe' and by the Swiss National
Science Foundation (SNF).

\begin{appendix}

\section{Heavy Mass Expansion and Reduction to Basic Two-Loop Integrals}
\label{Sec:HME}

\subsection{Heavy Mass Expansion}

Heavy mass expansion (HME) is an asymptotic expansion in small masses
and momenta \cite{Smirnov:1990rz,Smirnov:1994tg}. It applies in the
cases where the following assumptions can be made
\bit
\item each mass can be categorised as either large, ${M_i}$, or
small, ${m_i}$;
\item all external momenta are small compared to the large masses,
${k_i}\ll {M_i}$.  
\eit 
For a Feynman integral $\mathcal{F}$ of Feynman diagram, $\Gamma$ the
expansion is compactly written as
\bee
\mathcal{F}(\Gamma) \rightarrow \sum_\gamma
\mathcal{F}(\Gamma\backslash\gamma) * \mathcal{T}_{\lbrace k_i, m_i
\rbrace} \mathcal{F}(\gamma)
\eee 
where $\gamma$ represents the subgraphs over which the sum is
performed (a subgraph is derived by drawing all possible one-particle
irreducible graphs containing the heavy propagators). The operator
$\mathcal{T}$ performs a Taylor expansion on the subgraphs with
respect to the small parameters. $\mathcal{F}(\Gamma\backslash\gamma)$
represents the original graphs where the subgraphs have been shrunk to
a point.

When considering processes involving loops, we must pay special care
to the loop momenta $q_i$. Since these are unconstrained by momentum
and energy conservation, they can be both of the order of the small
parameters and of the large masses. In order to carry out the
expansion in a consistent manner, we must look at each of the momenta
regions, individually.

For the two-loop processes considered in this paper, there are four
such regions. Let us refer to the diagrams in \reffig{fig:2body2loop}
and denote the momentum running through the outer loop $q_1$, the
momentum running through the heavy loop $q_2$ and the heavy mass scale
$M$. Then these 4 regions are defined as follows:
\ben
    \item $ |q_1|,|q_2| \ll M $
    \item $ |q_1| \sim M, |q_2| \ll M $
    \item $ |q_1| \ll M, |q_2| \sim M $
    \item $ |q_1|,|q_2| \sim M $
\een
After the appropriate Taylor expansions, the first two regions reduce
to scaleless integrals, which vanish in dimensional regularisation.
This leaves only the third and fourth regions, which in practical
terms translate into the Taylor expansion of only the heavy loop and
the Taylor expansion of the whole diagram with respect to the small
parameters, respectively.
%
\FIGURE[t]{
\centerline{\epsfig{figure=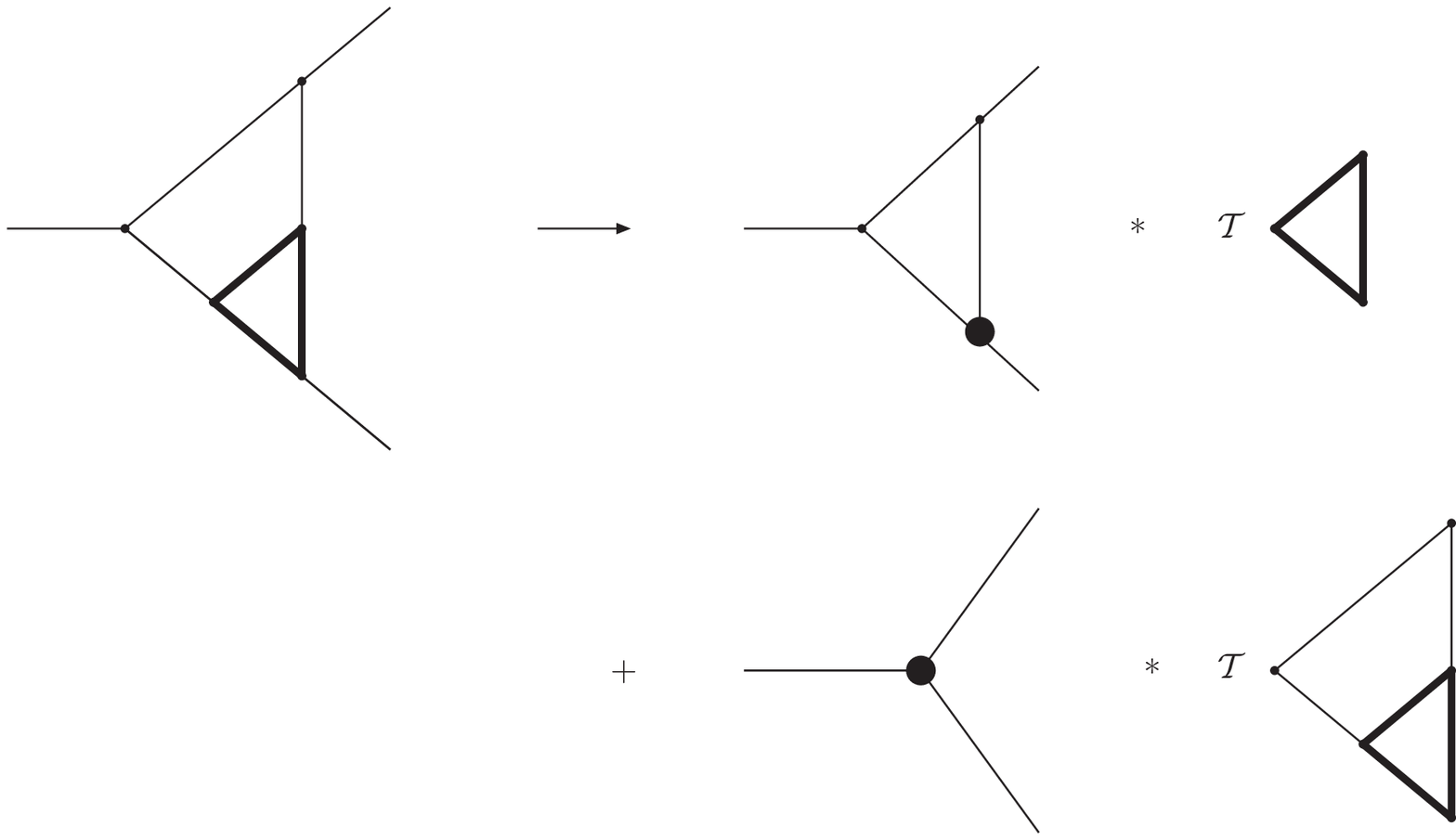, width=10cm}}
\caption{The leading term in the heavy mass expansion of the two-loop 2-body
decay of the stau NLSP into the axino LSP, where the propagators with
the large mass parameter $M$ are represented by thick lines and the
rest by thin lines.}
\label{fig:hme}}
%
\reffigure{fig:hme} gives a pictorial depiction of what is going on,
where the propagators with the large mass parameter $M$ are
represented by thick lines and the rest by thin lines.

\subsection{Tensor Reduction}

Using Passarino-Veltman reduction procedures, the tensor integrals were
simplified to scalar ones. Using the following notation,
\begin{align}
  T_0(n_1,n_2,n_3) &= \int d^D q_1 \, d^D q_2
    \frac{1}{(q_1^2-M_Q^2)^{n_1} (q_2^2-M_Q^2)^{n_2}(q_1+q_2)^{2n_3}} \, , \\
  T^{\mu_1 \dotsi \mu_n \nu_1 \dotsi \nu_n}(n_1,n_2,n_3) &= \int d^D q_1
    \, d^D q_2 \frac{q_1^{\mu_1}\dotsi q_1^{\mu_n}q_2^{\nu_1} \dotsi
    q_2^{\nu_n}}{(q_1^2-M_Q^2)^{n_1} (q_2^2-M_Q^2)^{n_2}(q_1+q_2)^{2n_3}} \, ,
\end{align}
we make use of the following relations in order to reduce our tensor
integrals:
\begin{align}
  T^{\mu_1\nu_1}(n_1,n_2,n_3) &=\frac{1}{D} g^{\mu_1\nu_1} q_1.q_2
  T_0(n_1,n_2,n_3), \\ 
  T^{\mu_1\mu_2\nu_1\nu_2}(n_1,n_2,n_3) &=\frac{1}{D^3+D^2-2D} \Bigl[ 
    \Bigl\lbrace (D+1)q_1^2q_2^2 - 2(q_1.q_2)^2\Bigr\rbrace
    g^{\mu_1\mu_2}g^{\nu_1\nu_2}\Bigr. \nonumber \\ 
  & \Bigl.+ \lbrace D(q_1.q_2)^2 - q_1^2 q_2^2 \rbrace
    (g^{\mu_1\nu_1}g^{\mu_2\nu_2} + g^{\mu_1\nu_2}g^{\mu_2\nu_1}) \Bigr]
    T_0(n_1,n_2,n_3) ,
\end{align}
where $g^{\mu\nu}$ is the metric tensor and $D$ is the number of dimensions.

\subsection{Integration by Part Identities}

After applying the heavy mass expansion and tensor reduction formulae,
we are left only with vacuum two-loop integrals and one-loop integrals
(the results of which are well-known). The two-loop integrals take the
form $T_0(n_1,n_2,n_3)$. These are simplified with the use of
integration by parts (IBP) identities, reducing the two-loop integrals
to master integrals.

Using the following identities,
\begin{align}
  \int d^D q_1 \, d^D q_2 \, \frac{\partial}{\partial q_1^\mu}
  \frac{q_1^\mu} {(q_1^2-M_Q^2)^{n_1}
    (q_2^2-M_Q^2)^{n_2}(q_1+q_2)^{n_3}} = 0 \\ 
  \int d^D q_1 \, d^D q_2 \, \frac{\partial}{\partial q_1^\mu} 
  \frac{q_2^\mu} {(q_1^2-M_Q^2)^{n_1}
  (q_2^2-M_Q^2)^{n_2}(q_1+q_2)^{n_3}} = 0 \\ 
  \int d^D q_1 \, d^D q_2 \, \frac{\partial}{\partial q_2^\mu} 
  \frac{q_1^\mu} {(q_1^2-M_Q^2)^{n_1}
  (q_2^2-M_Q^2)^{n_2}(q_1+q_2)^{n_3}} = 0 \\ 
  \int d^D q_1 \, d^D q_2 \, \frac{\partial}{\partial q_2^\mu} 
  \frac{q_2^\mu} {(q_1^2-M_Q^2)^{n_1}
  (q_2^2-M_Q^2)^{n_2}(q_1+q_2)^{n_3}} = 0
\end{align}
we reduced all our integrals to master integrals. This was implemented
using the Automated Integral Reduction program
\cite{Anastasiou:2004vj}. In fact, in the course of our calculation,
we only had one master integral, $T_0(1,1,0)$, where the two-loop
integral decouples into a product of two one-loop integrals.

\end{appendix}




\end{document}